\documentclass{aa}  
\usepackage{longtable,booktabs}
\usepackage[colorlinks=true,     linkcolor=blue, citecolor=blue, filecolor=blue, urlcolor=blue]{hyperref}
\usepackage{amsmath}
\usepackage{mwe}
\usepackage{tabularx}
\usepackage{natbib}
\bibpunct{(}{)}{;}{a}{}{,}
\setcitestyle{citesep={;}}
\usepackage{graphicx}
\usepackage{txfonts}
\usepackage{siunitx}
\usepackage{orcidlink}
\usepackage{latexsym}
\usepackage{amssymb}
\usepackage{lscape}
\usepackage{subfigure}
\usepackage{multicol}
\usepackage{multirow}
\usepackage{pdflscape}
\usepackage{textcomp}
\usepackage{caption,rotating}

\defcitealias{WolfC2024}{Wolf C. et al. 2024}
\defcitealias{WolfJ2024}{Wolf J. et al. (2024)}

\newcommand{\nqsos}{25} 

\makeatletter
\AtBeginDocument{\let\linenumbers\relax}
\makeatother

\begin{document} 

\title{Discovery and characterization of 25 new quasars at $4.6 < z < 6.9$ from wide-field multi-band surveys}

\author{Silvia~Belladitta\inst{1,2}\orcidlink{0000-0003-4747-4484}\, \and 
Eduardo~Ba\~nados\inst{1}\orcidlink{0000-0002-2931-7824}\, \and 
Zhang-Liang~Xie\inst{1,3}\orcidlink{0000-0002-2931-7824}\, \and 
Roberto~Decarli\inst{2}\orcidlink{0000-0002-2662-8803}\, \and
Silvia~Onorato\inst{4}\orcidlink{0009-0009-1715-4157}\, \and
Jinyi~Yang\inst{5,6}\orcidlink{0000-0001-5287-4242}\, \and
Manuela~Bischetti\inst{7,8}\orcidlink{0000-0002-4314-021X}\, \and
Masafusa~Onoue\inst{9,10,11}\orcidlink{0000-0003-2984-6803}\, \and
Federica~Loiacono\inst{2}\, \and
Laura N.~Martínez-Ramírez\inst{1,3,12,13}\orcidlink{0009-0003-5506-5469}
Chiara~Mazzucchelli\inst{14}\orcidlink{0000-0002-5941-5214}\, \and
Frederick B.~Davies\inst{1}\orcidlink{0000-0003-0821-3644}\, \and
Julien~Wolf\inst{1}\orcidlink{0000-0003-0643-7935}\, \and  
Jan-Torge~Schindler\inst{15}\orcidlink{0000-0002-4544-8242}\, \and 
Xiaohui~Fan\inst{5}\orcidlink{0000-0003-3310-0131}\, \and
Feige~Wang\inst{5,6}\orcidlink{0000-0002-7633-431X}\, \and
Fabian~Walter\inst{1}\orcidlink{0000-0003-4793-7880}\, \and
Tatevik~Mkrtchyan\inst{14}\, \and
Daniel~Stern\inst{16}\orcidlink{0000-0003-2686-9241}\, \and
Emanuele P.~Farina\inst{17}\orcidlink{0000-0002-6822-2254}\,
\and
Bram P.~Venemans\inst{4}\orcidlink{0000-0001-9024-8322}
}
 
\authorrunning{Belladitta et al.}

\institute{Max-Planck-Institut f{\"u}r Astronomie, K{\"o}nigstuhl 17, D-69117, Heidelberg, Germany
\and
INAF — Osservatorio di Astrofisica e Scienza dello Spazio, via Gobetti 93/3, I-40129, Bologna, Italy
\and
Fakult{\"a}t f{\"u}r Physik und Astronomie, Universit{\"a}t Heidelberg, Im Neuenheimer Feld 226, 69120 Heidelberg, Germany
\and
Leiden Observatory, Leiden University, P.O. Box 9513, 2300 RA Leiden, The Netherlands
\and
Steward Observatory, University of Arizona, 933 N. Cherry Ave., Tucson, AZ 85721, USA
\and
Department of Astronomy, University of Michigan, 1085 S. University Ave., Ann Arbor, MI 48109, USA
\and
Dipartimento di Fisica, Universit{\'a} di Trieste, Sezione di Astronomia, Via G.B. Tiepolo 11, I-34131 Trieste, Italy
\and
INAF—Osservatorio Astronomico di Trieste, Via G. B. Tiepolo 11, I-34131 Trieste, Italy  
\and
Kavli Institute for the Physics and Mathematics of the Universe (Kavli IPMU, WPI), The University of Tokyo Institutes for Advanced Study, The University of Tokyo, Kashiwa, Chiba 277-8583, Japan
\and
Center for Data-Driven Discovery, Kavli IPMU (WPI), UTIAS, The University of Tokyo, Kashiwa, Chiba 277-8583, Japan
\and
Kavli Institute for Astronomy and Astrophysics, Peking University, Beijing 100871, China  
\and
Instituto de Astrofísica, Facultad de Física, Pontificia Universidad Católica de Chile Av. Vicu\~na Mackenna 4860, 7820436 Macul, Santiago, Chile
\and 
Millennium Institute of Astrophysics (MAS), Nuncio Monseñor Sótero Sanz 100, Providencia, Santiago, Chile
\and
N{\'u}cleo de Astronomía de la Facultad de Ingeniería, Universidad Diego Portales, Av. Ejército Libertador 441, Santiago, Chile
\and
Hamburger Sternwarte, Universit{\"a}t Hamburg, Gojenbergsweg 112, D-21029 Hamburg, Germany 
\and 
Jet Propulsion Laboratory, California Institute of Technology, 4800 Oak Grove Drive, Pasadena, CA, 91109, USA
\and
International Gemini Observatory/NSF NOIRLab, 670 N A’ohoku Place, Hilo, Hawai'i 96720, USA
}

   \date{Received: 29 March 2025; Accepted: 19 May 2025}


  \abstract
{Luminous quasars at $z>4$ provide key insights into the early Universe. 
Their rarity necessitates wide-field multi-band surveys to efficiently separate them from the main astrophysical contaminants (i.e., ultracool dwarfs). 
To expand the sample of high-$z$ quasars, we conducted targeted selections using optical, infrared, and radio surveys, complemented by literature-based quasar candidate catalogs. 
In this paper, we report the discovery of \nqsos\ new quasars at $4.6<z<6.9$ (six at $z\geq6.5$), with $M_{1450}$ between $-$25.4 and $-$27.0. 
We also present new spectra of six $z>6.5$ quasars we selected, but whose independent discovery has already been published in the literature.
Three of the newly discovered quasars are strong radio emitters (L$_{1.4~\rm GHz}$$=0.09-1.0\times$10$^{34}$~erg~s$^{-1}$ Hz$^{-1}$).  
Among them, one source at $z=4.71$ exhibits typical blazar-like properties, including a flat radio spectrum, radio-loudness $\sim$1000, and multi-frequency variability. 
It is also detected by SRG/eROSITA X-ray telescope (f$_{\rm 0.2-2.3keV} \sim 1.3\times10^{-13}$~erg~s$^{-1}$~cm$^{-2}$).
In addition, for seven $6.3<z<6.9$ quasars we present near-infrared spectroscopy and estimate the central black hole mass from their C\,$\rm IV$ and Mg\,$\rm II$ broad emission lines.
Their masses (log[M$_{\rm BH,MgII}$]$=8.58-9.14~\rm M_{\odot}$) and Eddington ratios ($\lambda_{\rm Edd,MgII}=0.74-2.2$) are consistent with other \textit{z}$>$6 quasars reported in the literature.
A $z = 6.3$ quasar exhibits a velocity difference of approximately $9000$\,km~s$^{-1}$ between the C$\rm IV$ and Mg$\rm II$ emission lines, making it one of the most extreme C$\rm IV$ outflows currently known. 
Additionally, the sample includes three high-ionization broad absorption line quasars. One of these quasars shows potential evidence of an extremely fast outflow feature, reaching $48\,000$\,km~s$^{-1}$. 
}


\maketitle

\section{Introduction}
\label{sec:intro}
Quasars are among the most luminous, non-transient sources in the sky. 
They can be studied at large cosmological look-back times across the entire electromagnetic spectrum---with the current records at $z\sim7.6$ ($\sim$0.7 Gyr after the Big Bang, e.g., \citealt{Banados2018,Yang2020,WangFeige2021}).  
The evolution of the quasar population across cosmic time provides critical insights into the co-evolution of supermassive black holes (SMBHs) and their host galaxies. 
In particular, high-$z$ quasars reveal both the rapid growth mechanisms of the first SMBHs \citep[e.g.,][]{WangFeige2021} and the early development of their galactic hosts \citep[e.g.,][]{Neeleman2021,Wang2024}. 
Moreover, they provide key information on the chemical composition and metal enrichment of the intergalactic medium through intervening absorbers \citep[e.g.,][]{Becker2015a,Davies2023}, while also highlighting the densest cosmic environments \citep[e.g.,][]{Mignoli2020,Pudoka2024,Lambert2024} and they play a crucial role in our understanding of how active galaxies drive the cosmic re-ionization process \citep[e.g.,][for a recent review]{FanX2023ARA}.\\
In the past two decades, the exploitation of several optical and near-infrared (NIR) wide-area surveys, mostly covering the northern hemisphere, such as the Sloan Digital Sky Survey \citep[SDSS,][]{York2000} and the Panoramic Survey Telescope and Rapid Response System \citep[Pan-STARRS1, PS1,][]{Chambers2016} have enabled a drastic increase in the number of quasars discovered at $z>4$ \citep[e.g.,][]{Shen2011,Banados2016,Banados2023,Jiang2016,WangFeige2016,Caccianiga2019,Gloudemans2022,Belladitta2020,Belladitta2023}. 
The availability of recently released large-area sky surveys covering large fractions of the southern sky, such as the Dark Energy Survey \citep[DES,][]{Flaugher2005,Abbott2018}, the DESI Legacy Survey (DELS; \citealt{Dey2019}) and the SkyMapper (\citealt{Keller2007}) has made the discovery of high-$z$ quasars in the southern hemisphere possible \citep[e.g.,][]{Reed2017,Pons2019,Belladitta2019,WolfC2020,Onken2022,YangJ23,Ighina2023}.
The two most luminous high-$z$ quasars discovered so far have been identified in the southern sky. 
These are SMSS~J215728.21$-$360215.1 at $z$=4.75 \citep{WolfC2018} and SMSS~J052915.80$-$435152.0, at $z$=3.98 \citepalias{WolfC2024}. 
They have bolometric luminosities greater than 10$^{48}$~erg s$^{-1}$ and $z$-band magnitudes (AB system) in the Skymapper database of 17.11 and 16.04, respectively.\\
Most of the high-$z$ quasars discovered so far were selected using the Lyman-break technique (i.e., the dropout method, \citealt{Steidel1996}). 
Indeed, color-color selection techniques, which rely on multiwavelength broadband observations, are among the most commonly used methods to find high-$z$ quasars. 
The quasar flux at wavelengths shorter than the Ly$\alpha$ emission line (at rest-frame $\lambda_{\rm rf}$ = 1215.67~\AA) is absorbed by the intervening neutral medium, causing an extremely red $(r-i)$ or $(i-z)$ color if the source is at $z \sim$ 5 ($r$-dropouts) or at $\sim$ 6 ($i$-dropouts), respectively.
Recently, however, new selection methods based on machine learning techniques have led to the discovery of new high-$z$ quasars even in surveys that have been widely explored in the past \citep[e.g.,][see also \citealt{Calderone2024} for machine learning quasars selection at $z\sim2.5$]{Wenzl2021,YangDM2024,Byrne2024}.\\
Main results emerging from the discovery of all these high-$z$ quasars include: \textit{i)} extremely massive black holes ($>$10$^{8-9}$M$_{\odot}$) are observed in their centers \citep[e.g.,][]{Farina2022,Mazzucchelli2023}, comparable to the most massive black holes at any redshift, placing important constraints on the nature and growth of primordial black hole seeds \citep[e.g.,][]{Inayoshi2020}; \textit{ii)} they are found in gas-rich, massive, and highly star-forming host galaxies \citep[e.g.,][]{Decarli2018,Neeleman2021,Wang2024}; \textit{iii)} no significant redshift evolution for either broad UV emission lines or quasar continuum has been found \citep[e.g.,][]{Shen2019,Schindler2020,Yang2021,Jiang2024,Dodorico2023}; 
however, notable blueshifts have been identified in the C$\rm IV$$\lambda$1549 broad emission line (BEL), with shifts reaching up to $>$5000~km~s$^{-1}$. These blueshifts are significantly larger than those observed in lower redshift quasars, suggesting an evolution in quasar outflow properties over cosmic time \citep[e.g.,][]{Meyer2019}.
\textit{iv)} $\sim$10-15\% of these sources show a strong radio emission \citep[e.g.,][]{Banados2015,Liu2021,Gloudemans2021b,Keller2024}, i.e., they are classified as radio-loud (or jetted) Active Galactic Nuclei (AGN)\footnote{A source is considered to be radio-loud when the radio loudness (R) is $>$10, with R defined as the ratio of the 5~GHz and 4400\AA\ rest-frame flux densities: $R= S_{5\,\mathrm{GHz}}/S_{4400\AA}$ (\citealt{Kellermann1989}). The origin of the radio emission in radio-loud quasars is synchrotron radiation, which is produced by charged particles accelerated and collimated relativistically in a strong magnetic field, mostly along bipolar jets emitted from the central SMBH \citep[e.g.,][]{Bridle1984,Zensus1997}.}. 
High-$z$ radio-loud quasars are indispensable tools for studying the early evolutionary stage of the first jetted SMBHs \citep[e.g.,][]{Momjian2021}, their feedback on the host galaxy and the environment \citep[e.g.,][]{Hardcastle2020,Khusanova2022,Mazzucchelli2025}, and their contribution to the reionization epoch \citep[e.g.,][]{TorresAlba2020}.\\
Although the James Webb Space Telescope (JWST) has now spectroscopically confirmed more than 2000 galaxies \citep[e.g.,][]{CurtisLake2023,RobertsBorsani2024,Heintz2025..60H,DEugenio2025} and more than 100 UV-faint AGN (with bolometric luminosity around 10$^{41-44}$ erg~s$^{-1}$) at $z>4$ \citep[e.g.,][]{Harikane2023,Greene2024,MaiolinoXray2024,MaiolinoCensus2024,Taylor2024}, high redshifts quasars--being significantly more luminous than typical galaxies--serve as unparalleled tools for probing key properties of the Universe within its first billion years \citep[e.g.,][]{FanX2023ARA}. \\
To expand the population of high-$z$ quasars, we combine optical, infrared (IR), and radio photometric data sets to identify them throughout the sky. 
In addition, we present spectroscopic follow-up observations of high-$z$ quasar candidates available in published catalogs. 

\noindent
The paper is structured as follows: in Sect.~\ref{sec:selection}, we describe the several selections performed to identify high-$z$ radio-loud and radio-quiet quasar candidates; in Sect.~\ref{sec:observations}, we present the spectroscopic and photometric follow-up campaigns together with information about the data reduction. 
Results from these campaigns are described in Sec.~\ref{sec:results}, while in Sect.~\ref{sec:individualnotes}, we provide details and properties of some of the quasars presented in this work. 
Black hole mass estimates for a sub-sample of the quasars are reported in Sect.~\ref{sec:bhmass}.
Finally, we summarize the work in Sect.~\ref{sec:conc}. \\
The magnitudes reported in this work are all in the AB system (unless otherwise specified) and, when listed in the tables of the manuscript, have also been corrected for Galactic extinction using the extinction law provided by \cite{Fitzpatrick1999}, with R$_V$ = 3.1.
We used a flat $\Lambda$ cold dark matter ($\Lambda$CDM) cosmology with H$_0$ = 70~km~s$^{-1}$~Mpc$^{-1}$, \mbox{$\Omega_{\rm m} = 0.30$}, and $\Omega_{\Lambda}=0.70$.  
Radio spectral indices are given assuming S$_{\nu} \propto \nu^{-\alpha}$. 
All uncertainties are reported at 1$\sigma$.

\section{Candidates selection}
\label{sec:selection}
We identified the quasar candidates for further spectroscopic follow-up campaigns using several selection methods, as described below.

\subsection{[RPS1AW] Radio/PS1/AllWISE selection of \cite{Belladitta2023}}
The target PSO~J200$-$13 was selected as an $i$-dropout blazar candidate following the selection of \cite{Belladitta2023}. We refer to this selection with the acronym RPS1AW. 
For completeness, here we report a brief summary of the selection steps. 
We require detections in the NRAO VLA Sky Survey \citep[NVSS,][]{Condon1998} in the radio, PS1 in the optical, and the AllWISE Source Catalog (\citealt{Wright2010,Mainzer2011}) in the mid-infrared (MIR).  
From the entire NVSS catalog, we selected bright (S$_{1.4\rm GHz}$ $\geq$ 30~mJy) and compact objects to ensure a radio position accuracy better than $2\arcsec$. We then cross-matched these sources with the PS1 catalog using a maximum separation equal to $2\arcsec$. 
This impact parameter guarantees the recovery of more than 90\% of the real optical counterparts (\citealt{Condon1998}). 
We selected sources with $i_{\rm P1}<21.5$ outside the Galactic plane (|b|$\geq$~20$^\circ$) to minimize contamination from stars, and at Dec~$>-25^\circ$ to exclude optical false objects close to the declination limit of the Pan-STARRS survey. 
We additionally imposed the following criteria: \textit{(i)} no detection in $g_{\rm P1}$-band; \textit{ii)} drop-out: $r_{\rm P1} - i_{\rm P1}$ $\geq$1.2; \textit{iii)} blue continuum: $i_{\rm P1} - z_{\rm P1}$ $\leq$ 0.5; \textit{iv)} point-like sources: $i_{\rm P1}- i_{\rm Kron}$ $<$ 0.05; and \textit{v)} no detection in WISE (W2) or $i_{\rm P1}$ - $W2_{\rm Vega}$ $<$ 5. This last constraint has been placed to minimize the contamination by dust-reddened AGN at $z=1-2$ 
\citep[e.g.,][]{Caccianiga2019,Carnall2015}.
After applying these filters, 14 candidates remained;
5 of them are known radio quasars at $z=4.7-5.3$ (\citealt{Belladitta2023} and reference therein).
PSO~J200$-$13 was prioritized for spectroscopic follow-up, given its large $r_{\rm P1}-i_{\rm P1}\sim$ 1.6 color and strong radio emission (S$_{1.4\rm GHz} >$ 35 mJy). 
PS1 and WISE magnitudes of this object are reported in Table~\ref{tab:photomag}, together with NIR detections found in the VISTA Hemisphere Survey \citep[VHS,][]{McMahon2013} DR6 catalog.

\subsection{[PS1B23] PS1 $i$-dropout selection of \cite{Banados2023}}
\label{selectionbanados23}
The quasars PSO~J143$-$21 and PSO~J273$+$38 were recovered from the PS1 $i-$dropout selection described in  \cite{Banados2023}, requiring $z_{\rm P1} - y_{\rm P1}<0.5$ (see their Section 2.1). We use PS1B23 to allude to this selection. 
The main selection criteria are summarized here. 
We selected compact sources by requiring an absolute difference between the aperture and PSF magnitudes to be less than 0.3 in the $z_{\rm{P1}}$ or $y_{\rm P1}$ bands. Furthermore, we requested \textit{i)} no detection in $g_{\rm PS1}$, i.e., S/N$<3$; \textit{ii)} drop-out: $i_{\rm P1}-z_{\rm P1}>$2.0 or $i_{\rm P1,lim}-z_{P1}>$2.0, where $i_{\rm P1,lim}$ is the $3\sigma$ limiting magnitude for sources undetected in the $i_{\rm P1}$ band; \textit{iii)}  S/N($z_{\rm P1}$)$>10$, S/N($y_{\rm P1}$)$>5$ and S/N($r_{\rm P1}$)$<3$ or $r_{\rm P1}-z_{\rm P1}>2.2$. 
We then performed our own forced photometry in both the stacked and single-epoch PS1 images (see Sections 2.2 and 2.3 in \citealt{Banados2014}), and we finally visually inspected the stacked and single-epoch images to remove remaining obviously poor candidates and artifacts.\\
The PS1 magnitudes for PSO~J143$-$21 and PSO~J273+38 are shown in Table~\ref{tab:photomag}. 
For PSO~J143$-$21 we also list the $J$ and $K$ band measurements from VHS DR6.  
For PSO~J273+38 we report the $J$ and $K$ band magnitudes from the UKIRT Hemisphere Survey \citep[UHS,][]{Dye2018} DR2, together with CatWISE2020 (\citealt{Eisenhardt2020,Marocco2021}) W1 and W2 magnitudes.

\begin{figure}[h!]
    \centering
\includegraphics[width=0.99\columnwidth]{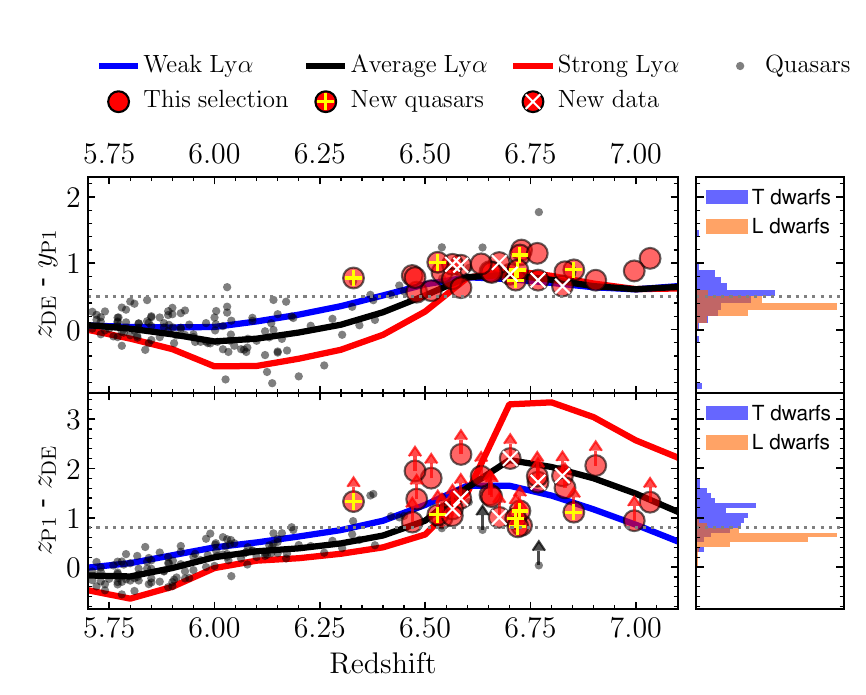}
    \caption{\small Redshift vs.\  $z_{\rm P1} - z_{\rm DE}$ (bottom) and $z_{\rm DE}-y_{\rm P1}$ (top) colors. The blue, black, and red solid lines represent the color tracks of composite quasar spectra from \cite{Banados2016}, illustrating weak, average, and strong Ly$\alpha$ emission lines, respectively.
    The dotted lines indicate the color cuts for the [DELS+PS1] selection, as described in Section \ref{selectionbanados23}. The red circles represent quasars that meet the selection criteria. Sources marked with a yellow cross are newly discovered quasars from this study, while those with a white `x' are known quasars for which we present new spectroscopy data. Lower limits correspond to sources undetected in $z_{\rm P1}$, for which we use their $3\sigma$ limiting magnitude. On the right, we show the histograms of the same colors for the L and T dwarfs compiled in \cite{Banados2016}.}
\label{fig:z66sel}
\end{figure}

\subsection{[DELS+PS1] $z\gtrsim 6.6$ quasar candidates from DELS and PS1}
\label{selectionbanadosnew}
We developed a selection aiming at $z\gtrsim6.6$ quasars combining PS1 and DELS, taking advantage of the different transmission of the $z_{\rm P1}$ and $z_{\rm DE}$ filters. \\
Given the sharp break in flux at $\lambda>9250$\,\AA\ for $z>6.6$ quasars, which is caused by absorption from intervening neutral Hydrogen, these quasars are largely undetected in the $z_{\rm P1}$-band. However, they are still strongly detected in the $z_{\rm DE}$-band and $y_{\rm P1}$ filters. 
Fig.~\ref{fig:z66sel} shows how effective the $z_{\rm P1} - z_{\rm DE}$ and $z_{\rm DE}-y_{\rm P1}$ colors can identify $z>6.6$ quasars, and at slightly lower redshifts if they have weak emission lines. 
On the other hand, given the smoother decline in the flux of L/T dwarfs (i.e., the main astrophysical contaminants in quasar candidates selection), they are much more likely to be detected in the $z_{\rm P1}$-band. 
We excluded sources with bad or low-quality detections in the $y_{\rm P1}$ band using the same quality flags described in Table 6 of \cite{Banados2014}. We also excluded objects flagged in the $z_{\rm DE}$ band (\texttt{anymask} and \texttt{allmask} not equal to 0). 
We did not consider sources with high Galactic extinction $E(B-V)>0.3$. 
We required S/N\,$>5$ in both $z_{\rm DE}$ and $y_{\rm P1}$ bands and non-detections (S/N$<3$) in all the other filters from both PS1 and DELS. Then, the color criteria are \textit{i)} $z_{\rm P1,3\sigma}- z_{\rm DE}>0.8$ and \textit{ii)} $z_{DE} - y_{\rm P1}>0.5$ (Fig.~\ref{fig:z66sel}). We crossmatched the L and T dwarfs compiled in \cite{Banados2016} with DELS within s 2\arcsec\ radius, and show their colors in the right panels of Fig.~\ref{fig:z66sel}. After applying the two color criteria from this selection, we remove 96\% of the L dwarfs and 91\% of the T dwarfs.
When we started this selection process, only a handful of $z\gtrsim 6.6$ quasars were known to meet our criteria \citep[e.g.,][]{Venemans2015,Mazzucchelli2017}. In addition, some of our candidates were independently identified by other works \citep[e.g.,][]{Pons2019,WangFeige2019}. \\
Here we present the discovery of six new quasars from our selection: PSO~J016$+$06, PSO~J037$-$08, PSO~J041$+$06, PSO~J067$-$14, PSO~J217$+$04 and PSO~J289$+$50. These quasars have redshifts ranging from 6.3 to 6.9.   
Additionally, 22 other quasars that meet our criteria have been published in the literature, with redshifts between 6.4 and 7.0 \citep{Venemans2015,Mazzucchelli2017,WangFeige2017,WangFeige2018,WangFeige2019,Pons2019,Banados2021,Banados2025,Yang2021}.
We present either independent discoveries or follow-up near-infrared (NIR) spectra (and, in two cases, follow-up NIR photometry) for six quasars already published in the literature: 
PSO~J062$-$09, 
PSO~J127$+$41, PSO~J129$+$49, 
PSO~J164$+$29, PSO~J162$-$01, and PSO~J354$+$21. 
The coordinates and photometry of the six new quasars presented in this work, along with the six previously known quasars that have new spectra, are reported in Table~\ref{tab:decalsmag}. 
In Fig.~\ref{fig:z66sel}, the new quasars are indicated by red circles with crosses, while the known quasars are shown with red circles featuring an “x.” 
Additionally, 16 other quasars meet our selection criteria (red circles in Fig.~\ref{fig:z66sel}); their coordinates and photometry are provided in Table ~\ref{tab:knownphotometry}.

\subsection{[PS1/AllWISE] $z>6.5$ quasar candidates from PS1 and AllWISE}
\label{sec:masafusa}
PSO~J335$-$15 was selected from the Pan-STARRS1 and ALLWISE catalogs.
The main steps are as follows: $i)$ S/N($y_{\rm P1}$)$>$7 and S/N($W1$)$>$3 or S/N($W2$)$>$3; $ii)$ compactness: requiring the absolute difference between the aperture and PSF magnitudes in $y_{\rm P1}$ band to be less than 0.3; $iii)$ dropout: $z_{\rm P1}-y_{\rm P1}>$1.5 and 
S/N($g_{\rm P1},r_{\rm P1}$)$<$3 or S/N($i_{\rm P1}$)$<$5. 
Then, forced photometry was performed on the  PS1 five-band images with SExtractor (\citealt{Bertin1996}) and required that the forced magnitudes also satisfy the aforementioned criteria. 
We then performed an SED fitting analysis of the initially selected $z$-dropout quasar candidates using optical, NIR, and ALLWISE photometry. The quasar SED models have a single power-law shape with broad emission lines. 
The slope index and emission line equivalent widths are based on a composite spectrum of low-redshift quasars (\citealt{VandenBerk2001,Selsing2016}) with some scatter. 
These templates are redshifted with a step of $dz=0.01$ from $z$=5 to 8 with IGM absorption taken into account at each redshift (\citealt{Madau1995}). 
The brown dwarf SED templates are based on \textsc{BT-Settl} model (\citealt{Allard2012}), for which we only consider their Solar metallicity models with effective temperatures 400~K to 6000~K.  
We apply these quasar and brown dwarf SED templates to find the best-fit SED solutions based on chi-square minimization to rank the candidates for follow-up spectroscopy.
As a final step, we did a visual inspection of all the targets to remove artifacts. 
PSO~J335$-$15 has a red $z - y = 1.68 \pm 0.08$ color without significant detection in bluer PS1 filters. Moreover, the public VHS J-band photometry $J=19.71 \pm0.09$ implies a flat $y-J=-0.10 \pm 0.10$ color, which is consistent with a quasar SED model at $z\sim6.5$.

\subsection{[DVCW] DES/VHS/CatWISE2020 $i$-dropout selection of \citetalias{WolfJ2024}}
The target PSO~J060$-$65 has been selected combining optical data from DES, NIR, MIR photometry from the VHS DR5 and CatWISE2020 (\textit{DVCW} acronym in the last column of Table~\ref{tab:specobslogproperties}). The selection steps are described in details in \citetalias{WolfJ2024}. 
Here, we summarize the main steps. 
First, a series of photometric cuts were imposed on the \textit{g,r,i,z,y} Kron magnitude (mag\_auto) of the catalog: $i)$ S/N($z$)$>$10, S/N($y$)$>$5 and S/N($g$)$<$3; $ii)$ mag\_auto\_i $-$ mag\_auto\_z $>$ 0.8 and mag\_auto\_z $-$ mag\_auto\_y $<$ 0.12 and mag\_auto\_r $>$ 22.5 or mag\_auto\_i $-$ mag\_auto\_z $>$ 2.2 and magerr\_auto\_r $>$ 0.36.
Then we adopted as a main filtering step a SED template fitting and photometric redshift computation using the code \texttt{Le PHARE} \citep[v2.2,][]{Arnouts11}, supplemented with a custom template library for AGN and galaxies (see Sect. 2.1.2 of \citetalias{WolfJ2024}). 
Then, we further extracted forced photometry on optical DES images of the candidates to identify artifacts and problematic blue-band PSF-matched photometry.  
We discarded candidates for which we measure a significant aperture flux in the $g$-band (i.e., with errors on $g$-band aperture magnitudes ap\_magerr\_g > 0.36 when a positive g-band flux is measured).
PSO~J060$-$65 and other 5 sources, published in \citetalias{WolfJ2024}, resulted to be among the best candidates from this selection. 
PSO~J060$-$65 photometric properties are reported in Table~\ref{tab:photomag}.

\subsection{[MQC] Radio candidates from the Million Quasar catalog of \cite{Flesch2023}}
\label{mqcsel}
The Million Quasars Catalogue (MQC\footnote{\url{http://quasars.org/milliquas.htm}}) has been available since its inception in 2009. 
It has collected the discoveries of quasars at all redshifts published in the literature over the years. 
In addition, the catalog also provides radio and X-ray information on the included sources \citep{Flesch2023}. 
Approximately 76\% of the objects have spectroscopic redshifts, and their spectra can be found in the literature. 
However, only photometric redshifts are available for the remaining sources. Thus, a spectroscopic follow-up is necessary to determine the true nature of these objects.\\
From MQC v8 (\citealt{Flesch2023}), we selected only objects with radio detection and with a photometric redshift larger than 4, resulting in 218 objects.  
We specifically focused on radio-detected targets to reduce contamination from ultracool dwarfs. 
We removed 10 candidates  (see Table~\ref{mqcdiscarted}) based on a cross-match with SIMBAD\footnote{\url{http://simbad.cds.unistra.fr/simbad/}} and with SDSS DR18 (\citealt{Almeida2023}).
We conducted a visual inspection using optical PS1 images and NIR images from VHS, UHS and the VISTA Kilo-Degree Infrared Galaxy Survey (VIKING, \citealt{Edge2013}).  We discarded 114 sources detected in the $g_{\rm P1}-$band, which are likely at a lower redshift. 
The remaining 97 catalog objects are good $g-$dropout (at $\sim3.5\leq z \leq4.5$) and $r-$dropout (with redshift between $\sim$4.5 and 5.5) candidates. 
In this paper, we present the spectroscopic confirmation of two quasars from this catalog: MQC~J021$+$19 and MQC~J133$-$02, indicated as \textit{MQC} in Table~\ref{tab:specobslogproperties}). 
Their photometry is reported in Table~\ref{tab:photomag}. 

\subsection{[YS23] Candidates from the DES survey of \cite{YangShen2023}}
\citet[hereafter YS23]{YangShen2023} provide a catalog of 1.4 million photometrically selected quasar candidates from DES, cross-matched with available NIR and unWISE MIR photometry (\citealt{Schlafly2019}). \\
From the YS23 catalog, we selected only objects with a probability of being a quasar larger than 98\% and a photometric redshift greater than 5. 
This resulted in a list of 134 sources. 
After a visual inspection of all optical images from DES and DELS, and NIR images from VHS, we discarded five sources (see Table~\ref{ys23discarted}). 
Then, we cross-matched the list of targets with the literature and we found 27 objects already published as high-$z$ quasars (\citealt{McGreer2013,Venemans2013,Reed2015,Menzel2016,WangFeige2016,YangJ2016,Reed2017,Banados2014,Banados2016,Banados2023,Ighina2023,YangDM2024}). 
Thus, the remaining $z>5$ quasar candidates are 102. 
In this paper we report the spectroscopic confirmation of twelve new quasars from this list of candidates, they are marked as \textit{YS23} in the last column of Table~\ref{tab:specobslogproperties}.
Their photometric properties are detailed in Table~\ref{tab:photomag}. Additionally, we provide information about these twelve quasars directly from the YS23 catalog, including their ID, photometric redshift, and quasar probability.

\section{Follow-up observations}
\label{sec:observations}
In this section we describe both the dedicated NIR photometric observations and the spectroscopic campaign for all the sources listed in Tables~\ref{tab:photomag} and ~\ref{tab:decalsmag}. 

\subsection{NIR Photometry}
\label{sec:photometry}
We obtained NIR follow-up images of all six new quasars discovered by the DELS+PS1 selection, and for the quasars PSO~J062$-$09, PSO~J162$-$01, and PSO~J354$+$21 (see Table~\ref{tab:decalsmag} and Sect.~\ref{selectionbanadosnew}). 
We also obtained NIR observations for the new quasar PSO~J335$-$15 from the PS1/ALLWISE selection (see Table~\ref{tab:photomag} and Sect.~\ref{sec:masafusa}). 
The imaging follow-up was performed ouwith the $J$, $H$ and $Ks$ filters.  
The observations have been carried out with the Son OF ISAAC (SofI, \citealt{Moorwood1998}) instrument mounted on the NTT at La Silla Observatory and the Nordic Optical Telescope near-infrared Camera (NOTCam\footnote{https://www.not.iac.es/instruments/notcam/}) mounted on the Nordic Optical Telescope at the Roque de Los Muchachos Observatory. 
A summary of these NIR observations is reported in Table~\ref{tab:nirfollowup}. 
A standard data reduction was performed with IRAF (Image Reduction and Analysis Facility, \citealt{Tody1993}). 
We obtained the zero points following standard procedures (e.g., see Section 2.6 in \citealt{Banados2014} or Section 3.3 in \citealt{Dodorico2023}). 
We also list photometry of the NIR surveys VHS, UHS, and VIKING in Tables~\ref{tab:photomag} and ~\ref{tab:decalsmag}.

\begin{table}[!h]
    \caption{\small Summary of the NIR follow-up observations.}
    \label{tab:nirfollowup}
    \begin{tiny}
    \begin{tabular}{p{1.75cm}p{1.85cm}p{0.5cm}rp{0.1cm}}
    \hline
    Quasar name & Telescope/ Instrument & Filter & Obs. date & T$_{\rm exp}$ \\
    (1) & (2) & (3) & (4) & (5) \\
\hline 
PSO~J016$+$06 &  NTT/SofI & $J$ & 2018 Dec. 23 & 5 \\
 & NTT/SofI & $H$ & 2020 Nov. 18 & 15 \\
 & NTT/SofI & $Ks$ & 2020 Nov. 18 & 15 \\
PSO~J037$-$08 &  NTT/SofI & $J$ & 2021 July 27 & 15 \\
PSO~J041$+$06 &  NTT/SofI & $J$ & 2020 Nov. 20 & 15 \\
PSO~J062$-$09 & NTT/SofI & $J$ & 2021 July 27  & 30  \\
PSO~J067$-$14 &  NTT/SofI & $J$ & 2021 July 27  & 15 \\ 
&  NTT/SofI & $H$ & 2023 Jan. 04  & 16 \\ 
&  NTT/SofI & $Ks$ & 2023 Jan. 04  & 10 \\ 
PSO~J162$-$01 &  NTT/SofI & $J$  & 2021 July 27 & 15 \\ 
&  NTT/SofI & $Ks$  & 2019 Dec. 15 &  15 \\ 
PSO~J217$+$04 &  NTT/SofI & $J$ & 2020 Feb. 06 & 30 \\
PSO~J289$+$50 & NOT/NOTCam & $J$ & 2020 July 10 & 27  \\
& NOT/NOTCam & $Ks$ & 2020 July 10 & 18 \\
PSO~J335$-$15 &  NTT/SofI & $J$ & 2021 July 28  & 5 \\
PSO~J354$+$21 &  NOT/NOTCam & $J$ & 2019 May 19  & 27  \\
            & NOT/NOTCam & $Ks$ & 2019 May 19 & 18 \\
\hline
    \end{tabular}
    \end{tiny}
    \tablefoot{\tiny Col (1): Quasar name;
    Col (2): Telescope and instrument used for the imaging follow-up; Col (3): Filter; Col (4): Observing date; Col (5): Exposure time in minutes.}
\end{table}

\subsection{Spectroscopy}
\label{sec:spectroscopy}
\begin{figure*}
{\includegraphics[width=0.5\textwidth]{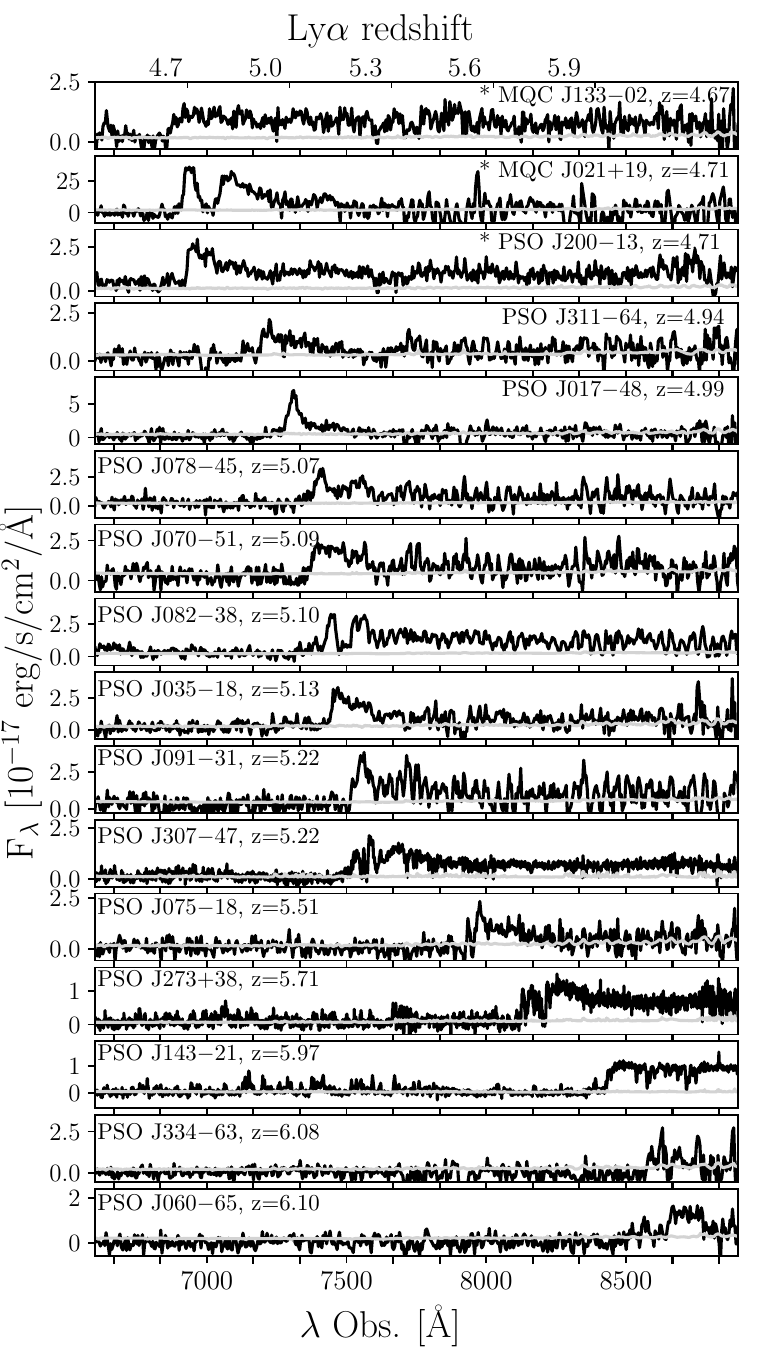}\hspace{0.05cm}
\includegraphics[width=0.5\textwidth]{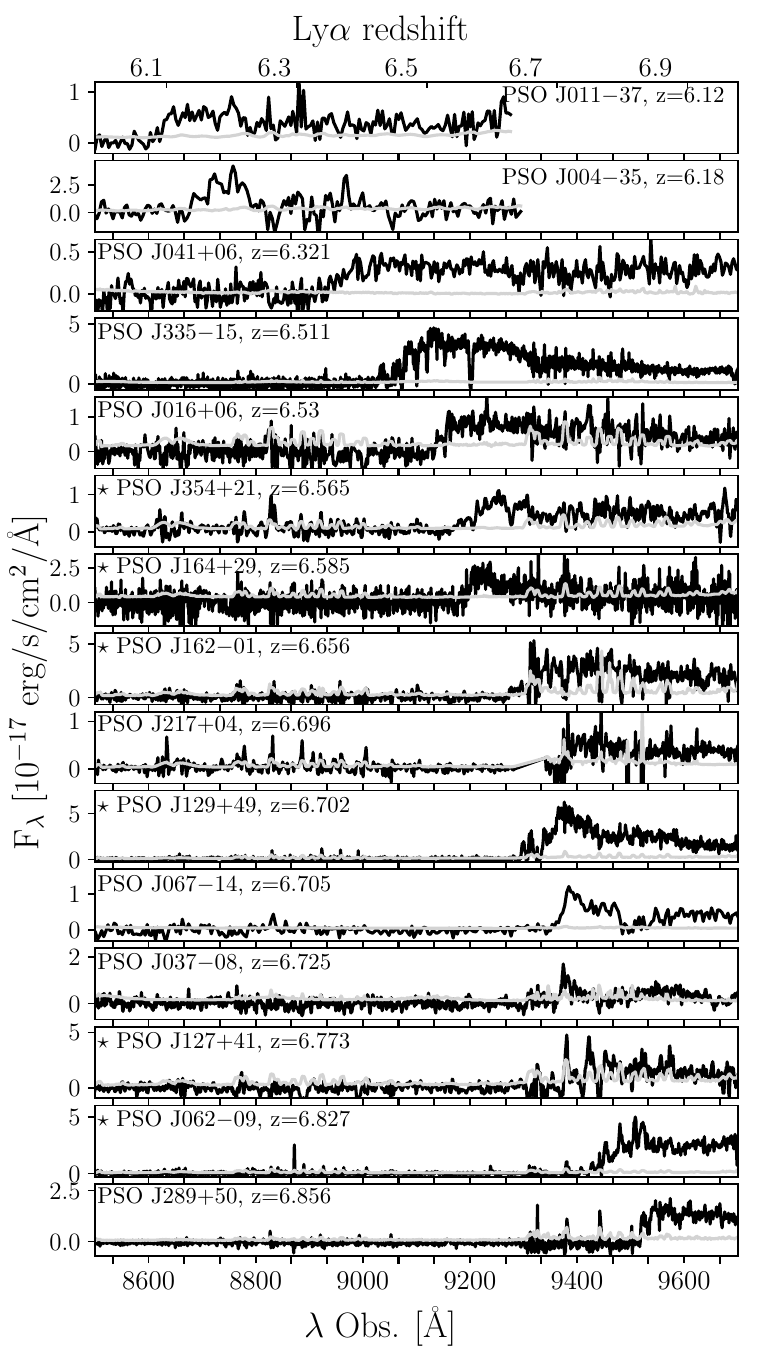}}
\caption{\small Newly discovered spectra (25 sources, $*$ marked radio-loud objects) and new spectra publication (6 objects, marked with a $\star$) for the quasars reported in this paper. Only the part of the spectrum that covers the Lyman-$\alpha$ break is shown here. The noise spectrum is reported in gray. Sorted by increasing redshift.}
\label{fig:spectra_lyman}
\end{figure*}

\begin{figure*}[!ht]
    \centering
    \includegraphics[width=0.9\textwidth]{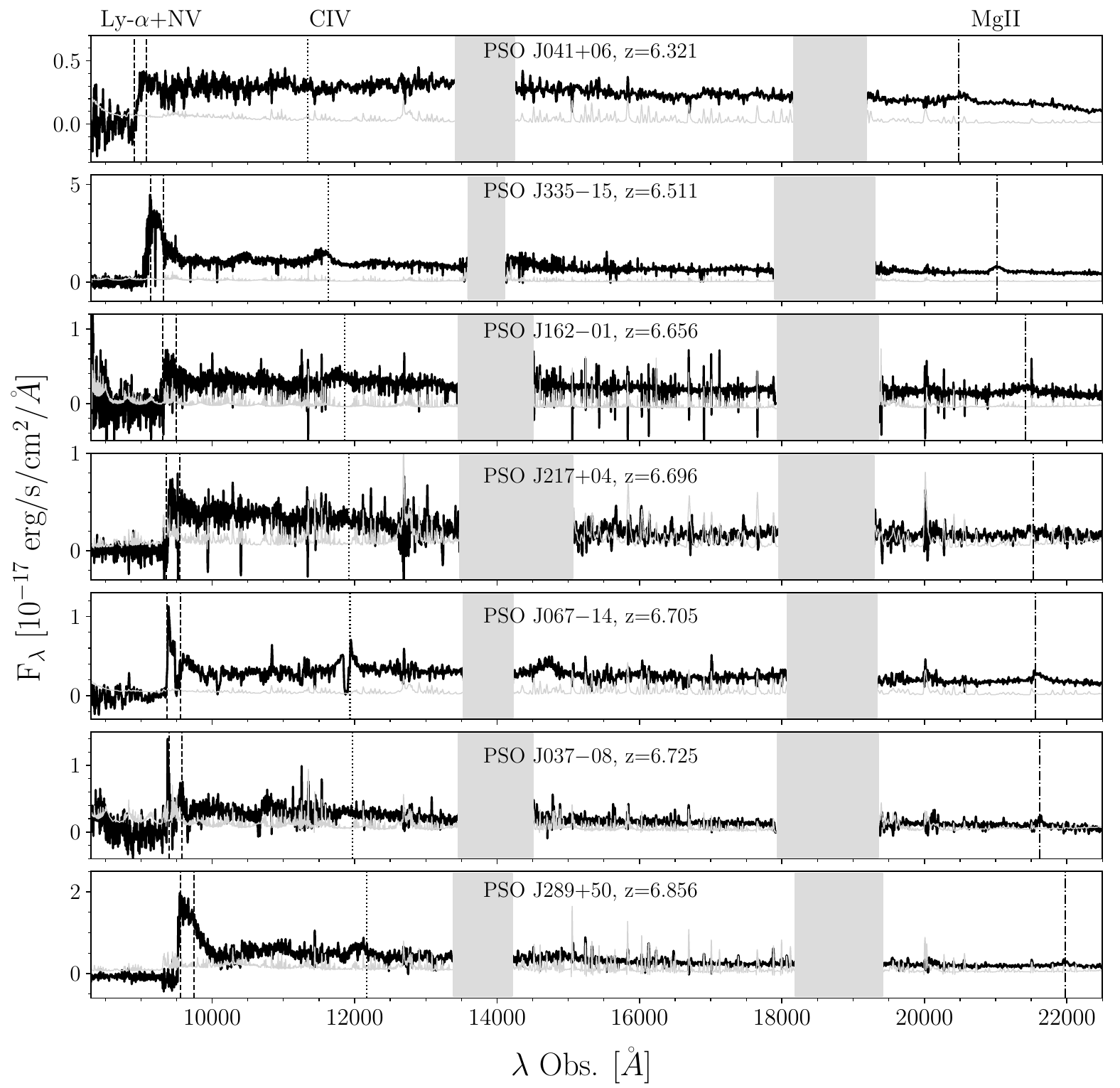}
    \caption{\small NIR follow-up spectra of a sub-sample of quasars reported in this paper (sorted by increasing redshift). We show the NIR spectra of the newly discovered quasars PSO~J041$+$06, PSO~J335$-$15, PSO~J067$-$14, PSO~J217$+$04, PSO~J037$-$08 and PSO~J289$+$50 and the newly published NIR spectrum for the already known quasar PSO~J162$-$01. 
    The noise spectra are reported in gray. 
    Shaded light gray areas highlight regions strongly affected by telluric absorption.
    On the top the location of the main emission lines according to the Mg$\rm II$ redshift are also shown: Lyman-$\alpha$$+$N$\rm V$ (dashed), C$\rm IV$ (dotted) and Mg$\rm II$ (dashed-dotted). }
    \label{fig:nirspectra}
\end{figure*}

The spectroscopic follow-up was carried out over different observing runs and different instruments: EFOSC2 (\citealt{Buzzoni1984}) at the New Technology Telescope (NTT) located at the Observatory of La Silla, the Folded-port Infrared Echellete (FIRE; \citealt{Simcoe2008,Simcoe2013}) spectrometer at the Magellan/Baade telescope and the Low Dispersion Survey Spectrograph (LDSS3, \citealt{Boutsia2017}) at the Magellan/Clay telescope both 
at Las Campanas Observatory, the near-infrared spectrograph at Gemini North (GNIRS; \citealt{Elias2006a,Elias2006b}), the Multi-object Double Spectrograph (MODS; \citealt{Pogge2010}) and the LBT Utility Camera in the Infrared (LUCI; \citealt{Seifert2003}) at the Large Binocular Telescope (LBT), the FOcal Reducer/low dispersion Spectrograph 2 (FORS2;
\citealt{Appenzeller1992}) at the Very Large Telescope (VLT) and the Near-Infrared Echellette Spectrometer (NIRES; \citealt{Wilson2004}) mounted on the Keck 2 telescope located at W. M. Keck Observatory. The details of the spectroscopic observations are summarized in Table~\ref{tab:specobslogproperties} , while spectra covering the Lyman-alpha break are shown in Fig.~\ref{fig:spectra_lyman}. 
In Fig.~\ref{fig:nirspectra} we plot the NIR spectra obtained for a subsample of sources.\\
Both LBT/MODS and LBT/LUCI observations have been carried out in binocular mode (red grating for MODS and G200 grating coupled with $zJ$+$HK$ filters for LUCI), except for PSO~J127$+$41 that was observed in monocular mode. 
All NTT/EFOSC2 observations were conducted using Grism $5$ (5200--9350$\,\AA$), except for quasars PSO~J004$-$35, PSO~J011$-$37, PSO~J334$-$63 for which Grism $16$ (6015--10320$\,\AA$) was used. 
However, the resulting spectra taken with Grism $5$ of PSO~J017$-$48, PSO~J035$-$18 and PSO~J075$-$18 showed wiggles in the flux, due to fringing effects in the red part of the EFOSC2 detector combined with the Grism $5$. 
Therefore, these three objects were observed again with Grism $16$, and the spectra shown in Fig.~\ref{fig:spectra_lyman} are the ones resulting from this last data analysis. \\
The targets PSO~J143$-$21 and PSO~J307$-$47 were recently re-observed with Magellan/LDSS3 because their NTT/EFOSC2 spectra was very noisy. 
Both observations are reported in the Table~\ref{tab:specobslogproperties}, but we highlight that the redshift is derived only from the best spectrum (Magellan/LDSS3)\footnote{The redshift obtained from the fit of the Magellan/LDSS3 spectra are consistent with the measurement previously obtained from the NTT/EFOSC2 spectra.} and that only the latter is shown in Figure~\ref{fig:spectra_lyman}.\\
All the spectra were reduced using standard routines, including bias subtraction, flat fielding, sky subtraction, wavelength calibration using exposures of arc lamps and flux calibration using exposures of spectrophotometric standard stars. 
The spectra were absolute flux calibrated to match the $z-$band (for all the sources at $z\leq$6) or one of the NIR magnitudes (mostly $J-$band) depending on their redshift.
We used the following reduction pipelines: FIREHOSE\footnote{\url{https://wikis.mit.edu/confluence/display/FIRE/FIRE+Data+Reduction}}, \texttt{PypeIt}\footnote{\url{https://pypeit.readthedocs.io}} \citep{Prochaska2020} and SIPGI\footnote{\url{https://pandora.lambrate.inaf.it/sipgi/}} (\citealt{Gargiulo2022}). 
Section 2.4 in \cite{Onorato2025} describes the coadding procedure of the LBT/MODS and Keck/NIRES spectra of PSO~J289$+$50 and of the LBT/MODS+LUCI spectra of PSO~J217$+$04. 
The coadd is done after the flux calibration procedure using directly the multi\_combspec routine of \texttt{PypeIt}, to get a final spectrum covering [8000, 24700]\AA\ / [8000,23900]\AA\ with pixel size of 40~km~s$^{-1}$ / 58~km~s$^{-1}$ for PSO~J289$+$50 and PSO~J217$+$04, respectively.\\
For PSO~J062$-$09, PSO~J127$+$41, PSO~J129$+$49, PSO~J162$-$01 and PSO~J354$+$21 we report for the first time unpublished spectra obtained at Magellan, LBT and VLT. 
The details on these observations are listed in Table~\ref{tab:specobslogproperties} and the spectra are shown in Fig.~\ref{fig:spectra_lyman}. 
In Fig.~\ref{fig:nirspectra} we also report the Magellan/FIRE unpublished NIR follow-up of PSO~J162$-$01. 

\begin{table*}[!ht]
\caption{\small Spectroscopic observations and properties of the newly discovered quasars reported in this paper.}
\label{tab:specobslogproperties}
\centering
\begin{small}
\begin{tabular}{p{1.8cm}p{2.4cm}p{1.75cm}p{0.7cm}p{1.6cm}p{1.6cm}p{2.8cm}p{0.5cm}p{0.5cm}}
\hline\hline
Quasar  &  Telescope/  & Obs. Date & Exp   & Selection  & $z$ & $z$-method &  $m_{1450}$ & $M_{1450}$ \\  
Name & Instrument &                & Time (s) &         &     &            &             &            \\
    (1)    & (2) & (3) &   (4)    & (5)    & (6) & (7) & (8) & (9) \\
\hline
 \multicolumn{9}{c}{New discoveries} \\
\hline
PSO~J004$-$35  & NTT/EFOSC2 &  2024/10/26 & 4200 & YS23                 & 6.18 & template (\textit{on25})  & 20.99 & $-$25.74 \\
PSO~J011$-$37 & NTT/EFOSC2  & 2024/10/26  & 4200  & YS23                &  6.12  & template   (\textit{xqr-30})  & 20.72 & $-$26.01 \\
PSO~J016$+$06 & VLT/FORS2    &   2020/11/09 & 1800 &  DELS$+$PS1 & 6.53 & template (\textit{xqr-30}) & 20.98 & $-$25.89 \\
PSO~J017$-$48 & NTT/EFOSC2 & 2024/01/25 2024/10/27 & 3600 3600     & YS23 &  4.99   & template  (\textit{on25})   & 20.70 & $-$25.69 \\
MQC~J021$+$19$^r$ &  NTT/EFOSC2 & 2023/11/30 & 3600 &  MQC  &  4.71   & template  (\textit{strong-Ly$\alpha$}) & 20.24 & $-$26.05  \\
 PSO~J035$-$18 & NTT/EFOSC2 & 2024/01/26 2024/10/27 & 3600 5400 &  YS23 &  5.13   & template (\textit{weak-Ly$\alpha$})   & 20.41 & $-$26.02 \\
PSO~J037$-$08 & Magellan/FIRE & 2018/12/31 2019/01/13 & 700 7200 &  DELS$+$PS1   &  6.725$\pm$0.002  & Mg$\rm II$   & 20.61 & $-$26.26\\
PSO~J041$+$06 & Gemini/GNIRS   &  2019/01/12 & 8400 & DELS$+$PS1 & 6.321$\pm$0.002 & Mg$\mathrm{II}$   & 20.87 & $-$25.91 \\
PSO~J060$-$65 & NTT/EFOSC2 & 2024/01/24 & 3600 &  DVCW &  6.10   & template (\textit{xqr-30})    & 20.71 & $-$26.01 \\
PSO~J067$-$14 &  Gemini/GNIRS & 2019/01/12 + /02/03-06-09 &    12000   & DELS$+$PS1 &  6.705$\pm$0.002    & Mg$\rm II$     & 20.89 & $-$25.99 \\
PSO~J070$-$51 & NTT/EFOSC2 & 2023/12/01 & 7200 &  YS23   &  5.09   & template   (\textit{xqr-30})            & 20.55 & $-$25.87 \\
PSO~J075$-$18 & NTT/EFOSC2 & 2023/11/29   2024/10/27 & 5400 1800 &  YS23 &  5.51   & template (\textit{weak-Ly$\alpha$})   & 20.97 &  $-$25.58 \\ 
PSO~J078$-$45 & NTT/EFOSC2 & 2023/11/30 & 3600 &   YS23    &  5.08   & template (\textit{on25})  & 20.86 & $-$25.56  \\
PSO~J082$-$38 & NTT/EFOSC2 & 2023/11/29 & 3600 &  YS23 &  5.10   & template  (\textit{xqr-30})     & 20.12 & $-$26.30 \\
PSO~J091$-$31 & NTT/EFOSC2 & 2023/11/29 & 3600 &  YS23 &  5.22   & template  (\textit{strong-Ly$\alpha$}) & 20.41 & $-$26.06 \\
MQC~J133$-$02$^r$ & NTT/EFOSC2 & 2022/12/27 &  5400 & MQC &  4.67   & template  (\textit{on25})   & 20.80 & $-$25.48 \\
PSO~J143$-$21 & NTT/EFOSC2 & 2022/12/26 &  3600 &  PS1B23 &  5.97  & template  (\textit{xqr-30}) & 20.55 & $-$26.13  \\
  & Magellan/LDSS3 & 2025/05/01 &  1800 &   &      &   &  & \\
PSO~J200$-$13$^r$ &  NTT/EFOSC2 & 2022/02/05 & 4300 & RPS1AW   &  4.71   & template  (\textit{xqr-30})    & 20.48 & $-$25.82  \\
PSO~J217$+$04 & Magellan/FIRE  & 2023/05/19   & 900  & DELS$+$PS1 & 6.696$\pm$0.002 & Mg$\rm II$  & 21.11 & $-$25.76 \\
    &  LBT/MODS$+$LUCI &  2023/06/07 + /06/13 &  7200 13800 &                     &                               &                     &         &                   \\
PSO~J273$+$38 & LBT/MODS & 2023/06/10 & 1200  &  PS1B23 &  5.71   & template   (\textit{xqr-30})            & 20.87 & $-$25.73 \\
PSO~J289$+$50  & LBT/MODS    &  2019/10/07    &  4800   &   DELS$+$PS1   &  6.856$\pm$0.002 & Mg$\rm II$                             & 20.70  & $-$26.21  \\
    &  Keck/NIRES &  2020/09/07   &  11520     &                     &                               &                     &         &                   \\
PSO~J307$-$47  & NTT/EFOSC2  & 2024/10/28 & 4200 & YS23 & 5.22  & template (\textit{on25})   & 20.92 & $-$25.54 \\
                & Magellan/LDSS3 & 2025/05/01 & 1200 &   &  \\
PSO~J311$-$64 &  NTT/EFOSC2  & 2024/10/27 & 3600 & YS23 &  4.94   & template  (\textit{xqr-30})  & 21.01 & $-$25.37 \\
PSO~J334$-$63  & NTT/EFOSC2  & 2024/10/27 & 5400 & YS23 &  6.08   & template  (\textit{xqr-30})  & 20.89 & $-$25.82 \\
PSO~J335$-$15 & Magellan/FIRE & 2019/05/09 & 18000 &  PS1/AllWISE   &  6.511$\pm$0.001  & Mg$\rm II$    & 19.85 & $-$26.98 \\
\hline
\multicolumn{9}{c}{New spectra publication} \\
\hline
PSO~J062$-$09  &  Magellan/LDSS3 & 2018/11/13 & 1200 &  DELS$+$PS1    & 6.827$\pm$0.006 & Mg$\rm II$  &   20.28    & $-$26.62  \\
PSO~J127$+$41  &  LBT/MODS & 2018/10/05 & 3600 &  DELS$+$PS1 & 6.773$\pm$0.007 & Mg$\rm II$  &   20.52    & $-$26.36 \\
PSO~J129$+$49 &  LBT/MODS & 2017/04/23 & 2040 & DELS$+$PS1 & 6.702$\pm$0.001 & Mg$\rm II$  &    20.24    & $-$26.63 \\
PSO~J162$-$01   &   VLT/FORS2 & 2016/02/16 & 5260 &  DELS$+$PS1  & 6.656$\pm$0.001 & Mg$\rm II$  &   21.03  & $-$25.83  \\
&  Magellan/FIRE & 2015/03/12 & 19508 &  & & & \\
PSO~J164$+$29 & LBT/MODS & 2019/03/18 & 7200  & DELS$+$PS1 & 6.585$\pm$0.005 & Mg$\rm II$  &    21.16  & $-$25.68 \\
PSO~J354$+$21 &   LBT/MODS & 2020/10/23 & 1800 & DELS$+$PS1 & 6.565$\pm$0.009 & Mg$\rm II$  &     21.18    & $-$25.66 \\
\hline
\end{tabular}
\end{small}
\tablefoot{\tiny Quasars sorted by right ascension. Col (1): Quasar name; the objects marked with $r$ are radio-loud;
Col (2): Telescope and instrument used for the spectroscopic follow-up; Col (3): Date of observations; Col (4): Exposure time in seconds; 
Col (5): Selection method as described in Sect. \ref{sec:selection}; Col (6): redshift; Redshift for the already published quasars (second part of the table) are from \cite{Yang2021}, except for PSO~J162$-$01 which has been directly measured from the Magellan/FIRE spectrum in this study; Col (7): method used to compute the redshift (and in parenthesis the best template); Col (8) and Col (9): apparent and absolute magnitude at 1450\AA\ rest-frame. For the objects PSO~J062$-$09, PSO~J129$+$49 and PSO~J354$+$21 these quantities have been taken from \cite{FanX2023ARA}.} 
\end{table*}

\section{New quasars at $4.6 < z < 6.9$}
\label{sec:results}
We present the discovery of 25 new quasars\footnote{In Appendix~\ref{app:dwarf}, we report the objects that turned out to be not high-$z$ quasars after spectroscopic follow-up campaigns (73 in total).} at $4.6 < z < 6.9$; three are radio-loud.
To estimate the redshifts, we followed the procedure of \cite{Banados2023}: we fit the spectra of the quasars with different quasar templates, to account for differences in the emission lines properties, especially Ly$\alpha$. 
We use the following four templates: 
\begin{itemize}
\item[1.] \textit{strong-Ly$\alpha$}, which is the median of the 10\% of the $z\sim6$ PS1 quasars spectra with the largest rest-frame equivalent width for the Ly$\alpha$+N$\rm V$ emission line, from \cite{Banados2016}.
\item[2.] \textit{weak-Ly$\alpha$}, which is the median of the 10\% of the $z\sim6$ PS1 quasars spectra with the smallest rest-frame equivalent width for Ly$\alpha$+N$\rm V$, from \cite{Banados2016}.
\item[3.] \textit{xqr-30}, which is the median of 42 $z\sim6$ high quality quasars spectra observed with X-Shooter and reported in \cite{Dodorico2023}.
\item[4.] \textit{on25}, which is the weighted mean of 33 $z>6.5$ spectra presented in \cite{Onorato2025}.
\end{itemize}

Then, we choose the best-fitting template with the minimum chi-square in the 1212--1400\,\AA\ rest-frame wavelength range. 
The results are reported in Table~\ref{tab:specobslogproperties}.
We consider 0.03 as the mean uncertainty on the redshift computed using the template method (with a maximum uncertainty of 0.05), as reported by \cite{Banados2016,Banados2023}. \\
For PSO~J037$-$08, PSO~J041$+$06, PSO~J067$-$14, PSO~J335$-$15, and PSO~J289$+$50 we use the redshift and the error estimated from the fit of the Mg$\rm II$ broad emission line (see Sect. \ref{sec:bhmass} for details).
For the already known quasars we listed the Mg$\rm II$-based redshift reported in \cite{Yang2021}, while for PSO~J162$-$01 we estimate the Mg\,II redshift from the Magellan/FIRE spectrum published in this study. \\
The apparent and absolute magnitudes at rest-frame wavelength 1450\AA\ ($m_{1450}$, $M_{1450}$) are quantities commonly used to characterize the UV emission in quasars.
The apparent and absolute magnitude at 1450\AA\ of the quasars PSO~J037$-$08, PSO~J041$+$06, PSO~J067$-$14, PSO~J162$-$01, PSO~J217$+$04,
PSO~J289$+$50 and PSO~J335$-$15 were derived directly from the fit of the spectra (described in Sect.~\ref{sec:analysis}). \\
In order to determine $m_{1450}$ in a consistent way for all the objects with only a visible spectrum available, we followed the approach of, e.g., \cite{Banados2016} and \cite{Mazzucchelli2017}. 
We assumed a power-law continuum slope $\alpha_{\nu}$ = 0.44 (\citealt{VandenBerk2001}).
Then we used the $z-$band magnitude ($\lambda_{\rm eff}=9760$\,\AA\ for DES and 8700\,\AA\ for PS1) to extrapolate the $m_{1450}$. 
The redshifts and magnitudes at 1450\,\AA\ are reported in Table~ \ref{tab:specobslogproperties}.


\subsection{Notes on individual sources}
\label{sec:individualnotes} 
Here we present additional notes on selected objects, sorted by right ascension. 

\subsubsection{MQC~J021$+$19, $z = 4.71$}
\begin{figure*}[!ht]
\centering
{\includegraphics[width=0.325\textwidth]
{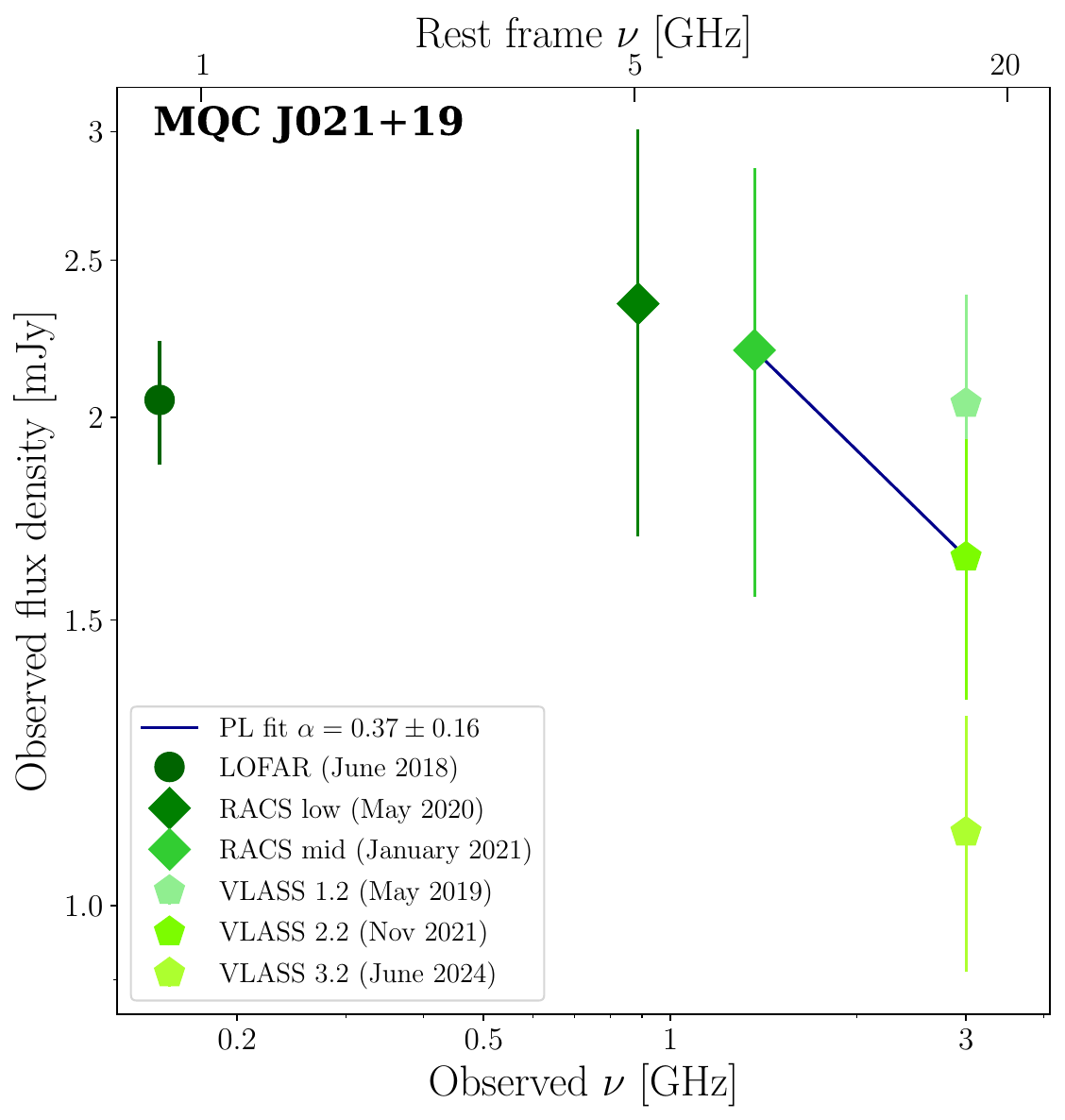}\hspace{0.1cm}
\includegraphics[width=0.32\textwidth]{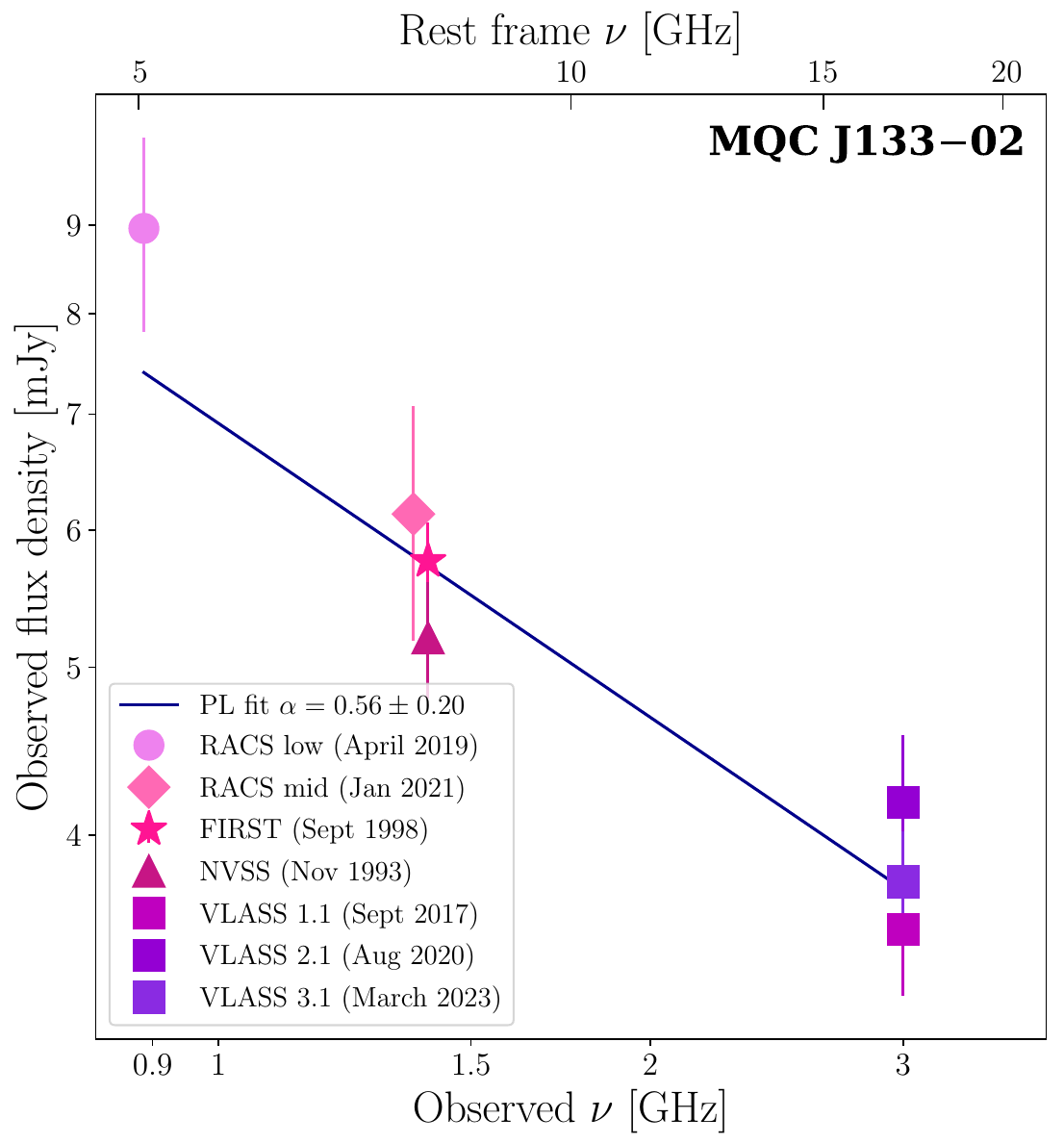}}\hspace{0.1cm}
\includegraphics[width=0.33\textwidth]{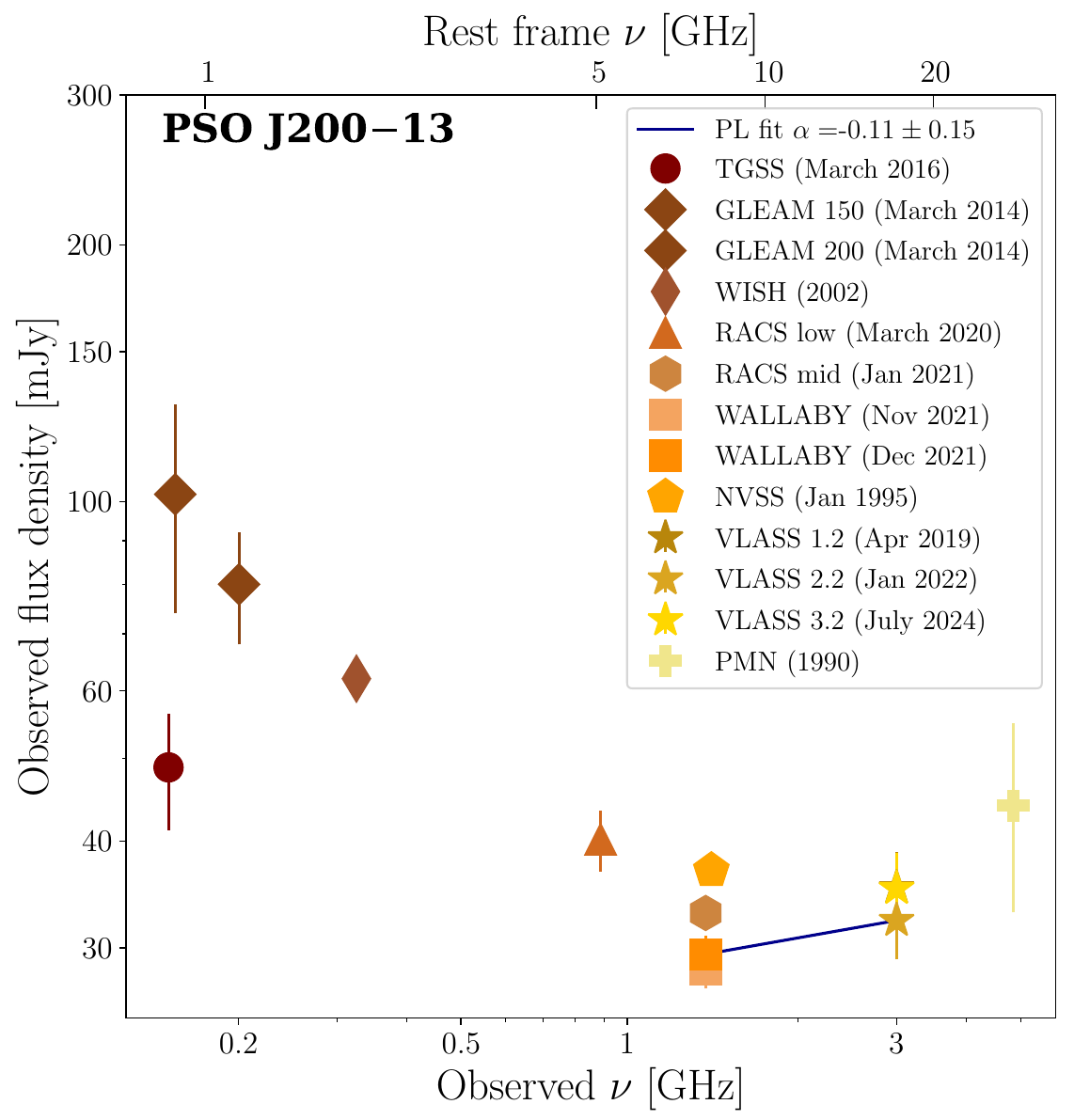}
\caption{\small Radio flux densities as a function of the observed (bottom x-axis) and rest-frame (top x-axis) frequency of MQC~J021$+$19 (\textit{left}), MQC~J133$-$02 (\textit{central}), and PSO~J200$-$13 (\textit{right}). The corresponding radio surveys are shown in the legend.
Due to the potential flux density variability of MQC~021$+$19, we report the spectral fit of the two values closest in time (RACS\_mid and VLASS 2.2). 
Similarly for PSO~J200$-$13 which shows clear evidence of variability on time-scales of years. We report the single power-law fit computed between WALLABY December 2021 and VLASS 2.2, the two values closest in time.}
\label{fig:radiospec}
\end{figure*}

This quasar was selected from the MQC v8 as a candidate at $z$$_{\rm phot}=4.9$. 
The NTT/EFOSC2 discovery spectrum confirms its high-$z$ nature, placing it at a redshift of $z=4.71$. 
The MQC reports a radio detection in the LOw-Frequency ARray (LOFAR) Two-metre Sky Survey (LoTSS, \citealt{Shimwell2022}) at 150~MHz, in the Rapid ASKAP Continuum Survey (RACS, \citealt{McConnell2020,Hale2021}) at 888~MHz and at 3~GHz in the VLA Sky Survey (VLASS, \citealt{Lacy2020}).
We also found a detection at 1.37~GHz (RACS-MID) on the CASDA website\footnote{\url{https://data.csiro.au/?searchBy=sci-domain}}.
In Table~\ref{tab:mqcradiodet} we listed the available integrated radio flux densities.
All these measurements have been quantified by performing a single Gaussian fit (since the source is unresolved) on the images, using the task IMFIT of the Common Astronomy Software Applications package (CASA, \citealt{McMullin2007}).
The source does not show clear signs of variability: the flux densities are all consistent within 1$\sigma$ and 2$\sigma$. 
However, there may be a decrease in VLASS flux density at 3~GHz from 2019 and 2024. 
Taking this potential decrease into account, to minimize possible variability issues, we used the radio data closest in time to compute the radio spectral index. 
These are RACS\_mid (1.367~GHz) and VLASS 2.2 (3~GHz). 
By assuming a single power-law for the continuum radio emission (S$_\nu$ $\propto$ $\nu^{-\alpha_r}$) we obtained $\alpha_r$ = 0.37$\pm$0.16 (see Fig.~\ref{fig:radiospec}, left panel).
Simultaneous radio data are necessary to assess the real shape of the radio spectral energy distribution.
Then, we computed the value of radio-loudness (R). 
Coincidentally, 5~GHz in the rest frame corresponds to an observed frequency of 0.875~GHz, equal to RACS\_low. 
Therefore we used the RACS flux density and the estimated spectral index to compute the 5~GHz rest frame radio flux density. 
The value of the flux density at 4400\,\AA\ rest frame was derived from the $i$-band de-reddened PS1 magnitude assuming the optical spectral index of \citet[$\alpha_{\nu}$ = 0.44]{VandenBerk2001}. 
We obtained R = 38$\pm$18, where the error takes into account the uncertainties on the flux densities, on the radio spectral index, and on the redshift of the source.\\
From the NTT/EFOSC2 spectrum shown in Fig.~\ref{fig:spectra_lyman} it is clear that there is an absorption blue-ward the N$\rm V$ line. 
This feature has a width of $\sim4800$~km~s$^{-1}$ and is likely associated with a N$\rm V$ Broad Absorption Line (BAL, \citealt{Weymann1991}) outflow with a maximum velocity of $\sim5000$~km~s$^{-1}$. 
A high S/N spectrum covering NIR wavelengths could confirm the presence of such an outflow in other atomic species (e.g., C$\rm IV$, Si$\rm IV$, Mg$\rm II$). 

\begin{small}
 \begin{table}[!ht]
 \caption{Summary of the archival radio observations of MQC~J021$+$19.}
 \label{tab:mqcradiodet}
 \centering
 \begin{tabular}{p{0.7cm}p{1.2cm}p{1.65cm}p{1.3cm}l}
 \hline\hline
 Obs. Freq. & S$_{\nu}$ & Survey & Resolution & Obs. date \\
 (GHz)     & (mJy)       &   & (arcsec) &      \\
 (1) & (2) & (3) & (4) & (5) \\
 \hline
 0.150       &  2.05$\pm$0.18 & LOFAR & 6 & 2018-06-12 \\
 0.8875      &  2.35$\pm$0.66 & RACS low  & 25   & 2020-05-02 \\
 1.3675      &  2.20$\pm$0.65 & RACS mid  & 16$\times$9 & 2021-01-01 \\
 3.0         &  2.04$\pm$0.34 & VLASS 1.2 & 2.5  & 2019-05-05 \\
 3.0         &  1.63$\pm$0.30 & VLASS 2.2 & 2.5  & 2021-11-07 \\
 3.0         &  1.10$\pm$0.20 & VLASS 3.2 & 2.5  & 2024-06-28 \\
 \hline
 \end{tabular}
 \tablefoot{Col (1): Observed frequency in GHz; Col (2): integrated flux density in mJy; VLASS flux densities have been corrected for systematics as explained at \url{https://science.nrao.edu/vlass/data-access/vlass-epoch-1-quick-look-users-guide}; the error on the RACS and VLASS measurements take into account both the statistical error and systematic error, which is 0.5~mJy for RACS (see Sect.~3.4.4 in \citealt{McConnell2020}), 8\% for VLASS 1.1 epoch and 3\% for VLASS 2.1 and onwards; Col (3): reference survey; Col (4): angular resolution in arcsec; Col (5): Date of the observation.} 
 \end{table}
 \end{small}

\subsubsection{PSO~J037$-$08, $z = 6.725$}
This quasar was discovered with a 700s spectrum with FIRE in longslit mode on Dec.\ 2018 and then a follow-up for 2 hours with FIRE in Echelle mode in Jan.\ 2019 (Table~\ref{tab:specobslogproperties}). 
Here we only show the higher S/N and resolution Echelle spectrum in both  Fig.~\ref{fig:spectra_lyman} and ~\ref{fig:nirspectra}.\\
The $J$-band magnitude estimate from our follow-up with NTT/SofI (see Table~\ref{tab:decalsmag}) is consistent with the detection in the VHS DR6: $J_{\rm VHS}=20.92\pm$0.14.\\
We note that this quasar is part of the JWST ASPIRE program \citep{ASPIREproposal} and its [CII] properties have been reported in \cite{Wang2024}. 
We used the Mg$\rm II$ BEL to estimate its redshift: $z=6.725\pm0.002$ (see Table~ \ref{tab:specobslogproperties}), which is consistent with the systemic redshift measured from the [CII] line ($6.7249\pm0.0002$, \citealt{Wang2024}). 
The difference between the two redshift values is 
$\Delta_v=(4\pm 78)\,\mathrm{km~s}^{-1}$.

\subsubsection{PSO~J041$+$06, $z=6.321$}
Figure~\ref{fig:nirspectra} shows the Gemini/GNIRS spectrum of this source, covering from the Lyman break to the Mg$\rm II$ broad emission line. 
The Mg$\rm II$ is clearly detected and sets the redshift of this quasar at $z=6.321$ (see Table~\ref{tab:specobslogproperties}). 
This is the lowest redshift quasar confirmed from the $z\gtrsim 6.6$ selection presented in Sect.~\ref{selectionbanadosnew}. Such a low redshift can be explained by a very weak Ly$\alpha$ line (see  Fig.~\ref{fig:z66sel}), which could be intrinsic or due to a proximate absorber truncating the emission (e.g., \citealt{Banados2019}).  
Conversely, the C$\rm IV$ is very weak, with an Equivalent Width (EW) $<10\AA$ (see more details in Sect.~\ref{sec:analysis}), making PSO~J041$+$06 a weak emission line quasar (WLQ; \citealt{Fan1999,DiamondStanic2009}). Furthermore, the C$\rm IV$ is blueshifted with respect to Mg$\rm II$, by $\sim$9000~km~s$^{-1}$, the largest ever measured in the early Universe (see Sect.~\ref{sec:civbluesfhit} for caveats and discussion).

\subsubsection{PSO~J067$-$14, $z = 6.705$}
\begin{figure}[h!]
    \centering
    \includegraphics[width=0.95\columnwidth]{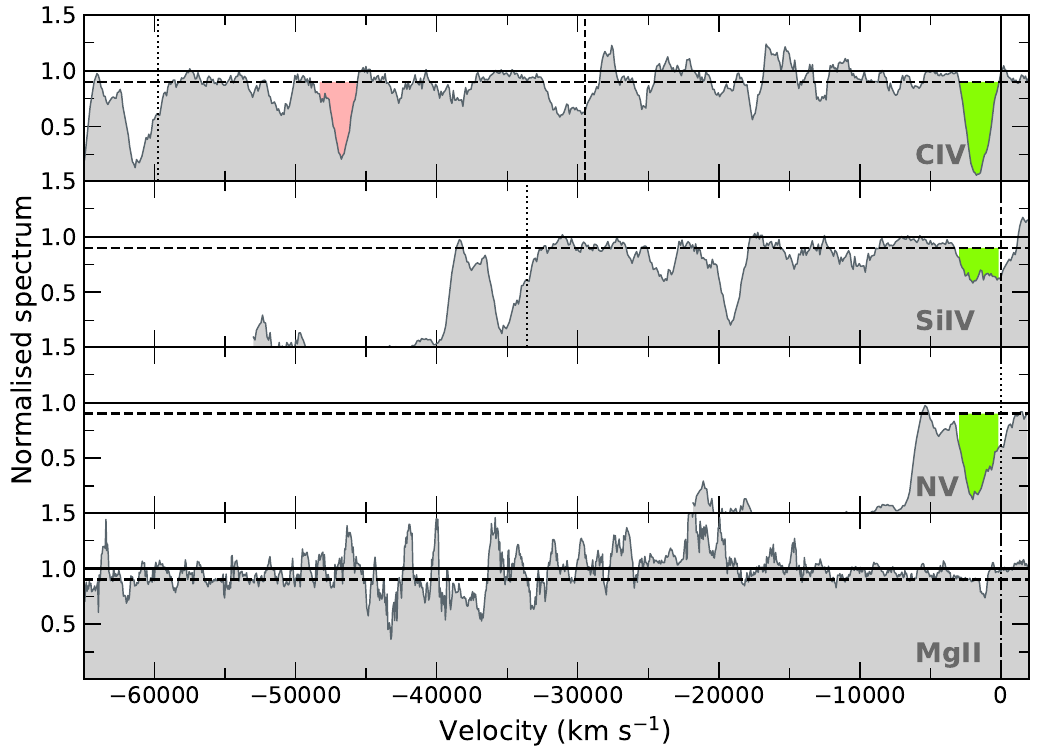}
    \caption{\small Normalised spectrum of PSO~J067$-$14, smoothed to 500~km~s$^{-1}$. The velocity axis in each panel is relative to the rest-frame wavelength of the ionic species indicated by the label. Vertical solid, dashed, dotted and dashed-dotted lines indicate the velocity associated with C$\rm IV$, Si$\rm IV$, N$\rm V$ and Mg$\rm II$ emission lines, respectively. The solid (dashed) horizontal line represents a flux level of 1.0 (0.9). BAL troughs, corresponding to a flux level $< 0.9$ \citep{Weymann1991}, are highlighted as green shaded areas. The red shaded area indicates a potential extremely high-velocity ($\sim48000$~km~s$^{-1}$) C$\rm IV$ BAL.}
    \label{fig:balPSOJ067}
\end{figure}

Similarly to PSO~J037$-$08, this quasar is part of the JWST ASPIRE program \citep{ASPIREproposal} and 
\cite{Wang2024} report the [CII] detection of PSO~J067$-$14, finding a redshift of $z_{\rm [CII]}=6.7142\pm0.0006$. 
The difference in ~km~s$^{-1}$ between the [CII]-based redshift and the one computed in this work from the Mg$\rm II$ line (6.705$\pm$0.002, see Table~\ref{tab:specobslogproperties}) is $\Delta_v=(359\pm 85)\,\mathrm{km~s}^{-1}$, which is in line with values estimated in quasars at similar redshift (e.g., \citealt{Schindler2020,Yang2021}).\\
As several absorption features are clearly visible in the spectrum, we used the method adopted in \cite{Bischetti22,Bischetti23} to detect and characterize possible BAL outflows in this quasar. 
The main steps include modeling the intrinsic quasar emission using a composite template spectrum that matches the continuum slope and equivalent width of the C$\rm IV$ emission line in PSO~J067$-$14. 
The composite template is normalized to the median continuum flux in the wavelength range rest-frame 1650--1750\,\AA\ (according to the Mg$\rm II$ redshift). 
A normalized spectrum is obtained by dividing the spectrum of PSO~J067$-$14 by the composite template, as shown in Figure~\ref{fig:balPSOJ067}. \\
We detected an absorption feature blue-ward of C$\rm IV$, tracing a BAL outflow with a width of $\sim2700$~km~s$^{-1}$, a maximum velocity of $\sim2900$~km~s$^{-1}$ and a balnicity index BI$\sim$1600~km~s$^{-1}$, that is a modified equivalent width of the BAL absorption, calculated according to Eq.~(1) in \cite{Bischetti23}, which follow the traditional BI definition of \cite{Weymann1991}. 
We consider that C$\rm IV$ optical depth is usually similar or larger than the Si$\rm IV$ depth in BAL quasars (e.g. \citealt{Gibson2009,Dunn2012}). 
This allows us to use the velocity range of the C$\rm IV$ BAL troughs to identify absorption associated with Si$\rm IV$ and N$\rm V$ as it has been done in the literature (e.g., \citealt{Bruni2019,Bischetti23}). Indeed, we identify an associated Si$\rm IV$ and N$\rm V$ absorption as highlighted in Figure~\ref{fig:balPSOJ067}. \\
We note an additional absorption feature at $\sim$1300\,\AA\ that might be interpreted as related to the CII $\lambda1335$\AA\ transition, which would make PSO~J$067-14$ as the second low-ionization broad absorption line (Lo-BAL) quasar identified at $z>6$ (\citealt{Bischetti23,Bischetti24}). 
However, the lack of strong Mg$\rm II$ absorption questions the above interpretation (see last panel of Fig.~\ref{fig:balPSOJ067}). 
Alternatively, the $\sim$1300\AA\ absorption feature might be associated with an extremely high-velocity ($\sim48000$~km~s$^{-1}$, red line in Figure~\ref{fig:balPSOJ067}) C$\rm IV$ BAL outflow. 
Similar velocities have been observed in other three $z\gtrsim6$ quasars by \cite{Bischetti22} and \cite{WangFeige2021}, but are rare (a few \%) in SDSS BAL quasars at $z<4$ (e.g., \citealt{Rodriguez20}).\\
Finally, the spectrum of PSO~J067$-$14 reveals the presence of an absorption system at $z_{\rm abs}=4.94$, traced by both the Mg$\rm II$ and Fe$\rm II$ transitions at 2374\AA\ and 2382\AA. 

\subsubsection{PSO~J082$-$38, $z = 5.10$}
Similarly to MQC~J021$+$19, this spectrum shows potential evidence of absorption blue-ward of N$\rm V$. 
As this absorption has a width of $\sim3600$~km~s$^{-1}$, PSO~J082$-$38 likely hosts a N$\rm V$ BAL outflow with a maximum velocity of $\sim$4200~km~s$^{-1}$. A higher S/N spectrum, also covering redder wavelengths, is necessary to confirm and properly characterize the BAL features also from other atomic species. 

\subsubsection{MQC~J133$-$02, $z = 4.67$}
This source is reported in the MQC with a z$_{\rm phot}=5.9$. 
However, given its detection in $r_{\rm P1}$ band, we considered that the photo-$z$ was overestimated. 
Indeed,  we measured $z=4.67$ from the NTT/EFOSC2 spectrum. \\
According to the MQC, MQC~J133$-$02 has a radio detection at 0.887~GHz in RACS and at 1.4~GHz in the Faint Images of the Radio Sky at Twenty-cm (FIRST, \citealt{Becker1994}) and in the NVSS.
We also retrieved the RACS-MID image at 1.37~GHz from CASDA.
On the RACS-MID image, we performed a single Gaussian fit using the CASA task IMFIT to quantify the flux density. 
MQC~J133$-$02 is clearly detected at 3~GHz in the VLASS 1.1, 2.1 and 3.1 epoch images. 
From a Gaussian fit performed in CASA we estimated the three radio flux densities. 
MQC~J133$-$02 is also marginally detected ($\sim$3$\sigma$) in the TIFR GMRT Sky Survey (TGSS, \citealt{Intema2017}) at 150~MHz, with a peak flux density of $\sim$11 mJy.  
Table~\ref{hmqradiodet} summarizes the radio detections available for this quasar, while Figure~\ref{fig:radiospec} (\textit{central panel}) shows the radio spectral energy distribution. 
No significant variability is detected between FIRST and NVSS at 1.4~GHz and between the three VLASS epochs (all measurements are consistent within 1$\sigma$).
By assuming a single power-law for the continuum radio emission, we obtained a radio spectral index ($\alpha_r$) between 0.8875 and 3.0~GHz of 0.56$\pm$0.20. 
Then, we computed the values of radio-loudness as already done for MQC~J021$+$19. 
Given that 5~GHz in the rest frame corresponds to a frequency of 0.880~GHz, we used the RACS flux density and the estimated spectral index to compute the 5~GHz rest frame radio flux density. 
The value of the flux density at 4400\,\AA\ rest frame was derived from the $z$-band de-reddened PS1 magnitude assuming the optical spectral index of \citet[$\alpha_{\nu}$ = 0.44]{VandenBerk2001}. 
We obtained R = 280$\pm$130, where the error takes into account the uncertainties on the flux densities, on the radio spectral index, and on the redshift of the source.
\begin{table}[!ht]
\caption{\small Summary of the archival radio observations of MQC~J133$-$02.}
\label{hmqradiodet}
\centering
\begin{small}
\begin{tabular}{p{0.7cm}p{1.2cm}p{1.65cm}p{1.3cm}l}
\hline\hline
 Obs. Freq. & S$_{\nu}$ & Survey & Resolution & Obs. date \\
 (GHz)     & (mJy)       &   & (arcsec) &      \\
 (1) & (2) & (3) & (4) & (5) \\
 \hline
 0.150       &  $\sim$11$^a$ & TGSS  & 25 & 2016-03-15 \\
 0.8875      &  8.96$\pm$1.13 & RACS      & 25   & 2019-04-29 \\
 1.3675      &  6.13$\pm$0.93 & RACS      & 25 & 2021-01-18 \\
 1.4         &  5.2$\pm$0.4 & NVSS      & 45   & 1993-11-15 \\
 1.4         &  5.76$\pm$0.30  & FIRST     & 5   & Sept. 1998 \\
 3.0         &  3.53$\pm$0.30 & VLASS 1.1 & 2.5  & 2017-09-30 \\
 3.0         &  4.18$\pm$0.39 & VLASS 2.1 & 2.5  & 2020-08-16 \\
 3.0         &  3.76$\pm$0.26   & VLASS 3.1 & 2.5  & 2023-03-05 \\
\hline
\end{tabular}
\tablefoot{Col (1): Observed frequency in GHz; Col (2): integrated flux density in mJy; $a$: the TGSS value is the flux density of the brightest pixel; Col (3): reference survey; Col (4): angular resolution in arcsec; Col (5): Date of the observation.}
\end{small}
\end{table}

\subsubsection{PSO~J143$-$21, $z = 5.97$}
PSO~J143$-$21 was part of the candidates selected by \cite{Banados2023}, but its spectroscopic confirmation occurred after that publication was completed.  
Therefore, it satisfied all the selection criteria reported in Sect.~\ref{selectionbanados23} of \cite{Banados2023}.
However, its color in the latest version of PS1 data release (DR2) is $i_{\rm P1}-z_{\rm P1} = 1.9$, and would not have been selected.\\
The analysis of the new spectrum obtained recently with  Magellan/LDSS3 show three distinct absorption systems, identified by the Mg$\rm II$ doublet and Fe$\rm II$ transitions at 2586\AA/2600\AA. These systems are located at redshifts z$_{\rm abs}=2.6624$, 2.6584 and 2.4650 respectively.

\subsubsection{PSO~J162$-$01, $z = 6.656$}
The discovery of this quasar has been already reported by \cite{WangFeige2017}.
In this work we show unpublished spectra, taken with VLT/FORS2 and Magellan/FIRE (see Table~\ref{tab:specobslogproperties} for details).
The Mg$\rm II$-based redshift estimated by our analysis ($z = 6.656\pm0.001$, see Table~\ref{tab:specobslogproperties}) is consistent within 2$\sigma$ with the one reported in \cite{Farina2022} (6.640$^{+0.008}_{-0.007}$).\\
This quasar is detected in the $YJHKs$ bands of VIKING DR5: $Y = 20.99\pm$0.10, $J = 21.44\pm$0.12, $H = 20.49\pm$0.11, $Ks = 20.43\pm$0.12. 
These values are consistent with the photometry reported for this source in Table~\ref{tab:decalsmag}.\\
Finally, the Magellan/FIRE spectrum reveals the presence of two absorption systems traced by both the Mg$\rm II$ doublet and Fe$\rm II$ transitions at 2586\AA/2600\AA: z$_{\rm abs}=3.497$ and 3.746, respectively.

\subsubsection{PSO J164$+$29, $z=6.585$ }
At the optical coordinates of this quasar we found a $\sim$4.5$\sigma$ radio detection in the LOFAR LoTSS survey (144~MHz). 
The quasar is not detected in any other radio survey that covers this part of the sky (e.g., RACS, FIRST, NVSS, VLASS).
We measured the flux density directly from the LOFAR LoTSS image by using the CASA software. 
The source is unresolved, with a peak flux density of $(451 \pm 97)\,\mu$Jy.
This flux density corresponds to a radio luminosity at 144~MHz equal to 3.2$\pm$0.6$\times$10$^{41}$~erg~s$^{-1}$.
Then we computed the value of R, by assuming a radio spectral index of 0.29 (typical of high redshift quasars detected in LoTSS, see \citealt{Gloudemans2022}) and by computing the flux density at 4400\,$\AA$\ starting from the W1 magnitude (see Table~\ref{tab:decalsmag}, i.e., 4400$\AA$\ rest frame corresponds to 33374$\AA$\ in the observed frame, which matches the W1 filter). 
We obtained a value of 21$\pm$10, which allow us to define PSO~J164$+$29 as a radio-loud source.
Additional radio data are needed to compute the actual radio spectral index of the source and then a more precise value of R.

\subsubsection{PSO~J200$-$13, $z = 4.71$}
PSO~J200$-$13\footnote{During the revision process of this work, PSO~J200$-$13 was independently reported by \cite{Ighina2025}.} is a powerful radio source, with a flux density of 36$\pm$1.2~mJy at 1.4\, GHz (NVSS). 
This quasar is also detected in the following radio surveys (see Table~\ref{j200radiodet}): TGSS at 150~MHz, the GaLactic and Extragalactic All-sky MWA Survey (GLEAM, \citealt{Wayth2015,HurleyWalker2017}) from 76 to 231~MHz, the Westerbork in the Southern Hemisphere (WISH) Survey (\citealt{DeBreuck2002}) at 325~MHz, RACS at 888~MHz, the Widefield ASKAP L-band Legacy All-sky Blind surveY (WALLABY, \citealt{Koribalski2020}) at 1.37~GHz, the VLASS at 3~GHz and the Parkes-MIT-NRAO (PMN) survey (\citealt{WrightA1994}) at 4.85~GHz. 
WALLABY and VLASS flux densities have been computed from a single Gaussian fit (CASA IMFIT) on the images.
\cite{Massaro2014} included PSO~J200$-$13 in the LOw frequency Radio CATalog of flat spectrum sources (LORCAT), due to the availability of NVSS and WISH flux densities ($\alpha_{low}=0.37\pm0.05$). 
Fig.~\ref{fig:radiospec} (\textit{right panel}), shows that PSO~J200$-$13 radio emission is characterized by variability on time scales of few years (observed frame, i.e., hundreds of days in the rest frame). 
Because of the different angular resolutions of the radio surveys in which PSO~J200$-$13 is detected, the measured flux difference could also be due to the presence of other nearby sources that, in some cases, are not resolved because of a large radio beam. 
Therefore, we checked that there were no contaminating sources, and we can confirm that the flux density measured in all radio images is coming from the quasar. 
To compute the source's radio spectral index, we use the closest flux densities in time (the WALLABY Dec.~2021 and VLASS2.2 observed in Jan.~2022, see Fig.~\ref{fig:radiospec}) to minimize variability effects.
We obtained a flat spectral index: $\alpha_{1.3 \rm GHz}^{3 \rm GHz}$ = $-$0.11 $\pm$ 0.15. 
Then, we computed the values of radio-loudness, following the same procedure for MQC~J021$+$19 and MQC~133$-$02. 
We started from the $z$-band de-reddened magnitude to compute the value of the flux density at 4400\AA\ rest frame (by assuming again $\alpha_{\nu} = 0.44$). 
Given that 5~GHz in the rest frame corresponds to a frequency of 0.873~GHz, we used the RACS flux density and the estimated spectral index to compute the 5~GHz rest frame radio flux density. 
We obtained R = 830$\pm$70. 
The high value of radio loudness, the flat radio spectral index, and variability indicate that this source could be classified as a blazar, i.e. a radio-loud AGN with the relativistic jet pointed towards the Earth, in agreement with the classification presented in \cite{Massaro2014}.
Simultaneous radio observations on a wide range of frequencies are necessary to definitively assess the radio spectral shape of this source. 
Furthermore, PSO~J200$-$13 has a clear detection at X-ray frequencies in the eROSITA-DE Data Release 1 (DR1, \citealt{Merloni2024}). 
The source is identified as 1eRASS~J132206.5-132350, and the separation between optical PS1 and eRASS coordinate is 4\farcs4 (i.e. well within the 16$''$ PSF of eROSITA, \citealt{Merloni2024}). 
The net counts in the 0.2-2.3 keV bands are 19.22, with 4.73 expected background counts. 
The catalog also reported an estimated X-ray flux in the same band of 1.26$\pm$0.31$\times$10$^{-13}$ erg~s$^{-1}$~cm$^{-2}$ (see \citealt{Merloni2024} for more details). 
More information about the X-ray eROSITA properties of this source will be presented in Hämmerich et al.\ (in prep.) and Sbarrato et al.\ (in prep.). 

 \begin{table}[!ht]
 \caption{\small Summary of the archival radio observations of PSO~J200$-$13.}
 \label{j200radiodet}
 \centering
 \begin{small}
 \begin{tabular}{p{0.8cm}p{1.5cm}p{1.65cm}p{1.3cm}l}
 \hline\hline
 Obs. Freq. & S$_{\nu}$ & Survey & Resolution & Obs. date \\
 (GHz)     & (mJy)       &   & (arcsec) &      \\
 (1) & (2) & (3) & (4) & (5) \\
 \hline
 0.150       &  48.8$\pm$7.6   & TGSS  & 25 & 2016-03-15 \\
 0.1542      &  102.0$\pm$28.0 & GLEAM$^a$ & 100 & 2014-03-10 \\
 0.2005      &  98.59$\pm$10.81  & GLEAM$^a$ & 100   & 2014-03-10 \\
 0.325       &  62$^b$       & WISH     & 54     &   2002\\
 0.8875      &  40.17$\pm$3.31 & RACS low      & 15   & 2020-03-26 \\
 1.3675      &  32.94$\pm$0.57 &  RACS mid & 10 & 2021-01-16 \\
 1.3675      &  28.28$\pm$1.43 &  WALLABY &  30 & 2021-11-23 \\
1.3675       &  29.47$\pm$1.48  &  WALLABY & 30 & 2021-12-09 \\
 1.4         &  36.96$\pm$0.90 & NVSS      & 45   & 1995-01-15 \\
 3.0         &  35.25$\pm$3.54 & VLASS 1.2 & 2.5  & 2019-04-25 \\
 3.0         &  32.28$\pm$3.24 & VLASS 2.2 & 2.5  & 2022-01-31 \\
 3.0         &  35.19$\pm$3.53 & VLASS 3.2 & 2.5  & 2024-07-22 \\
 4.85        &  44.0$\pm$11.0  & PMN & 250 & 1990 \\
 \hline
 \end{tabular}
 \tablefoot{ Col (1): Observed frequency in GHz; Col (2): integrated flux density in mJy; $a$: the GLEAM flux density refer to the wide band from 139 to 170 MHz and from 170 to 231 MHz; $b$: the WISH value is reported without an uncertainty in the catalog; the error on the WALLABY measurements takes into account both the statistical uncertainty from the Gaussian fit the a 5\% of systematic error as reported by \cite{Glowacki23}; Col (3): reference survey; Col (4): angular resolution in arcsec; Col (5): Date of the observation.}
 \end{small}
 \end{table}

\subsubsection{PSO~J217$+$04, $z = 6.72$}
This source was first considered a promising quasar candidate from 900\,s \textit{Magellan}/FIRE longslit observation carried out in May 2023, identified from the 2D spectrum.  
However, these data were not of sufficient quality for publication. 
Therefore, a follow-up with LBT/MODS (covering the wavelength range from $\sim$6000 to $\sim$10000$\AA$) and LBT/LUCI (covering the NIR up to about 24000$\AA$) was carried a month later (see Table~\ref{tab:specobslogproperties}). 
In this paper we report in Fig.~\ref{fig:nirspectra} only the coadd of the LBT spectra. \\
The low value ($<10$\AA) of C$\rm IV$ EW (see Table~\ref{bhmasses}) allow us to classify PSO~J217$+$04 as a WLQ, similar to PSO~J041$+$06.

\section{Analysis of the NIR spectra: emission line properties and black hole masses}
\label{sec:analysis}
\begin{figure*}[!ht]
    \centering
{\includegraphics[width=0.3\linewidth]{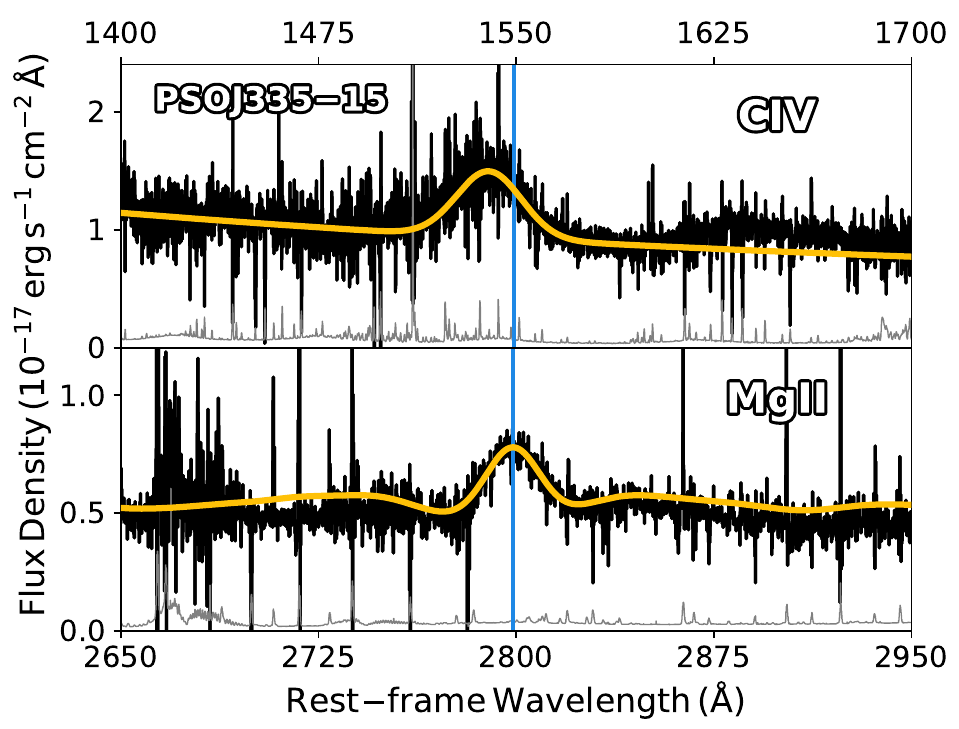}\hspace{0.01cm}
\includegraphics[width=0.3\linewidth]{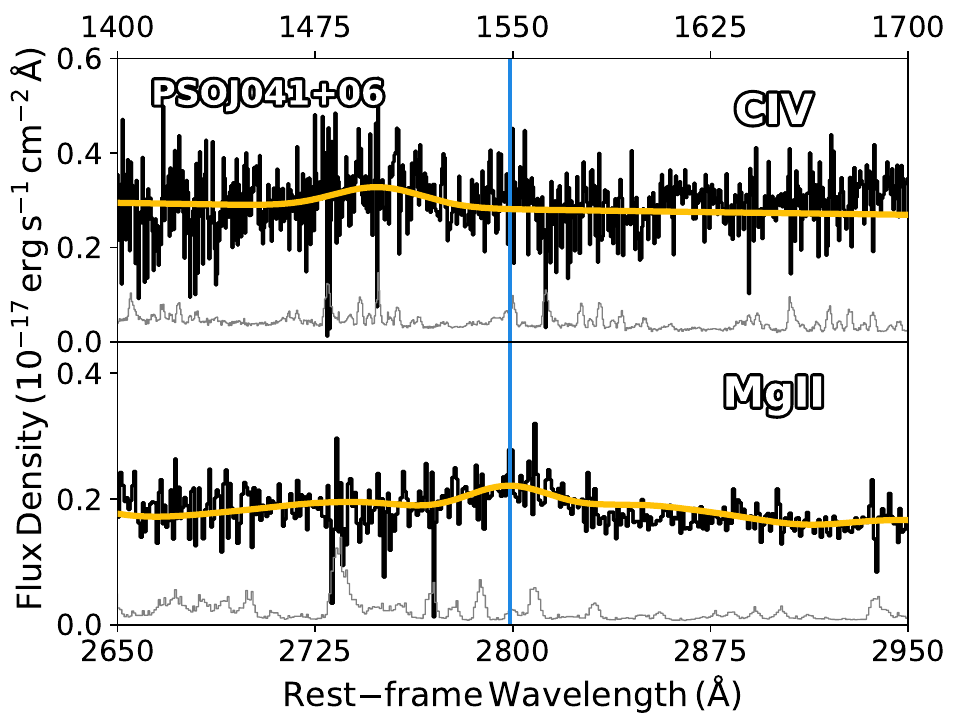}\hspace{0.01cm}
\includegraphics[width=0.3\linewidth]{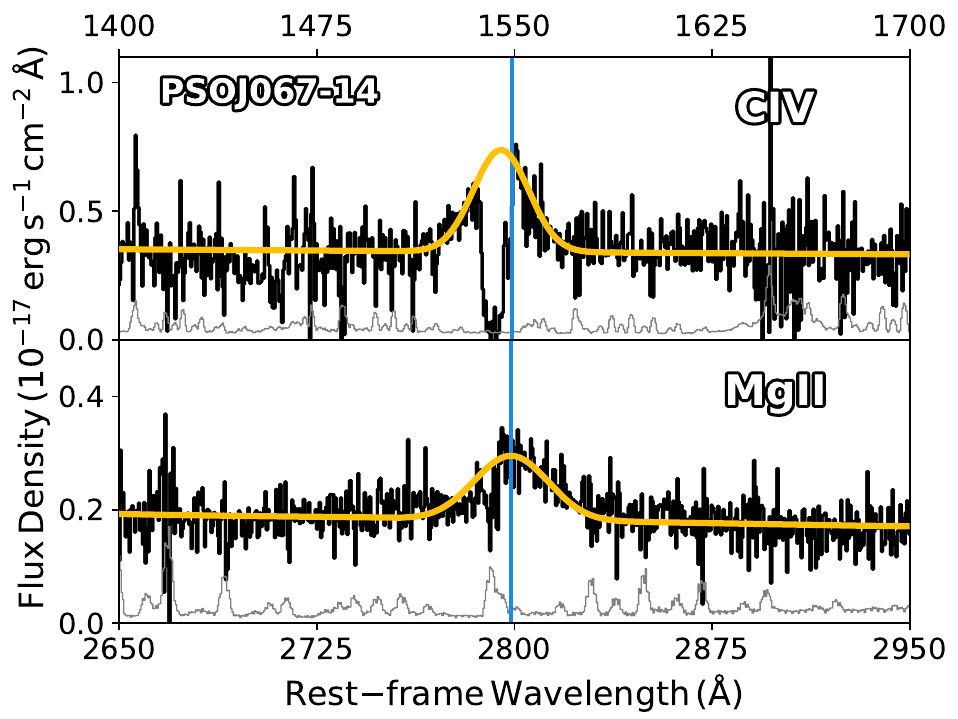}\hspace{0.01cm}
\includegraphics[width=0.3\linewidth]{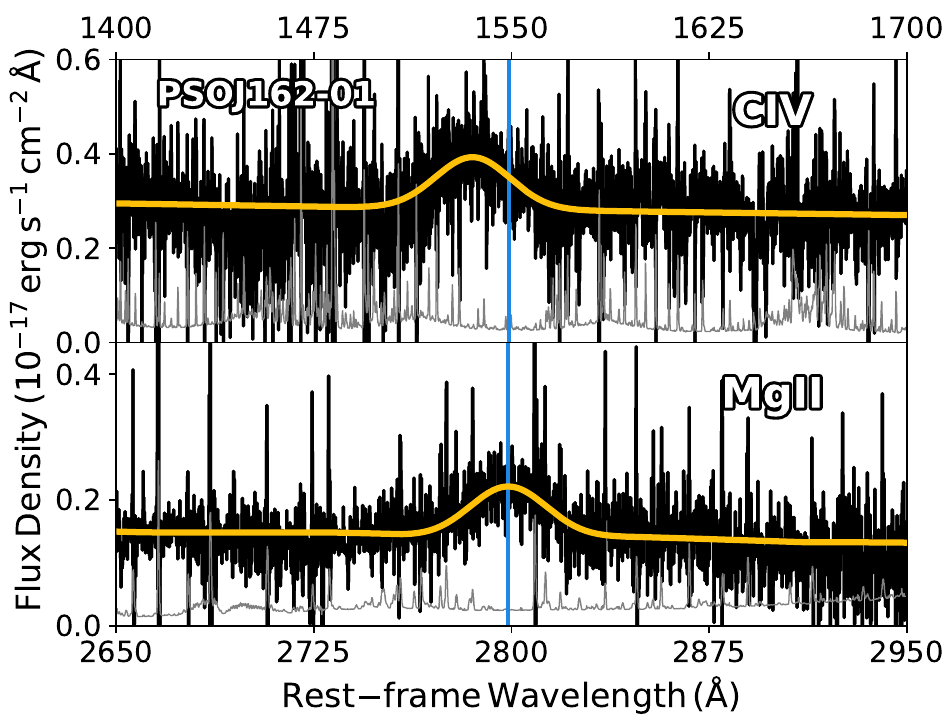}\hspace{0.01cm}
\includegraphics[width=0.3\linewidth]{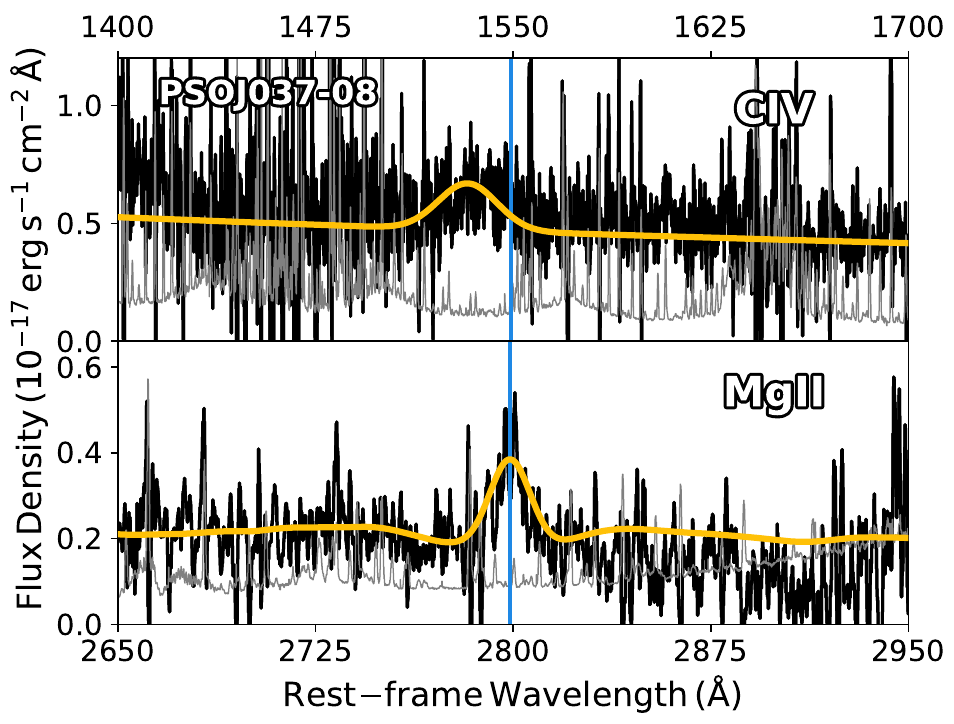}\hspace{0.01cm}
\includegraphics[width=0.3\linewidth]{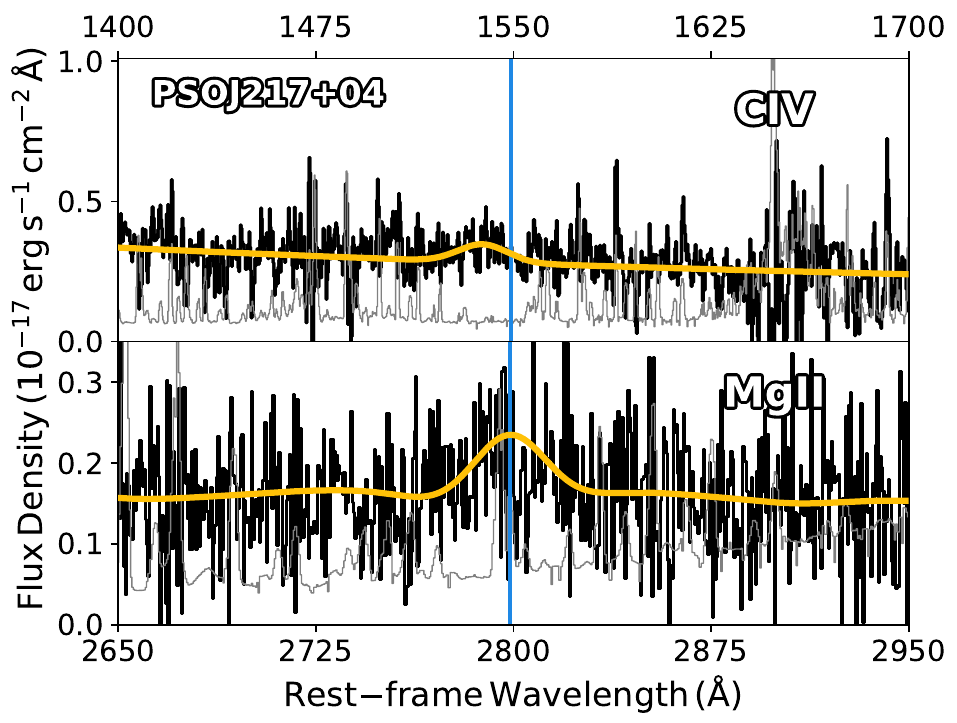}\hspace{0.01cm}
\includegraphics[width=0.3\linewidth]{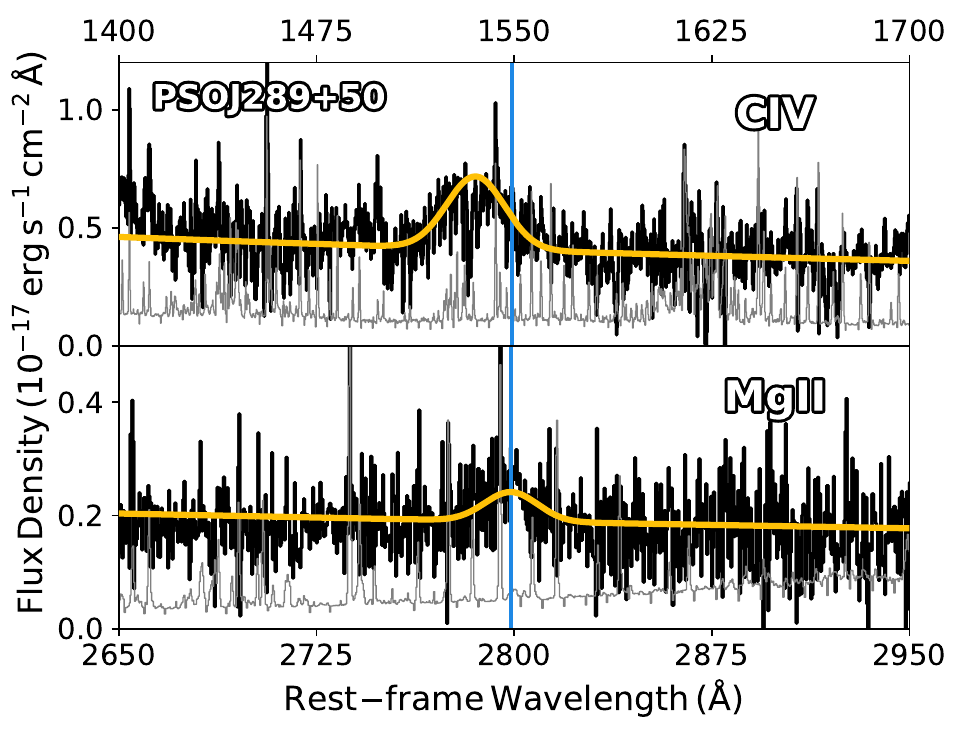}\hspace{0.01cm}
}
\caption{\small Close-up views of the spectral fit of the C$\rm IV$ and Mg$\rm II$ BELs. 
The orange curves depict the fitted models, comprising Gaussian components for the emission lines and a continuum model (see text for details). 
The transparent blue vertical line indicates the position of the lines based on the redshift reported in Table~\ref{tab:specobslogproperties} (i.e., the Mg$\rm II$-based redshifts). The same redshifts were used to draw the rest-frame wavelength $x-$axes.
To fit the C$\rm IV$ line in the spectra of PSO~J067$-$14 and PSO~J289$+$50, the absorption features have been masked.}
\label{sculptorfit}
\end{figure*}

In this section we report the analysis of the spectra shown in Fig.~\ref{fig:nirspectra} for the study of the C$\rm IV$ and Mg$\rm II$ BELs and the consequent estimation of black hole masses. \\
We analyzed the NIR spectra of PSO~J037$-$08, PSO~J041$+$06, PSO~J067$-$14, PSO~J162$-$01, PSO~J217$+$04, PSO~J289$+$50 and PSO~J335$-$15 (see Fig.~\ref{fig:nirspectra}) using the Sculptor software\footnote{\url{https://sculptor.readthedocs.io/en/latest/}} (\citealt{Schindler2022}) to derive the properties of these two BELs and compute black hole masses. 
We followed an approach used in several studies in the literature (e.g., \citealt{Mazzucchelli2017,Vito2022,Farina2022}).
First, we subtracted the continuum emission – which is described by a power law ($f_{pl} \propto (\lambda/2500\AA)^{\alpha_{\lambda}}$), an iron pseudo-continuum template, and a Balmer pseudo-continuum – from the spectra. 
We modeled the Fe II contribution with the empirical template of \cite{Vestergaard2001}, which is used in the derivation of the scaling relation that we later consider for estimating the black hole mass of the quasars. 
To perform the continuum fit, we chose a region of the quasar continuum free of broad emission lines and of strong spikes from residual atmospheric emission.
We then subtracted the entire pseudo-continuum model from the observed spectra, and modeled the two BELs with Gaussian functions.
For PSO~J067$-$14 and PSO~J289$+$50 we masked the absorption feature in the middle of the C$\rm IV$ to fit the line properly.
The results of this fitting procedure are shown in Fig.~\ref{sculptorfit} and Table~\ref{bhmasses}.

\subsection{The C$\rm IV$--Mg$\rm II$ velocity shift}
\label{sec:civbluesfhit}
\begin{figure}
    \centering
    \includegraphics[width=0.5\textwidth]{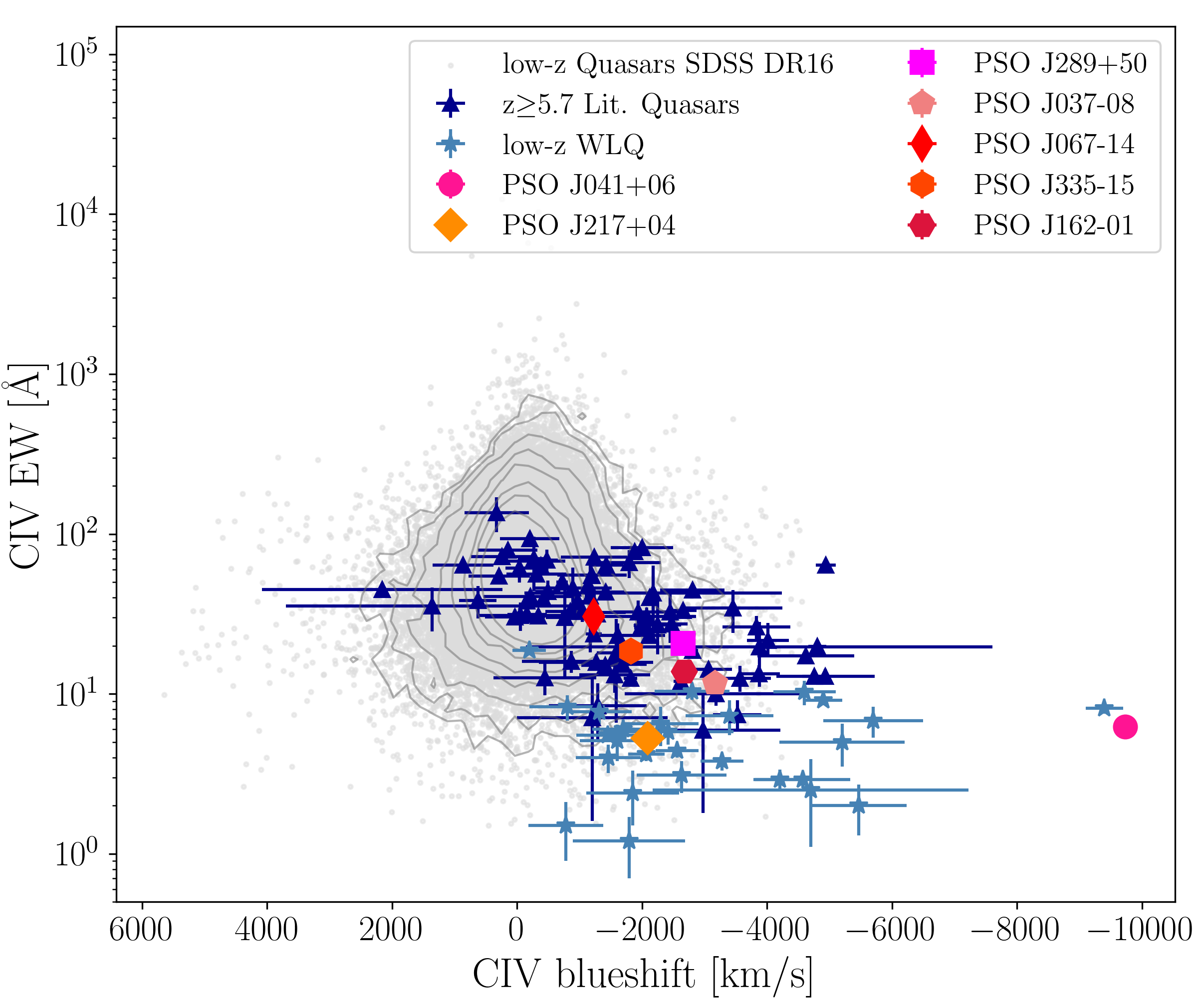}
    \caption{\small C$\rm IV$ EW as a function of the C$\rm IV$--Mg$\rm II$ velocity shift for the quasars analyzed in this work. Dark blue triangles represent $z>5.7$ quasars from \cite{Shen2019}, \cite{Schindler2020} and \cite{Yang2021}, while light blue stars shows the WLQs samples of \cite{Wu2011}, \cite{Luo2015} and \cite{Plotkin2015}.  
    The gray contours and gray dots show the low-$z$ comparison sample at 0.35$< z <$2.25 from the SDSS DR16 catalog of \cite{Wu2022}.
 Our sample generally follows the high-$z$ quasars distribution. The only exception is the quasar PSO~J041$+$06 which shows a velocity shift $>$9000 km~s$^{-1}$ and a EW value of 6\AA~, typical of a WLQ. The only object similar to PSO~J041$+$06 is J1219$+$1244 at $z=$2.233.}
    \label{fig:civshiftew}
\end{figure}

Velocity shifts between emission lines in quasars spectra were first identified decades ago \citep[e.g.,][]{Gaskell1982} and remain a widely debated topic for quasars at any cosmic time 
\citep[e.g.,][]{Shen2016,Coatman2017,Ge2019,Meyer2019,Schindler2020,Onoue2020,Yang2021,Stepney2023}.
Broad high-ionization lines, such as C$\rm IV$ and Si$\rm IV$, are known to exhibit particularly large blueshifts compared to lower-ionization lines (e.g., Mg$\rm II$). 
The origin of these shifts is often attributed to non-gravitational effects, such as radiation-driven outflows, most likely originating in disk winds (e.g., \citealt{Richards2011} and reference therein). \\
Strong C$\rm IV$--Mg$\rm II$ blueshifts ($>$3000 km~s${-1}$) are commonly observed in a large fraction of $z\geq 6$ quasars \citep[e.g.,][]{Mortlock2011,Derosa2014,Mazzucchelli2017,Banados2018,Meyer2019,Shen2019} and recent studies pointed out an evolution with redshift (up to $z\sim$7) of the C$\rm IV$--Mg$\rm II$ shift \citep[e.g.,][]{Meyer2019,Schindler2019,Yang2021}, implying a change of quasar-driven winds properties over cosmic time. 
Large C$\rm IV$ blueshifts translate into asymmetrical broad-line profiles, 
suggesting that non-virial motions have a significant effect on the observed emission velocity profile. 
To date, the strongest C$\rm IV$ velocity shifts have been observed in the so-called WLQs, which exhibit a rest frame equivalent width (REW) values smaller than 10\AA\ \citep[e.g.,][]{DiamondStanic2009}.\\
In this work we measured the C$\rm IV$--Mg$\rm II$ velocity shift for the quasars reported in Fig.~\ref{fig:nirspectra}. 
We computed the C$\rm IV$ redshift from the peak of the BEL.  
Then we computed the velocity shift between C$\rm IV$ and Mg$\rm II$ by using the following equation: 
\begin{equation}
    \Delta_{\rm CIV-MgII} = \frac{z_{\rm CIV} - z_{\rm MgII}}{(1+z_{\rm MgII})} 
    \times c 
\end{equation}
The velocity shift values are reported in Table~\ref{bhmasses}, together with the estimated REW. 
Figure~\ref{fig:civshiftew} shows our results on the so-called \textit{C$\rm IV$ plane}, where we also report the C$\rm IV$ blueshift and REW of $z>5.7$ quasars from the literature \citep{Shen2019,Schindler2020,Yang2021} and of low-$z$ quasars from the SDSS DR16 (grey dots and contours).
These low-$z$ sources have been selected from the catalog of \cite{Wu2022} to be at 0.35$<z<$2.25 (where the Mg$\rm II$ BEL falls in the SDSS wavelength range) and to have a valid measure of the virial black hole mass, bolometric luminosity, Eddington ratio and emission lines properties. 
We also display low-$z$ WLQs from the works of \cite{Wu2011}, \cite{Luo2015}, and \cite{Plotkin2015}.   
As can be noticed from Fig.~\ref{fig:civshiftew}, all the quasars analyzed in this work show a similar shift to that of the bulk of the high-$z$ population, except for PSO~J041$+$06, which is an outlier in this plot. 
In fact, this quasar has a C$\rm IV$--Mg$\rm II$ shift $>$9000 km~s$^{-1}$, the highest ever recorded at $z>6$. 
The only low-$z$ quasar we could find on a similar C$\rm IV$ parameter space is J1219$+$1244 at $z=2.233$, which is an X--ray weak WLQ from the sample of \cite{Wu2011}, highlighting the exceptional rarity and uniqueness of such WLQs with wind-dominated Broad Line Regions (BLRs). \\
While the C$\rm IV$ line is detected in the spectrum, its derived properties remain uncertain due to the low S/N and resolution of the data. 
A higher S/N and resolution spectrum is needed (e.g, with VLT/X-Shooter) to better characterize the C$\rm IV$ properties. 

\subsection{Estimates of the black hole mass}
\label{sec:bhmass}
For un-obscured, type-1 quasars, the most common technique used to compute the mass of the central black hole is the single epoch (SE) method.
Based on the assumption that the gas in the BLR is virialized, the line-width of BELs is used to trace the velocity distribution of the gas. 
When combined with the luminosity of the accretion disk continuum (i.e., a proxy of the distance of the gas from the central SMBH), M$_{\rm BH}$ can be estimated from the virial relation \citep[e.g.,][]{Vestergaard2006,Vestergaard2009,Shen2008,Shen2011}. 
This method made it possible to estimate the mass of SMBHs hosted by high-$z$ quasars mainly using C$\rm IV$ and Mg$\rm II$ BELs that fall in the wavelength range of most optical/NIR instruments mounted on ground-based telescopes \citep[e.g.,][]{Jiang2007,Wu2015,Mazzucchelli2017,Mazzucchelli2023,Shen2019,Onoue2019,WangFeige2021,Diana2022,Farina2022,Vito2022,Belladitta2023}.\\
For the Mg$\rm II$ line, we used the single epoch scaling relation presented by \cite{Vestergaard2009}, for a direct comparison with other estimates in the literature:
\begin{equation}
\label{eqmass}
\rm M_{\text{BH}} = 10^{6.86} \left[ \frac{\text{FWHM~(Mg\rm II)}}{1000 \, \text{km/s}} \right]^2 
\left[ \frac{\lambda L_{\lambda} (3000 \, \text{\AA})}{10^{44} \, \text{erg/s}} \right]^{0.5}
\end{equation}

For the C$\rm IV$, we used the scaling relation of \cite{Vestergaard2006}, which is the most frequently used in the literature:
\begin{equation}
\label{eqmass2}
\rm M_{\text{BH}} = 10^{6.66} \left[ \frac{\text{FWHM (C\rm IV$_{\rm corr}$)}}{1000 \, \text{km/s}} \right]^2 
\left[ \frac{\lambda L_{\lambda} (1350 \, \text{\AA})}{10^{44} \, \text{erg/s}} \right]^{0.53}, \end{equation}
where FWHM~(C \rm IV$_{\rm corr}$) is the width of the C$\rm IV$ line corrected by potential blueshift effects according to \cite{Coatman2017}:
\begin{equation}
\label{blueshift}
    \rm FWHM~(C \rm IV_{corr}) = \frac{FWHM~(C\rm IV_{measured})}{0.36 \times \frac{C\rm IV_{blueshift}}{1000\, km \,s^{-1}} + 0.61 }
\end{equation}

\begin{figure}[!h]
    \centering
    \includegraphics[width=0.5\textwidth]{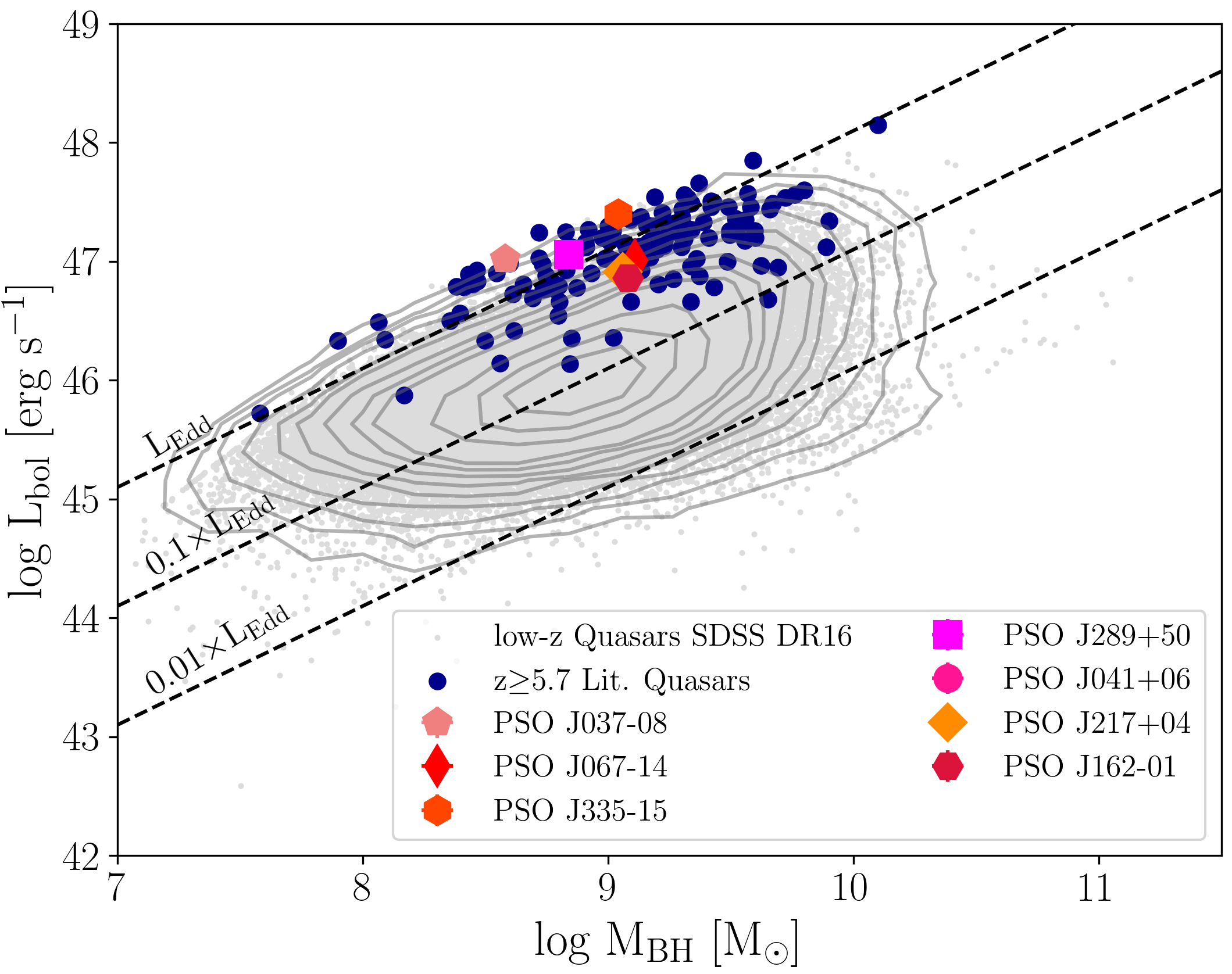}
    \caption{\small Bolometric luminosities versus Mg$\rm II$ black hole masses of seven quasars reported in this paper. We also show $z\geq6$ quasars already published in the literature (dark blue dots, \citealt{FanX2023ARA,Mazzucchelli2023}) and quasars at 0.35$<z<$2.25 from the SDSS DR16 catalog (grey contours and grey points, \citealt{Wu2022}).
    Dotted black lines show the location of constant accretion rate at 0.01, 0.1 and 1 times the Eddington luminosity. 
    Our quasars reside in the same parameter space as the population of $z>6$ quasars.}
    \label{fig:bhmass}
\end{figure}

In Eq.~(\ref{eqmass}) and Eq. (\ref{eqmass2}) $\lambda L_{\lambda}(3000\,\AA)$ and $\lambda L_{\lambda}(1350\,\AA)$ are the monochromatic luminosity at 3000\,\AA\ and 1350\,\AA\ derived from the best-fit model of the continuum. 
Table~\ref{bhmasses} reports the estimated values of black hole masses, including only statistical uncertainties. 
In addition to this statistical error, we must also consider the intrinsic scatter of the used scaling relation: 0.36-0.43~dex for C$\rm IV$ from \cite{Vestergaard2006} and 0.55~dex for Mg$\rm II$ from \cite{Vestergaard2009}.\\
The value of the black hole masses allowed us to derive the Eddington ratio, which quantifies how fast the black hole is accreting with respect to the Eddington limit: $\lambda_{\rm Edd}$ = L$_{\rm bol}$/L$_{\rm Edd}$.
Here L$_{\rm Edd}$ is the Eddington luminosity, the maximum luminosity beyond which radiation pressure will overcome gravity\footnote{L$_{Edd}$ = 1.26 $\frac{M_{\rm BH}}{M_{\odot}}$ $\times$ 10$^{38}$ erg s$^{-1}$}, and L$_{\rm bol}$ is the bolometric luminosity, which is the total energy produced by the quasar per unit of time-integrated over all wavelengths.
We derived the bolometric luminosity by following the equation in \cite{Shen2008}:
\begin{equation}
    L_{\rm bol} = 5.15\pm1.26 \times \lambda L_{\lambda}(3000\,\AA)
\end{equation}
The values are reported in Table~\ref{bhmasses} together with the corresponding values of the Eddington ratio. 
Figure~\ref{fig:bhmass} shows that the quasars in our sample are located close to the line of Eddington luminosity, indicating that their central SMBHs are accreting close to the Eddington limit. 
The quasar PSO~J037$-$08 has an Eddington ratio larger than 2, which suggests a super-Eddington accretion regime. 
This is not uncommon at these redshifts, since other few quasars have been reported to be super-Eddington in the literature (see e.g., \citealt{Banados2021,Yang2021}).\\
The values of M$_{\rm BH}$ and $\lambda_{\rm Edd}$ of the newly discovered high-$z$ quasars in this work are in line with those derived for $z\geq6$ quasars already reported in the literature (e.g., \citealt{Mazzucchelli2023}), as shown in Fig.~\ref{fig:bhmass}.  
These values are also in line with those derived by H$_{\beta}$~$\lambda4861$ for $z>6$ quasars recently observed with JWST (e.g., \citealt{Loiacono2024,Eilers2023}).
For comparison, we also show in Fig.~\ref{fig:bhmass} the distribution of L$_{bol}$ and M$_{\rm BH}$ of low-$z$ quasars (same sources as Fig.~\ref{fig:civshiftew}). 

\section{Summary and conclusions}
\label{sec:conc}
In this work, we reported the discovery of 25 quasars at $4.6<z<6.9$, with 1450\,\AA\ rest-frame absolute magnitudes between $-$25.4 and $-$27.0. 
These objects have been selected from a number of criteria (Sect.~\ref{sec:selection}) and spectroscopically confirmed through dedicated follow-up campaigns at several ground-based facilities (Sect.~\ref{sec:spectroscopy}).
We further presented unpublished spectra of six quasars at $z>6.5$ already known in the literature.

Three of the new quasars (MQC~J021$+$19, MQC~J133$-$02 and PSO~J200$-$13) show intense radio emission, with a  1.4\,GHz  luminosity in the range 0.09--1.0$\times$10$^{34}$ erg~s$^{-1}$~Hz$^{-1}$. 
All of them are classified as radio-loud, with a radio-loudness in the range $\sim$30--850. 
They are compact at all frequencies at the resolution of the available radio images. 
Overall, the radio properties of these 3 new radio-loud quasars (spectral indices, radio-loudnesses, radio luminosities) are within the range of values of other radio-loud quasars at $z > 4.5$ discovered so far (e.g., \citealt{Belladitta2019,Gloudemans2022,Ighina2024}). The quasar PSO~J200$-$13 shows properties that are typical of blazars, i.e. a sub-class of radio-detected quasars with the relativistic jet pointed toward the Earth: flat radio spectrum ($\alpha_r$=$-$0.11), log(R)$>$2.5 and variability at several radio frequencies at different time scales. 
It is also detected in the X-ray by the eROSITA-DE DR1 (f$_{\rm 0.2-2.3keV} \sim 1.3\times10^{-13}$~erg~s$^{-1}$~cm$^{-2}$).

We further reported the radio identification of the quasar PSO~J164$+$29. 
It is detected only in the LOFAR LoTSS at 144\,MHz, and this makes the measured radio loudness uncertain (R$=$21$\pm$10), due to a strong dependence on the assumption of the radio spectral index. Additional multi-frequency radio data will be essential to confirm the radio-loud classification of this quasar.

{For seven sources, we also reported NIR spectroscopic follow-up to unveil the C$\rm IV$ and Mg$\rm II$ broad emission lines (see Fig.~\ref{fig:nirspectra}) and computed single epoch black hole masses and Eddington ratios. 
Their values are in agreement with the typical M$_{\rm BH}$ and $\lambda_{\rm Edd}$ of the population of high-$z$ quasars reported in the literature (e.g., \citealt{Farina2022,Mazzucchelli2023}).

The NIR spectral analysis of the quasar PSO~J041$+$06, classifies it as a WLQ (with C$\rm IV$ EW of 6\,\AA).
Moreover, the source shows the largest C$\rm IV$-Mg$\rm II$ velocity shift ever measured at a $z>6$: $\sim$9000~km~s$^{-1}$, suggesting the presence of strong outflows on the BRL-scale.  
WLQs with a wind-dominated BLR are rare sources: by exploiting the SDSS DR16 catalog and other works in the literature, we were able to find only one analog at low-$z$. 

The quasar PSO~J067$-$14 shows strong absorption features in the NIR spectrum blueward the C$\rm IV$, the N$\rm V$ and the Si$\rm IV$ emission lines. 
These features have a maximum velocity of $\sim$$2900$~km~s$^{-1}$, which classifies the source as a high ionization BAL (HiBAL) quasar. 
An additional absorption feature at $\sim$1300\AA~(rest frame) might be associated with an extremely high-velocity ($v\sim$0.16c) C$\rm IV$ BAL outflows.

Two other quasars (MQC~J021$+$19 and PSO~J082$-$38) could be classified as HiBALs thanks to the absorption feature blue-ward the N$\rm V$ clearly visible in their optical spectra. 
However, a NIR follow-up is necessary to confirm this classification by detecting the absorption feature in other atomic species. 
The strong radio emission of MQC~J021$+$19 classifies it as a radio-BAL, which are rare sources among the BAL population ($<10-20$\%, e.g., \citealt{Hewett2003,Morabito2019}).

The discovery and characterization of the quasars reported in this work came after extensive spectroscopic follow-up campaigns at several ground-based telescopes. 
A list of all observed objects that turned out to be non-quasars at high$-z$ (73 sources) is given in the Appendix. 
Most of these rejected candidates (51) are ultracool dwarfs, which remain the major source of astrophysical contamination in the search for quasars using primarily color selection and the dropout technique. 

Another source of astrophysical contamination is galaxies at low redshift, where the D4000\,\AA\ Calcium Break mimics the Lyman-$\alpha$ break. 
This is especially true when selecting quasar candidates from radio surveys (see Table~\ref{rejected}). 
However, ultracool dwarfs can also be contaminants in radio selections (see Table~\ref{rejected}), especially if objects with radio flux density at $\sim$mJy are selected, as also demonstrated by \cite{Gloudemans2023}. 

The majority of the newly discovered quasars reported in this paper are located in the southern hemisphere. 
This will provide the ideal ground for follow-up investigations using state-of-the-art (e.g., the Very Large Telescope, the Atacama Large Millimiter Array) and future (e.g., the Extremely Large Telescope, the Square Kilometer Array) ground-based facilities focusing on properties such as the metallicity of the gas surrounding the central SMBHs, intergalactic medium along the line of sight, mass, star formation rate, gas content, and internal dynamics of host galaxies, as well as their large-scale environment. 

The discovery and characterization of the quasars reported in this paper came from the in-depth exploration of surveys already widely used in past years, confirming that these data sets can still be exploited for the discovery of quasars in the bright tail of their luminosity function. 
With the advent of wide surveys, such as the European Space Agency (ESA) \textit{Euclid} (\citealt{Scaramella2022,Mellier2024}) or the Legacy Survey of Space and Time (LSST) by the Vera C.\ Rubin Observatory (\citealt{Ivezic2019}), fainter quasars will eventually be discovered (e.g., \citealt{Barnett2019,Nanni2022,TeeWL2023,Lenz2024}), promising fundamental progress in our knowledge of the faint-end of the quasar luminosity function in the re-ionization epoch.

\begin{landscape}
\begin{table}
\caption{\small Broad emission lines and black holes properties derived from the fit of the NIR spectra shown in Fig.~\ref{fig:nirspectra}.}
\label{bhmasses}
\begin{small}
\begin{tabular}{p{1.3cm}cccccccccccc}
\hline\hline
Quasar Name  & FWHM C$\rm IV_{corr}$ & FWHM Mg$\rm II$ &  $z_{C\rm IV}$ & C$\rm IV_{blueshift}$ & CIV EW & log $\lambda L_{3000}$ & log $\lambda L_{1350}$ & log L$_{bol}$ & log M$_{\rm BH, C \rm IV}$ & log M$_{\rm BH, Mg \rm II}$ & $\lambda_{\rm Edd, C \rm IV}$ & $\lambda_{\rm Edd, Mg \rm II}$ \\ 
 &   km s$^{-1}$ & km s$^{-1}$ &  & km s$^{-1}$ & \AA & erg s$^{-1}$ & erg s$^{-1}$ & erg s$^{-1}$ & M$_{\odot}$ & M$_{\odot}$ & & \\
 (1)     & (2) & (3) & (4)  & (5) & (6) & (7)  & (8) & (9) & (10) & (11) & (12) & (13) \\
\hline
PSO~J037$-$08 & 2900$\pm$95 & 1925$\pm$70 & 6.644$\pm$0.002 & 3170$\pm$40 & 11.5$\pm$0.3 & 46.31$\pm$0.01 & 46.48$\pm$0.01 & 47.02$\pm$0.01 & 8.90$\pm$0.02 & 8.58$\pm$0.02 & 1.05$\pm$0.07 & 2.18$\pm$0.15 \\
PSO~J041$+$06 & 1900$\pm$20 & 5013$\pm$5 & 6.084$\pm$0.003 & 9730$\pm$25 & 6.3$\pm$0.1  & 46.34$\pm$0.02 & 46.09$\pm$0.02 & 46.80$\pm$0.02 & 8.46$\pm$0.03 & 9.31$\pm$0.02 & 1.75$\pm$0.03 & 0.25$\pm$0.02 \\
PSO~J067$-$14 & 4570$\pm$10 & 3555$\pm$5 & 6.685$\pm$0.001 & 1225$\pm$3 & 30.45$\pm$0.05 & 46.30$\pm$0.01 & 46.35$\pm$0.01 & 47.01$\pm$0.01 & 9.22$\pm$0.01 &    9.11$\pm$0.01 & 0.49$\pm$0.01 & 0.63$\pm$0.01  \\
PSO~J162$-$01 & 4140$\pm$10 & 3735$\pm$5 & 6.588$\pm$0.001 & 2673$\pm$3 & 13.79$\pm$0.02 & 46.15$\pm$0.01 & 46.30$\pm$0.01 & 46.86$\pm$0.01 & 9.11$\pm$0.01 & 9.08$\pm$0.01 & 0.45$\pm$0.02 & 0.48$\pm$0.02 \\
PSO~J217$+$04 & 3134$\pm$85 & 3546$\pm$70 & 6.642$\pm$0.002  & 2093$\pm$32 & 5.3$\pm$0.1 & 46.20$\pm$0.01 & 46.28$\pm$0.01 & 46.91$\pm$0.01 & 8.86$\pm$0.03 & 9.06$\pm$0.03 & 0.89$\pm$0.05 & 0.56$\pm$0.02 \\  
PSO~J289$+$50 & 3220$\pm$35 & 2520$\pm$26 & 6.786$\pm$0.002 & 2655$\pm$15 & 20.65$\pm$0.14 & 46.35$\pm$0.01 & 46.44$\pm$0.01 & 47.06$\pm$0.01 & 8.97$\pm$0.01 & 8.84$\pm$0.01 & 0.98$\pm$0.02 & 1.33$\pm$0.04   \\
PSO~J335$-$15 & 5648$\pm$6 & 2617$\pm$2 & 6.465$\pm$0.001 & 1819$\pm$4 & 18.61$\pm$0.02  & 46.78$\pm$0.01 & 46.69$\pm$0.01 & 47.40$\pm$0.01 & 9.43$\pm$0.01 & 9.04$\pm$0.01 & 0.74$\pm$0.01 & 1.82$\pm$0.01\\
\hline
    \end{tabular}
    \end{small}
    \tablefoot{Col (1): Quasar name; Col (2-3) C$\rm IV$ and Mg$\rm II$ full-width at half maximum in km~s$^{-1}$; Col (4) C$\rm IV$ redshift estimation from the peak of the emission line; Col (5) C$\rm IV$ blueshift in km~ s$^{-1}$; Col (6) EW of the C$\rm IV$ BEL; Col (7-8): monochromatic luminosity at rest-frame 1350\AA\ and 3000\AA\; Col (9): bolometric luminosity; Col(10-13) black hole masses and Eddington ratio values, derived from both C$\rm IV$ and Mg$\rm II$ BELs. The uncertainty in the BH mass values reported in the table does not include the uncertainty of the SE method ($\sim$0.40~dex for C$\rm IV$ and 0.55~dex for Mg$\rm II$, see Section~\ref{sec:bhmass}).}
\end{table} 
\end{landscape}

\bibliographystyle{aa} 
\bibliography{ref2} 

\begin{appendix}

\section{Broad-band photometry of the newly reported quasars}
In this section, we present the AB magnitudes — corrected for Galactic extinction as described at the end of Sect.~\ref{sec:intro} — from optical to NIR bands, of all 25 newly identified quasars discussed in this paper. 
We divided the sources into two tables: quasars selected from DELS+PS1 (see Sect. \ref{selectionbanadosnew}) are listed in Table~\ref{tab:decalsmag}, while all others are reported in Table~\ref{tab:photomag}. \\
Only for the sources selected from $YS23$ we report the information taken from the original catalog: ID, z$_{phot}$, z$_{phot}$ range and probability to be a true quasar at that redshift.

\begin{landscape}
\begin{table}
\caption{\small Photometric properties (AB mag corrected for Galactic extinction) of the newly discovered quasars reported in this paper, except for the sources selected from DELS$+$PS1, whose magnitudes are reported in Table~\ref{tab:decalsmag}. 
}
\label{tab:photomag}
\centering
\begin{tiny}
\scriptsize
\begin{tabular}{p{2.45cm}p{0.85cm}p{0.85cm}p{0.9cm}p{0.9cm}p{0.9cm}p{0.9cm}p{0.9cm}p{0.9cm}p{0.9cm}p{0.9cm}p{0.9cm}p{2.2cm}p{1.0cm}p{0.3cm}p{1.1cm}p{0.5cm}}
\hline\hline
Quasar Name  & RA & DEC & $r$ & $i$  & $z$ & $y$ & $J$  & $H$ & $K$ & $W1$ & $W2$ & ref. surveys/follow-up &  Cat. ID & Cat. $z_{phot}$ & Cat. $z_{phot}$ range & Cat. p$_{qso}$ (\%) \\  
    (1)     & (2) & (3) & (4)  & (5) & (6) & (7)  & (8) & (9) & (10) & (11) & (12) & (13) & (14) & (15) & (16) & (17) \\
\hline
PSO~J004.1261$-$35.9420 & 4.12608 &	$-$35.94206 & &  & 20.87$\pm$0.03 & 21.01$\pm$0.11 & 20.20$\pm$0.07 & & 19.84$\pm$0.15 & 20.23$\pm$0.07 & 20.16$\pm$0.15 & DES,VHS,WISE   & 1052413943 & 6.05 & 5.7-6.25 & 99.01 \\
PSO~J011.1707$-$37.1195 & 11.17069 &	$-$37.11954 &  & 
24.28$\pm$0.33 &
20.77$\pm$0.02 &
20.73$\pm$0.06 & 
20.37$\pm$0.08 & & 
19.89$\pm$0.15 &
20.11$\pm$0.06 & 
20.35$\pm$0.16 & 
DES,VHS,WISE  & 1093239958 & 6.25 & 6.15-6.95  & 99.52 \\
PSO~J017.4482$-$48.7625 & 17.44821 & $-$48.76258  &  22.33$\pm$0.04 & 20.59$\pm$0.01 & 20.65$\pm$0.03 & 20.83$\pm$0.10 & 20.73$\pm$0.14 & 20.14$\pm$0.13 & 19.93$\pm$0.15 & 20.19$\pm$0.06 & 20.12$\pm$0.12 & DES,VHS,WISE  & 1142387513 & 5.0  & 4.45-5.3 & 99.75  \\
MQC~J021.3712$+$19.1450$^r$ &	21.37128 & $+$19.14503 &  21.56$\pm$0.10 & 19.83$\pm$0.01 & 20.21$\pm$0.03 & 19.59$\pm$0.04 & 19.62$\pm$0.09 &  & 19.08$\pm$0.09 & 18.25$\pm$0.05 & 18.49$\pm$0.09 & PS1,UHS,WISE    & & &  & \\
PSO~J035.2605$-$18.9787 & 35.26053 & $-$18.97877 & 22.12$\pm$0.03 & 20.46$\pm$0.01 & 20.37$\pm$0.02 & 20.28$\pm$0.06 & 20.30$\pm$0.06 &  & 19.77$\pm$0.13 & 20.37$\pm$0.08 & 19.95$\pm$0.12 & DES,VHS,WISE  & 1267014001 & 5.0 & 4.4-5.65 &  98.51 \\
PSO~J060.0381$-$65.7335 & 60.03819 & $-$65.73352 &    & 23.11$\pm$0.10 & 20.47$\pm$0.02 & 20.72$\pm$0.10 & 20.38$\pm$0.08 & & 20.18$\pm$0.24 & 20.17$\pm$0.05 & 20.06$\pm$0.08 & DES,VHS,CatWISE  & & &  & \\
PSO~J070.9368$-$51.3424 & 70.936859 & $-$51.34243 &   22.28$\pm$0.03 & 20.77$\pm$0.01 & 20.54$\pm$0.02 & 20.49$\pm$0.07 & 20.73$\pm$0.20 & 20.27$\pm$0.20 & 20.51$\pm$0.29 & 20.17$\pm$0.05 & 20.09$\pm$0.09 & DES,VHS,WISE & 1500056114 & 5.05 & 4.3-5.5  & 99.93  \\
PSO~J075.5805$-$18.7999 &	75.58052 & $-$18.79990 & 23.40$\pm$0.11 & 21.34$\pm$0.03 & 20.96$\pm$0.04 & 20.84$\pm$0.11 & 20.79$\pm$0.14 & & & 20.37$\pm$0.07 & 20.04$\pm$0.11 & PS1,DES,VHS,WISE  & 1523525887 & 5.35 & 4.95-5.85 & 98.88 \\
PSO~J078.3185$-$45.1393 &	78.31854 & $-$45.13936 &  22.72$\pm$0.05 & 20.79$\pm$0.01 & 20.81$\pm$0.03 & 20.71$\pm$0.08 & 20.52$\pm$0.11 & & 20.26$\pm$0.24 & 20.19$\pm$0.05 & 20.36$\pm$0.11 & DES,VHS,WISE   & 1540630274 &  5.0  & 4.9-5.3 &  98.24 \\ 
PSO~J082.2068$-$38.5674 & 82.20688 & $-$38.56745 &   22.27$\pm$0.03 & 20.36$\pm$0.01 & 20.16$\pm$0.01 & 20.06$\pm$0.04 & 19.89$\pm$0.07 & & 19.13$\pm$0.13 & 18.82$\pm$0.02 & 18.74$\pm$0.03 & DES,VHS,WISE     & 1558678317 & 5.05 & 4.83-5.3 & 98.39  \\
PSO~J091.3857$-$31.9921 & 91.38576 & $-$31.99210 &    22.23$\pm$0.04 & 20.59$\pm$0.01 & 20.37$\pm$0.02 & 20.21$\pm$0.05 & 20.35$\pm$0.12 & & 20.09$\pm$0.24 & 19.95$\pm$0.05 & 20.08$\pm$0.11 & DES,VHS,WISE   & 159854908 & 5.15 & 4.7-5.7 & 99.73  \\
MQC~J133.2622$-$02.7132$^r$ & 133.26225 &  $-$2.71322  & 22.06$\pm$0.10 & 20.87$\pm$0.03 & 20.77$\pm$0.06 & 21.31$\pm$0.20 & & & & 20.75$\pm$0.13 & 21.29$\pm$0.41 & PS1,CatWISE    & & &  &  \\
PSO~J143.6025$-$21.8856 & 143.60254 & $-$21.88560  & & 22.45$\pm$0.16 & 20.56$\pm$0.06 & 20.57$\pm$0.11 & 20.01$\pm$0.16 &  & 19.46$\pm$0.18 & & &  PS1,VHS   & & &  &  \\	
PSO~J200.5269$-$13.3985$^r$ & 200.52692 & $-$13.39853  & 22.24$\pm$0.13 & 20.67$\pm$0.03 & 20.46$\pm$0.03 & 20.49$\pm$0.07 &  20.17$\pm$0.13 & 19.87$\pm$0.13 & & 19.62$\pm$0.05 & 19.39$\pm$0.07 & PS1,VHS,AllWISE   & & &  & \\ 
PSO~J273.2793$+$38.8379  & 273.27936  & $+$38.83790  &  23.26$\pm$0.24 & 22.52$\pm$0.10 & 20.79$\pm$0.05 & 20.87$\pm$0.12 & 20.27$\pm$0.15 & & 19.54$\pm$0.17 & 20.01$\pm$0.06 & 20.05$\pm$0.10 & PS1,UHS,CatWISE   & & &  & \\
PSO~J307.7455$-$47.3164 & 307.74553 & $-$47.31643 & 22.97$\pm$0.06 & 21.20$\pm$0.02 & 20.88$\pm$0.03 & 20.60$\pm$0.07 & 20.52$\pm$0.15 & 20.47$\pm$0.20 & 20.39$\pm$0.26 & 20.27$\pm$0.07 & 20.23$\pm$0.15 & DES,VHS,WISE    & 891420573 & 5.25 & 4.45-5.8 & 99.29 \\
PSO~J311.8092$-$64.9840 & 311.80923 & $-$64.98403 & 22.72$\pm$0.08 & 21.05$\pm$0.03 & 20.96$\pm$0.04  & 21.03$\pm$0.16 & 20.87$\pm$0.17 &   &   & 20.90$\pm$0.12 &   & DES,VHS,WISE   & 878441541 & 5.0 & 4.45-5.75 &  98.94 \\ 
PSO~J334.6905$-$63.0657 & 334.69056 & $-$63.06573 &        &  & 21.22$\pm$0.04 & 21.30$\pm$0.12  & 20.79$\pm$0.12  & & 20.51$\pm$0.27   & 20.63$\pm$0.09  &   & DES,VHS,WISE    & 956655457 & 6.15 & 5.65-6.35 & 99.28 \\
PSO~J335.6172$-$15.6807 & 335.61728 & $-$15.68079  &   &  & 21.25$\pm$0.06 & 19.57$\pm$0.04 & 19.84$\pm$0.02 & & 19.19$\pm$0.12 & 19.04$\pm$0.04 & 19.02$\pm$0.04 & PS1,NTT/SofI,CatWISE    & & &  & \\
\hline
\end{tabular}
\end{tiny}
\tablefoot{\tiny Col (1): Quasar name according to PS1 convention if selected from PS1; for the rest of the paper we always use a shortened name; the objects marked with $r$ are radio-loud; Col (2)-(3): coordinates in degrees; Col (4)-(7): optical magnitudes from PS1 or DES; Col (8)-(10): NIR magnitudes from VHS, UHS or the NTT/SofI follow-up; Col (11)-(12): MIR magnitudes from WISE; Blanks in the $r-$ and $i-$band or in the WISE columns refer to non-detections (SN$<3$), while for NIR bands a blank space indicates that no detection and/or image was found at that specific band. Col (13): reference surveys/follow-up campaign; Col (14)-(17): only for the sources selected from $YS23$: information taken from the catalog (ID, $z_{phot}$, $z_{phot}$ range, and probability to be a true quasar at that redshift).}
\end{table}

\begin{table}
\caption{\small Photometric properties (AB mag corrected for Galactic extinction) of the six new quasars discovered by DELS$+$PS1 selection (first part) and the six already know in the literature for which we report unpublished spectroscopic observations (second part).}
\label{tab:decalsmag}
\centering
\begin{tiny}
\scriptsize
\begin{tabular}{lcccccccccccc}
\hline\hline
Quasar Name  & RA & DEC & $z_{\rm P1,lim}$ & $z_{\rm DE}$ & $y_{\rm P1}$  & $J$  & $H$ & $K$ & $W1$ & $W2$ & ref. surveys/follow up & ref. \\  
    (1)     & (2) & (3) & (4)  & (5) & (6) & (7)  & (8) & (9) & (10) & (11) & (12) & (13) \\
\hline
\multicolumn{13}{c}{New discoveries} \\
\hline
PSO~J016.7097$+$06.1237 & 16.70975 &	$+$6.12368 & 23.06  & 22.02$\pm$0.14 & 21.01$\pm$0.14 & 21.27$\pm$0.08 & 21.12$\pm$0.2 & 20.75$\pm$0.19 & 20.52$\pm$0.12 & 20.44$\pm$0.10 &  DELS DR7,NTT/SofI & 1 \\ 
PSO~J037.3968$-$08.1397 & 37.39686 & $-$8.13970  & 23.16  & 22.03$\pm$0.05 & 20.89$\pm$0.16 & 21.12$\pm$0.09 & & & 20.60$\pm$0.13 & 20.77$\pm$0.34 &  DELS DR5,NTT/SofI & 1 \\ 
PSO~J041.4618$+$06.6525 & 41.46186 &	$+$6.65254 & 22.88 & 21.52$\pm$0.05 & 20.74$\pm$0.12 & 20.07$\pm$0.08 & 20.071$\pm$0.16 & 20.28$\pm$0.21 & 20.89$\pm$0.15 &  &  DELS DR7,NTT/SofI,UKIDSS & 1 \\ 
PSO~J067.6819$-$14.7614 & 67.6819  & $-$14.7614 & 22.5 & 21.52$\pm$0.08 & 20.77$\pm$0.21 & 20.71$\pm$0.09  & 20.23$\pm$0.05 & 19.86$\pm$0.04 & 19.90$\pm$0.07 & 19.60$\pm$0.12 &  DELS DR5,NTT/SofI & 1 \\ 
PSO~J217.0714$+$04.9084 & 217.07142 & $+$4.90844  & 23.15 & 22.35$\pm$0.11 &  21.45$\pm$0.17 & 21.05$\pm$0.05 & & & 20.26$\pm$0.08 & 20.23$\pm$0.17 &  DELS DR7,NTT/SofI & 1 \\ 
PSO~J289.3749$+$50.0537 & 289.37498 &	$+$50.05371 & 23.0 & 21.92$\pm$0.07 &  21.02$\pm$0.18 & 20.59$\pm$0.05 & & 20.47$ \pm$0.06 & 21.19$\pm$0.11 &  & DELS DR6,NOT/NOTCam & 1   \\  
\hline
\multicolumn{13}{c}{New spectra publication} \\
\hline
PSO~J062.8693$-$09.1305 &  62.86929 & $-$9.13050 & 22.53 & 20.68$\pm$0.03  &  20.04$\pm$0.06 & 19.92$\pm$0.04 &  & 19.52$\pm$0.21  & 19.55$\pm$0.05  & 19.46$\pm$0.10	& DELS DR5,NTT/SofI,VHS DR6 & 2,3  \\ 
PSO~J127.3832$+$41.2945 & 127.38319 & 41.29456 & 23.04 & 21.32$\pm$0.04 & 20.58$\pm$0.12 & 20.25$\pm$0.15 & & & 20.37$\pm$0.09 & 19.89$\pm$0.14 & DELS DR6,UHS & 3 \\
PSO~J129.4076$+$49.4834 & 129.40766 &	49.48343  & 22.88 & 20.69$\pm$0.02 & 19.85$\pm$0.07 & 20.18$\pm$0.17 & & & 20.06$\pm$0.07 & 19.26$\pm$0.07 & DELS DR6,UHS & 3 \\  
PSO~J162.0795$-$01.1612 & 162.07951 & $-$1.16123 & 22.96 & 21.96$\pm$0.07 & 20.95$\pm$0.15 &
20.82$\pm$0.10 &  20.48$\pm$0.08 & 20.57$\pm$0.11 &  20.47$\pm$0.10 & 19.95$\pm$0.14 & DELS DR7,UKIDSS,NTT/SofI &  4 \\ 
PSO~J164.5321$+$29.5115 & 164.53218 & $+$29.51159  & 22.96 & 21.57$\pm$0.06 & 20.59$\pm$0.09 &  & & & 21.06$\pm$0.16 & 20.94$\pm$0.33 & DELS DR7& 5 \\  
PSO~J354.5293$+$21.7328 & 354.52930 & 21.73282 & 23.03 & 21.86$\pm$0.07  & 
20.88$\pm$0.14& 20.77$\pm$0.17	 &  & 20.72$\pm$0.18 & 21.01$\pm$0.14  & 20.75$\pm$0.26 & DELS DR7,NOT/NOTCam & 5 \\  
\hline
\end{tabular}
\end{tiny}
\tablefoot{\tiny Col (1): Quasar name according to PS1 convention; for the rest of the paper we always use a shortened name; Col (2)-(3): coordinates in degrees; Col (4)-(11): optical, NIR and MIR magnitudes; the $H$ and $K-$bands magnitude of PSO J041$+$06 and the the $H-$band of PSO~J162$-$01 come from the UKIRT Infrared Deep Sky Survey (UKIDSS, \citealt{Lawrence2007}); In the NIR bands a blank space indicates that no detection and/or no image was found at that specific band. Col (12): reference surveys/follow-up campaign; Col (13): references paper: 1=This work, 2=\cite{Pons2019}, 3=\cite{WangFeige2019}, 4=\cite{WangFeige2017}, 5=\cite{Yang2021}. }
\end{table}
\end{landscape}

\section{DELS$+$PS1  candidates already reported as high-z quasars in the literature}
\label{app:delsps1published}
In this Appendix we list the candidates selected from the combination of both DELS and PS1 data (described in Section 2.3) that have been already published in the literature. 
These are 16 quasars at redshift between 6.4 and 7.0.
In Table~\ref{tab:knownphotometry} we report quasar name, coordinates, redshift, photometric information and reference papers. 
The photometric properties include new NIR imaging follow-up data for the quasar PSO~J137$+$16 and PSO~J027$-$28.
This latter was observed with th $J$, $H$ and $Ks$ filters of the NTT/SofI instrument on November 20, 2020 (900s in $J$ and $H$ bands and 1200s in $Ks$ band). 
We observed PSO~J137$+$16 in $H$ and $Ks$-band filters with NTT/SofI on January 23, 2023, for a total exposure time of 1800 and 1200s, respectively. 
The images were reduced with the same procedure described in Sect. \ref{sec:photometry}.\\
We further report the radio detection of the quasar PSO~J184$+$45. 
It is clearly detected (5$\sigma$) in the LOFAR LoTSS survey.
From a Gaussian fit performed by using the CASA software, we measured a peak flux density of $325 \pm 62\,\mu$Jy.
We calculated the value of R following the same steps applied to PSO~J164$+$29 (described in the main text, section 4.1.9).
We estimated a value of 15$\pm$5, which classifies the source as radio loud. 
However, as for PSO~J164$+$29, PSO~J184$+$45 is not detected in other published radio surveys, therefore additional multi-band radio data will be essential to reinforce this classification.

\begin{landscape}
\begin{table}
	\caption{\small Properties of the 16 quasars selected from DELS$+$PS1 already reported in the literature.}
 	\label{tab:knownphotometry}
\begin{tiny}
\scriptsize
\begin{tabular}{p{1.2cm}p{0.7cm}p{0.8cm}p{1.2cm}p{0.4cm}p{0.6cm}p{0.5cm}p{0.9cm}p{0.9cm}p{0.9cm}p{0.9cm}p{0.9cm}p{0.9cm}p{0.9cm}p{4.0cm}p{2.7cm}}\hline\hline
Quasar Name & RA    & Dec   &   $z$ & m$_{1450}$ & M$_{1450}$ & $z_{\rm P1,lim}$ & $z_{\rm DE}$ & $y_{\rm P1}$  & $J$  & $H$ & $K$ & $W1$ & $W2$ & ref. surveys/follow up & References \\  
(1)       & (2)      & (3)     &   (4)   & (5)                 &  (6)               & (7)                       & (8)                  & (9)                   & (10)   &  (11) &  (12) &  (13) &   (14)   &    (15)                           & (16)  \\
\hline
PSO~J009$-$15  & 9.65040 & $-$15.45656 & 7.021$\pm$0.005 & 19.93 & $-$27.10 &  22.96 & 21.65$\pm$0.08 & 20.58$\pm$0.11 & 19.67$\pm$0.07 & 19.52$\pm$0.06 &  19.37$\pm$0.05 & 19.64$\pm$0.06 &  19.79$\pm$0.16 & DELS DR5,UKIRT/WFCAM & 1 \\  
PSO~J011$+$09 & 11.38986 & $+$09.03249 & 6.4695$\pm$0.0002 & 20.85 & $-$25.96 & 22.25 &  21.35$\pm$0.04 &  20.54$\pm$0.10 & 20.77$\pm$0.07 & 20.62$\pm$0.24 &  20.74$\pm$0.34 &  20.46$\pm$0.11 & 20.91$\pm$0.37 &  DELS DR8,MPG2.2m/GROND,RC20 & 2, 3 \\ 
PSO~J027$-$28 & 27.21960 & $-$28.44422 & 6.54$\pm$0.03 & 20.79  & $-$26.09  & 22.68	& 21.68$\pm$0.03   &   20.82$\pm$0.14   &   21.86$\pm$0.15  &  21.45$\pm$0.16  & 20.88$\pm$0.14  &  20.08$\pm$0.06  &  20.14$\pm$0.16  & DELS DR8,NTT/SofI & 4, This work  \\  
PSO~J062$-$01 & 62.53771 & $-$01.65552 & 6.995$\pm$0.001 & 21.33 & $-$25.60 & 22.95  & 22.04$\pm$0.06  & 21.15$\pm$0.21	 & 20.67$\pm$0.07  & 20.80$\pm$0.13  & 20.26$\pm$0.10   & 20.44$\pm$0.08  & 20.15$\pm$0.15 & DELS DR10,NTT/SofI  & 5   \\ 
PSO~J129$+$39 & 129.94534 & $+$39.00319  & 6.9046$\pm$0.0003  & 20.63   &  $-$26.29 & 23.05 &  20.99$\pm$0.03 & 20.25$\pm$0.09 & 20.36$\pm$0.20 & & &  19.19$\pm$0.03 & 19.05$\pm$0.07 & DELS DR6,UHS & 6, 7 \\
PSO~J137$+$16 & 137.55678 & $+$16.94163 &  6.719$\pm$0.005 & 21.3   & $-$25.57  & 22.84 &  22.01$\pm$0.07  & 20.81$\pm$0.16 & 21.03$\pm$0.13  &  20.80$\pm$0.09 & 20.44$\pm$0.08 &  21.15$\pm$0.19 & 20.30$\pm$0.20  &  DELS DR6,UKIRT/WFCAM,NTT/SofI & 6, 7, This work \\  
PSO~J140$+$00  & 140.33566 & $+$00.12303 & 6.5654$\pm$0.0002 &  22.05 & $-$24.79  & 22.68 & 21.65$\pm$0.04 & 20.88$\pm$0.16	& 21.25$\pm$0.09 & 20.86$\pm$0.15 & 20.39$\pm$0.09 & 19.79$\pm$0.05 & 19.32$\pm$0.08 & DELS DR8,RC20 &  7, 8 \\  
PSO~J140$+$04  & 140.94633 & $+$04.04844 & 6.612$\pm$0.002 & 20.67 & $-$26.18  & 23.04 & 21.21$\pm$0.02 & 20.22$\pm$0.09 & 20.11$\pm$0.08 & 19.87$\pm$0.13 & 19.42$\pm$0.07 & 19.21$\pm$0.03 & 19.06$\pm$0.06  & DELS DR8,RC20 &  7, 8 \\  
PSO~J166$+$21 & 166.08997 & $+$21.57467  & 6.766$\pm$0.005  & 20.21   &  $-$26.67 & 22.94 &  21.10$\pm$0.03 &  19.95$\pm$0.06 & 19.93$\pm$0.12 & & &  19.99$\pm$0.06 & 19.90$\pm$0.13 & DELS DR7,UHS & 6, 7  \\  
PSO~J167$-$13  & 167.64159 & $-$13.49602 & 6.508$\pm$0.001 & 21.25 & $-$25.57   &  22.86 &	21.07$\pm$0.06 & 20.49$\pm$0.11 & 21.17$\pm$0.09 &	& 20.21$\pm$0.26 & & &  DELS DR3,VIKING,RC20 & 9  \\  
PSO~J172$+$18 & 172.35569 & $+$18.77342 & 6.823$^{+0.003}_{-0.001}$  & 21.08   & $-$25.81  & 23.22 &  21.61$\pm$0.05 & 20.74$\pm$0.09 & 20.88$\pm$0.11 & 21.35$\pm$0.24  & 21.06$\pm$0.18  & 21.25$\pm$0.20 & & DELS DR7,NOT/NOTCam & 10 \\  
PSO~J173$+$50  & 173.78721 & $+$50.1925 & 6.58$\pm$0.03  & 20.65 & $-$26.19   & 22.99 &  20.71$\pm$0.03 &  20.08$\pm$0.08 &  20.43$\pm$0.17 & & &  19.93$\pm$0.05 & 19.77$\pm$.010 & DELS DR8,RC20 & 6 \\  
PSO~J184$+$45 & 184.11489 & $+$45.31968 & 6.648$\pm$0.003  & 21.27   & $-$25.59  & 23.00 & 21.54$\pm$0.07 & 20.67$\pm$0.09 &  21.01$\pm$0.13 &  &  &  20.05$\pm$0.06	 & 19.80$\pm$0.10 &  DELS DR6,UKIRT/WFCAM & 6, 7  \\ 
PSO~J247$+$24  &  247.29708 & $+$24.12770 & 6.476$\pm$0.004 & 20.28 & $-$26.53  & 22.70 & 	 20.76$\pm$0.03 &	19.98$\pm$0.07 &	20.19$\pm$0.09 &  & & 19.37$\pm$0.03 & 19.29$\pm$0.06 & DELS DR8,CAHA 3.5 m/Omega2000 & 2  \\ 
PSO~J261$+$19  & 261.03643 & $+$19.02861 & 6.480$\pm$0.001 & 21.12   & $-$25.70 & 22.86 &  21.49$\pm$0.05  & 20.93$\pm$0.13 & &  & 20.25$\pm$0.23  & 20.57$\pm$0.09 &  20.38$\pm$0.17 & DELS DR7,UHS & 2, 7 \\  
PSO~J338$+$29  & 338.22982 & $+$29.50898 & 6.655$\pm$0.003  & 20.78   & $-$26.08 & 22.50  & 21.10$\pm$0.05 & 20.24$\pm$0.09 &  20.30$\pm$0.14 & & & 20.46$\pm$0.08  & 21.07$\pm$0.33 & DELS DR7,UHS& 7, 9\\  
\hline
	\end{tabular} \vspace{-0.17cm}
    \end{tiny}
\tablefoot{\tiny Quasars sorted by right ascension. Col (1): Quasar name according to PS1 convention; Col (2) and Col (3): Coordinates in degrees; Col (4): Redshift measured from the Mg$\rm II$ broad emission line; except for PSO~J027$-$28 and PSO~J173$+$50; Col (5)-(6): apparent and absolute magnitude at 1450\AA\ rest-frame taken from the review of \cite{FanX2023ARA}, except for PSO~J027$-$28, PSO~J062$-$01 and PSO~J172$+$18; Col (7)-(14): optical, NIR and MIR magnitudes (RC20 refers to \cite{RossCross2020} which contains NIR photometry from VHS, VIKING or UKIDSS surveys); Col (15): Reference surveys/follow-up campaign; Col (16): Reference papers for spectra and quasar properties: 1=\cite{WangFeige2018}, 2=\cite{Mazzucchelli2017}, 3=\cite{EilersAC2020}, 4=\cite{YangJ2020}, 5=\cite{Banados2025}, 6=\cite{WangFeige2019}, 7=\cite{Yang2021}, 8=\cite{Matsuoka2018} 9=\cite{Venemans2015}, 10=\cite{Banados2021}}

\end{table}
\end{landscape}

\section{Rejected candidates}
\subsection{Candidates discarded from MQC and from YS23}
\label{app:mqcys23discar}
In Table~\ref{mqcdiscarted} we report the sources discarted from our list of candidates selected from the MQC because of a further cross-match with SIMBAD and SDSS DR18: two objects are true quasars at $z>4$ already published, two sources are classified as galaxies, and for the remaining 6 the SDSS spectral classification is unclear (and beyond the scope of this paper), but they are not quasars at $z>4$. \\
Instead, visual inspection of the optical and NIR images of the YS23 candidates led us to discard 5 sources (see Table~\ref{ys23discarted}): 4 objects with no signal at all in the optical bands and 1 object that is only detected as a bright and extended source in $i-$band Decals.
We note that for each of these discarded sources from the YS23 catalog, the probability of being high-$z$ quasars reported in the catalog is greater than 98.57\%.

\begin{table}[!ht]
\caption{\small Candidates discarded from MQC selection.}
\label{mqcdiscarted}
\centering
\begin{small}
\begin{tabular}{p{2.1cm}p{1.2cm}p{1.0cm}p{0.53cm}p{0.9cm}p{0.4cm}}
\hline\hline
Name  & RA      & Dec   &  z  & Obj Class & Ref. \\
(1)   & (2)     & (3)   & (4) & (5) & (6)  \\
\hline
J130649+615233 & 196.70701  & 61.87586 &  4.25 & quasar & 1 \\	
J103720+421326 & 159.33672  & 42.22392 & 4.2    & quasar  & 2 \\ 	
J091515+333840 & 138.81547  & 33.64465 & -$^a$ & unknown & 3 \\
J021303+003811 & 33.26575   & 0.63663  & 0.7 & galaxy & 4 \\ 
J091618+005134 & 139.07809  & 0.85960  & 1.53 & galaxy & 3 \\
J090515+352624 & 136.31625  & 35.44011 &  -$^a$   & unknown & 3 \\
J114926+150817 & 177.35931  & 15.13811 & -$^a$   & unknown & 3 \\
J092323+271512 & 140.84719  & 27.25342 & -$^a$   & unknown & 3 \\
J163132+445849 & 247.88476  & 44.98035 & -$^a$   & unknown & 3 \\
J234806+022828 & 357.02508  & 2.47447 & -$^a$   & unknown & 3 \\
\hline
\end{tabular}
\end{small}
\tablefoot{\tiny Col (1): Object name as reported in the MQC; Col (2,3): Coordinates in degrees; Col (4): redshift; objects marked with $a$ are misclassified as high-$z$ quasar on SDSS spectra; Col (5) Object class as reported from the reference paper/database; Col (6): References: 1=\cite{Schindler2019}; 2=\cite{Lyke2020}; 3=SDSS DR18 Navigate Tool\footnote{\url{https://skyserver.sdss.org/dr18/VisualTools/navi}}; 4=\cite{Ahumada2020}}
\end{table}

\begin{table}[!ht]
\caption{\small Candidates discarded from YS23 selection.}
\label{ys23discarted}
\centering
\begin{small}
\begin{tabular}{cccc}
\hline\hline
Cat. ID  & RA      & Dec   &  Comment  \\
(1)   & (2)     & (3)   & (4)    \\
\hline
1238878646 & 31.49982 &  3.09267  &  no signal \\
1081134814 & 8.53176  &  1.29155  &  no signal \\
1336715064 & 41.85627 &  2.21061  &  no signal \\
1326787943 & 40.74876 &  $-$2.84063   &  no signal \\
979248393  & 342.59386 &  $-$2.32348  & bright object \\ 
   &   &    & only in $i-$band \\
\hline
\end{tabular} 
\end{small}
\tablefoot{\tiny Col (1): Object ID as reported in the YS23 catalog; Col (2,3): Coordinates in degrees; Col (4): Comment from visual inspection.}
\end{table}

\subsection{Spectroscopically rejected candidates}
\label{app:dwarf}
Table~\ref{rejected} lists 73 candidates whose spectra showed they were not high-$z$ quasars. 
Objects with a CLASS value of 1 are cool dwarfs, while sources with a CLASS value of 2 and 3 are more likely low-z galaxies and unknown-type objects, respectively.
A thorough classification of these objects is beyond the scope of this work. \\
We note that two sources (PSO~J027$-$45 and PSO~J047$-$60) have been classified as ultra cool dwarf by \cite{Dalponte23}.\\
The radio candidate ILT~J0010$+$3239 as been classified as a M7 dwarf by \cite{Gloudemans2023}.\\
Also, three objects from the YS23 catalog, with a probability of being $z>6$ quasars larger than 99\%, turned out to be contaminants after a dedicated spectroscopic follow-up with NTT/EFOSC2: YS~J041$-$29 belongs to CLASS number 3, YS~J049$-$34 is a cool dwarf (CLASS 1) and YS~J066$-$28 turned out to be a low$-z$ galaxy (CLASS 2). 

\begin{table*}
\caption{\small Candidates Spectroscopically Confirmed to not be high-$z$ Quasars}
\label{rejected}
\centering
\begin{tiny}
\begin{tabular}{p{2.0cm}p{0.75cm}p{0.85cm}p{1.3cm}p{0.9cm}r|p{1.8cm}p{0.8cm}p{0.85cm}p{1.3cm}p{0.9cm}r}
\hline\hline
Rejected & RA          & Dec       & Telescope  & Date & CLASS & Rejected & RA & Dec & Telescope & Date & CLASS \\  
Candidate &  (J2000) & (J2000) & Instrument &         &             & Candidate &  (J2000) & (J2000) & Instrument & \\ 
(1) & (2) & (3) & (4) & (5) & (6) & (7) & (8) & (9) & (10) & (11) & (12)  \\
\hline
ILT~J0010$+$3239$^g$       &  02.5364 & $+$32.6504 & LBT/MODS & 2023/11/11 & 1 & PSO J003$-$03 & 03.0822 & $-$03.7675 & LBT/MODS & 2023/11/12 & 1   \\
PSO~J008$+$04                  & 08.3993 & $+$04.9334 & LBT/MODS & 2023/11/13 & 1    & PSO~J010$+$46 &  10.0308 & $+$46.4924 & LBT/MODS & 2023/11/12 & 3   \\
PSO~J012$+$36                  & 12.0151 & $+$36.5762 & LBT/MODS & 2023/11/13 & 1 & PSO~J027$-$45$^a$  & 27.9230  & $-$45.5284  & NTT/EFOSC2 & 2024/01/24 & 1   \\
ILT~J0157$+$3042               & 29.3863 & $+$30.7092 & LBT/MODS & 2023/11/13 & 2 &  PSO~J039$+$04      & 39.9339 &  $+$04.4616      & LBT/MODS & 2023/11/13 &  3   \\
MALS~J0240$+$1719          & 40.0717 & $+$17.3222 & LBT/MODS & 2024/12/05 &  1 &  YS~J041$-$29$^b$  & 41.1159  & $-$29.0798  & NTT/EFOSC2 & 2022/12/27 & 3  \\
PSO~J047$-$60                   & 47.0409 & $-$60.8559   & NTT/EFOSC2 & 2024/01/27 & 1 & YS~J049$-$34$^b$  & 49.9160  & $-$34.3035  & NTT/EFOSC2 & 2022/12/27 & 1 \\
PSO~J060$+$03                  & 60.2691  & $+$03.9825  & NTT/EFOSC2 & 2023/12/01 & 3 & YS~J066$-$28$^b$ & 66.5230  & $-$28.7629  & NTT/EFOSC2 & 2023/11/29 & 2 \\ 
PSO~J069$-$26                   & 69.6370  & $-$26.8314  & NTT/EFOSC2 & 2023/12/02 & 1 & PSO~J099$+$57      & 99.7872 & $+$57.5060      & LBT/MODS & 2023/11/12 & 1 \\
PSO~J117$-$69                   & 117.2528 & $-$69.6421  & NTT/EFOSC2 & 2024/01/24 & 1 & PSO~J112$+$33 & 112.1840 & $+$33.2182 & LBT/MODS & 2020/12/25 & 1 \\
PSO~J139$-$04$^a$          & 139.7676 & $-$04.2978  & NTT/EFOSC2 & 2024/01/24 & 1 & PSO~J145$-$27 &  145.5390 & $-$27.2683 & NTT/EFOSC2 & 2024/01/25 & 1 \\ 
PSO~J153$+$47                 & 153.3546 & $+$47.6594 & LBT/MODS & 2023/04/19 & 1 & PSO~J155$+$23 & 155.3345 &	$+$23.3206 & LBT/MODS & 2023/04/01 & 1 \\
MALS~J1044$-$1412         & 161.0521 & $-$14.2127  & NTT/EFOSC2 & 2024/01/24 & 2 & PSO~J162$-$12     & 162.3426 & $-$12.6113  & NTT/EFOSC2 & 2022/02/04 & 1 \\
NVSS~J163$-$02               & 163.6297 & $-$02.0645 & NTT/EFOSC2 & 2022/02/04 & 3 & PSO~J170$-$28     & 170.6294 & $-$28.7200  & NTT/EFOSC2 & 2022/02/05 & 1 \\
PSO~J172$+$16                & 172.0984 & $+$16.0280  & NTT/EFOSC2 & 2022/02/04 & 1 & ILT~J1140$+$4202  & 175.0467  & $+$42.0335    & LBT/MODS   & 2023/06/13 & 3 \\
PSO~J181$+$30               & 181.2624  & $+$30.4606   & LBT/MODS & 2023/06/10 & 1 & PSO~J182$-$29     & 182.2378 & $-$29.9385  & NTT/EFOSC2 & 2022/02/05 & 1 \\
PSO~J182$-$05              &  182.3792  & $-$05.2293    & LBT/MODS & 2023/06/10 & 1 & PSO~J182$+$11 & 182.7458 & $+$11.1742 & LBT/MODS & 2023/04/01 & 3 \\
PSO~J182$+$83             & 182.7718 & $+$83.6176   & LBT/MODS & 2023/03/05 & 1 & PSO~J184$+$64      &  184.1091  & $+$64.8172  & LBT/MODS & 2023/06/08 & 1 \\
PSO~J186$+$80             & 186.1962 & $+$80.4875    &  LBT/MODS & 2023/04/20 & 1 & PSO~J186$+$60      &  186.2609  & $+$60.5096  & LBT/MODS & 2023/06/08 & 1 \\
PSO~J187$+$08              &  187.1222  & $+$08.6590   & LBT/MODS & 2023/06/10  & 1 & MQC~J1239$+$1617  &  189.8990  & $+$16.2973  & NTT/EFOSC2 & 2024/01/24 & 1 \\
PSO~J195$-$07               & 195.3068 & $-$07.2171   & NTT/EFOSC2 & 2022/02/04 & 1 & PSO~J195$+$86    &  195.6339 & $+$86.8951 & LBT/MODS & 2023/04/01 & 1 \\
PSO~J195$-$10               &  195.6990  & $-$10.0122   & LBT/MODS & 2023/06/10 & 3 & RACS~J198$+$40     & 198.7974 & $+$40.1505     & LBT/MODS & 2023/06/10 & 2 \\
ILT~J1326$+$5429$^c$   & 201.5518 & $+$54.4944   & LBT/MODS & 2023/06/08 & 1 & RACS~J206$+$40     & 206.9889 & $+$40.3283  & LBT/MODS & 2023/06/10 & 2 \\
ILT~J1407$+$4252   &	211.9176 & $+$42.8708 & LBT/MODS & 2023/06/07 & 2 & ILT~J1431$+$4635  &	217.8540 & $+$46.5862 & LBT/MODS & 2023/06/07 & 1 \\
PSO~J222$+$05      & 222.3130 &  $+$05.7496 & LBT/MODS & 2023/03/28 & 1 & NVSS~J225$-$11      & 225.2889  & $-$11.9584 & NTT/EFOSC2 & 2022/02/04 & 2 \\
ILT~J1504$+$6258  & 226.1189 & $+$62.9802  & LBT/MODS & 2023/06/13 & 1 & PSO~J227$-$17     & 227.0505 & $-$17.3678 & LBT/MODS & 2023/04/01 & 1 \\
PSO~J233+05          &  233.2398 &  $+$05.4933  & LBT/MODS & 2023/06/08 & 1 & PSO~J234$-$09      & 234.5534 & $-$09.4578 & LBT/MODS & 2023/04/21 & 1 \\
RACS~J234$+$40     & 234.6786 & $+$40.4477   & LBT/MODS & 2023/06/10 & 1 & PSO~J245+18        &  245.1485 &  $+$18.6299  & LBT/MODS & 2023/06/08 & 1 \\
ILT~J1623$+$5427   & 245.8994 & $+$54.4526 & LBT/MODS & 2023/06/13 & 3 & ILT~J1629$+$4156  & 247.4075 & $+$41.9488 & LBT/MODS & 2023/06/08 & 3 \\
PSO~J247$+$10      &  247.8238 &  $+$10.2805  & LBT/MODS & 2023/06/09 & 3 & PSO~J250$+$00  & 250.5826  & $+$00.9124 & LBT/MODS & 2023/04/21 & 1 \\
PSO~J253$+$00      &  253.1887 &  $+$00.5648   & LBT/MODS & 2023/06/08 & 3 & PSO~J253$+$07      &  253.5664 &  $+$07.4414   & LBT/MODS & 2023/06/08 & 1 \\
PSO254$+$25          & 254.7461 & $+$25.6797     & LBT/MODS & 2023/04/18 & 1 & PSO~J262$+$20      &  262.6972 &  $+$20.9801  & LBT/MODS & 2023/06/08 & 1 \\
PSO~J266$+$54      &  266.5850 &  $+$54.9565  & LBT/MODS & 2023/06/09 & 1 & PSO~J269$+$19      &  269.8055 &  $+$19.6993  & LBT/MODS & 2023/06/10 & 1 \\
RACS~J279$+$40    &  279.1253 & $+$40.6443   & LBT/MODS & 2023/06/13 & 1 & PSO~J281$+$58 & 281.8065 & $+$58.8070 & LBT/MODS & 2020/06/06 & 1 \\
PSO~J290$+$58      &  290.3967 &  $+$58.9075  & LBT/MODS & 2023/06/08 & 1 & PSO~J331$-$09      &  331.9305  & $-$09.3771 & LBT/MODS & 2023/05/28 & 1 \\
PSO~J333$+$29      &  333.0793 & $+$29.2715 & LBT/MODS & 2022/10/20 & 1 & PSO~J350$+$18      &  350.3733 & $+$18.8181 & LBT/MODS & 2023/11/11 & 1 \\
PSO~J351$+$07      & 351.4983 & $+$07.0055 & LBT/MODS & 2023/11/13 & 1 & PSO~J353$-$20      & 353.0240 & $-$20.3663 & LBT/MODS & 2023/11/12 & 1 \\
PSO~J359$+$28      & 359.2352 & $+$28.8194 & LBT/MODS & 2023/11/12 & 1  &  & & & & &  \\
\hline
    \end{tabular}
    \end{tiny}
   \tablefoot{\small Rejected candidates sorted by R.A. Col (1) and Col (7): Object name; sources named $ILT$ have been selected from the LOFAR LoTSS DR2 catalog (\citealt{Shimwell2022}); $RACS$ objects are from the RACS 888 MHz radio catalog (\citealt{McConnell2020}); sources named $NVSS$ have been selected from the NVSS catalog of \cite{Condon1998}); $MQC$ objects have been selected from the Million Quasar Catalog v8 of \cite{Flesch2023}; sources named $MALS$ were selected from the MeerKAT Absorption Line Survey DR1 catalog (\citealt{Deka2024}); $a$: for the selection method of these sources we refer the reader to Mkrtchyan et al. (in prep.); $b$: candidates in the \cite{YangShen2023} catalog with p$_{qso}$ large than 99\%; $c$: two nearby ($<1''$) objects; $g$: candidate already rejected by \cite{Gloudemans2022}; Col (2), Col (3), Col (8) and Col (9): Coordinates in degrees; Col (4) and Col (10): instrument and telescope used for the spectroscopic follow-up; Col (5) and Col (11) dates of the observation; Col (6) and Col (12): Spectroscopic class: 1=brown dwarf, 2=low-$z$ galaxy, 3=unknown.}
\end{table*} 

\begin{acknowledgements}
We thank the anonymous referee for their valuable suggestions, which improved the quality and clarity of this paper.
L.F. acknowledges support from the INAF 2023 mini-grant \textit{Exploiting the powerful capabilities of JWST/NIRSpec to unveil the distant Universe} and from the INAF GO 2022 grant \textit{The birth of the giants: JWST sheds light on the build-up of quasars at cosmic dawn}.
C.M. acknowledges support from Fondecyt Iniciacion grant 11240336 and the ANID BASAL project FB210003.
E.P.F. is supported by the international Gemini Observatory, a program of NSF NOIRLab, which is managed by the Association of Universities for Research in Astronomy (AURA) under a cooperative agreement with the U.S. National Science Foundation, on behalf of the Gemini partnership of Argentina, Brazil, Canada, Chile, the Republic of Korea, and the United States of America.
This work is based on observations collected at the European Southern Observatory under ESO programmes 0102.A-0233(A), 0104.A-0609(A), 105.203B.001, 0106.A-0540(A), 108.226E.002, 110.23RN.001, 110.23RN.002, 112.25VZ.001 and 112.25VZ.002 and observations collected at the Nordic Optical Telescope under program ID P59-015 and P61-003.

This paper includes data from the LBT. The LBT is an international collaboration among institutions in the United States, Italy, and Germany. The LBT Corporation partners are: The University of Arizona on behalf of the Arizona university system; Istituto Nazionale di Astrofisica, Italy; LBT Beteiligungsgesellschaft, Germany, representing the Max Planck Society, the Astrophysical Institute Potsdam, and Heidelberg University; The Ohio State University; The Research Corporation, on behalf of The University of Notre Dame, University of Minnesota and University of Virginia. 

This work includes data gathered with the 6.5 m Magellan Telescopes located at Las Campanas Observatory, Chile.
This paper is based on observations collected with the Magellan/LDSS3 under the programme allocated by the Chilean Telescope Allocation Committee (CNTAC) no:CN2025A-26.

Some of the data presented in this paper were obtained at the W.M. Keck Observatory, which is operated as a scientific partnership among the California Institute of Technology, the University of California and the National Aeronautics and Space Administration. 
The Observatory was made possible by the generous financial support of the W.M. Keck Foundation. 

Based on observations obtained at the international Gemini Observatory (under program: GN-2018B-Q-202), a program of NSF's NOIRLab, which is managed by the Association of Universities for Research in Astronomy (AURA) under a cooperative agreement with the National Science Foundation on behalf of the Gemini Observatory partnership: the National Science Foundation (United States), National Research Council.

The Pan-STARRS1 Surveys (PS1) and the PS1 public science archive have been made possible through contributions by the Institute for Astronomy, the University of Hawaii, the Pan-STARRS Project Office, the Max-Planck Society and its participating institutes, the Max Planck Institute for Astron- omy, Heidelberg and the Max Planck Institute for Extra- terrestrial Physics, Garching, The Johns Hopkins University, Durham University, the University of Edinburgh, the Queen’s University Belfast, the Harvard-Smithsonian Center for Astro- physics, the Las Cumbres Observatory Global Telescope Network Incorporated, the National Central University of Taiwan, the Space Telescope Science Institute, the National Aeronautics and Space Administration under Grant No. NNX08AR22G issued through the Planetary Science Division of the NASA Science Mission Directorate, the National Science Foundation Grant No. AST-1238877, the University of Maryland, Eotvos Lorand University (ELTE), the Los Alamos National Laboratory, and the Gordon and Betty Moore Foundation.

The Legacy Surveys consist of three individual and complementary projects: the Dark Energy Camera Legacy Survey (DECaLS; Proposal ID 2014B-0404; PIs: David Schlegel and Arjun Dey), the Beijing-Arizona Sky Survey (BASS; NOAO Prop. ID 2015A-0801; PIs: Zhou Xu and Xiaohui Fan), and the Mayall z-band Legacy Survey (MzLS; Prop. ID 2016A-0453; PI: Arjun Dey). DECaLS, BASS and MzLS together include data obtained, respectively, at the Blanco telescope, Cerro Tololo Inter-American Observatory, NSF’s NOIRLab; the Bok telescope, Steward Observatory, University of Arizona; and the Mayall telescope, Kitt Peak National Observatory, NOIRLab. Pipeline processing and analyses of the data were supported by NOIRLab and the Lawrence Berkeley National Laboratory (LBNL). The Legacy Surveys project is honored to be permitted to conduct astronomical research on Iolkam Du’ag (Kitt Peak), a mountain with particular significance to the Tohono O’odham Nation.

NOIRLab is operated by the Association of Universities for Research in Astronomy (AURA) under a cooperative agreement with the National Science Foundation. LBNL is managed by the Regents of the University of California under contract to the U.S. Department of Energy.

This project used data obtained with the Dark Energy Camera (DECam), which was constructed by the Dark Energy Survey (DES) collaboration. Funding for the DES Projects has been provided by the U.S. Department of Energy, the U.S. National Science Foundation, the Ministry of Science and Education of Spain, the Science and Technology Facilities Council of the United Kingdom, the Higher Education Funding Council for England, the National Center for Supercomputing Applications at the University of Illinois at Urbana-Champaign, the Kavli Institute of Cosmological Physics at the University of Chicago, Center for Cosmology and Astro-Particle Physics at the Ohio State University, the Mitchell Institute for Fundamental Physics and Astronomy at Texas A\&M University, Financiadora de Estudos e Projetos, Fundacao Carlos Chagas Filho de Amparo, Financiadora de Estudos e Projetos, Fundacao Carlos Chagas Filho de Amparo a Pesquisa do Estado do Rio de Janeiro, Conselho Nacional de Desenvolvimento Cientifico e Tecnologico and the Ministerio da Ciencia, Tecnologia e Inovacao, the Deutsche Forschungsgemeinschaft and the Collaborating Institutions in the Dark Energy Survey. The Collaborating Institutions are Argonne National Laboratory, the University of California at Santa Cruz, the University of Cambridge, Centro de Investigaciones Energeticas, Medioambientales y Tecnologicas-Madrid, the University of Chicago, University College London, the DES-Brazil Consortium, the University of Edinburgh, the Eidgenossische Technische Hochschule (ETH) Zurich, Fermi National Accelerator Laboratory, the University of Illinois at Urbana-Champaign, the Institut de Ciencies de l’Espai (IEEC/CSIC), the Institut de Fisica d’Altes Energies, Lawrence Berkeley National Laboratory, the Ludwig Maximilians Universitat Munchen and the associated Excellence Cluster Universe, the University of Michigan, NSF’s NOIRLab, the University of Nottingham, the Ohio State University, the University of Pennsylvania, the University of Portsmouth, SLAC National Accelerator Laboratory, Stanford University, the University of Sussex, and Texas A\&M University.

BASS is a key project of the Telescope Access Program (TAP), which has been funded by the National Astronomical Observatories of China, the Chinese Academy of Sciences (the Strategic Priority Research Program “The Emergence of Cosmological Structures” Grant \# XDB09000000), and the Special Fund for Astronomy from the Ministry of Finance. The BASS is also supported by the External Cooperation Program of Chinese Academy of Sciences (Grant \# 114A11KYSB20160057), and Chinese National Natural Science Foundation (Grant \# 12120101003, \#11433005).

The Legacy Survey team makes use of data products from the Near-Earth Object Wide-field Infrared Survey Explorer (NEOWISE), which is a project of the Jet Propulsion Laboratory/California Institute of Technology. NEOWISE is funded by the National Aeronautics and Space Administration.

The Legacy Surveys imaging of the DESI footprint is supported by the Director, Office of Science, Office of High Energy Physics of the U.S. Department of Energy under Contract No. DE-AC02-05CH1123, by the National Energy Research Scientific Computing Center, a DOE Office of Science User Facility under the same contract; and by the U.S. National Science Foundation, Division of Astronomical Sciences under Contract No. AST-0950945 to NOAO.

This project used public archival data from the Dark Energy Survey (DES). Funding for the DES Projects has been provided by the U.S. Department of Energy, the U.S. National Science Foundation, the Ministry of Science and Education of Spain, the Science and Technology FacilitiesCouncil of the United Kingdom, the Higher Education Funding Council for England, the National Center for Supercomputing Applications at the University of Illinois at Urbana-Champaign, the Kavli Institute of Cosmological Physics at the University of Chicago, the Center for Cosmology and Astro-Particle Physics at the Ohio State University, the Mitchell Institute for Fundamental Physics and Astronomy at Texas A\&M University, Financiadora de Estudos e Projetos, Funda{\c c}{\~a}o Carlos Chagas Filho de Amparo {\`a} Pesquisa do Estado do Rio de Janeiro, Conselho Nacional de Desenvolvimento Cient{\'i}fico e Tecnol{\'o}gico and the Minist{\'e}rio da Ci{\^e}ncia, Tecnologia e Inova{\c c}{\~a}o, the Deutsche Forschungsgemeinschaft, and the Collaborating Institutions in the Dark Energy Survey.
The Collaborating Institutions are Argonne National Laboratory, the University of California at Santa Cruz, the University of Cambridge, Centro de Investigaciones Energ{\'e}ticas, Medioambientales y Tecnol{\'o}gicas-Madrid, the University of Chicago, University College London, the DES-Brazil Consortium, the University of Edinburgh, the Eidgen{\"o}ssische Technische Hochschule (ETH) Z{\"u}rich,  Fermi National Accelerator Laboratory, the University of Illinois at Urbana-Champaign, the Institut de Ci{\`e}ncies de l'Espai (IEEC/CSIC), the Institut de F{\'i}sica d'Altes Energies, Lawrence Berkeley National Laboratory, the Ludwig-Maximilians Universit{\"a}t M{\"u}nchen and the associated Excellence Cluster Universe, the University of Michigan, the National Optical Astronomy Observatory, the University of Nottingham, The Ohio State University, the OzDES Membership Consortium, the University of Pennsylvania, the University of Portsmouth, SLAC National Accelerator Laboratory, Stanford University, the University of Sussex, and Texas A\&M University.
Based in part on observations at Cerro Tololo Inter-American Observatory, National Optical Astronomy Observatory, which is operated by the Association of Universities for Research in Astronomy (AURA) under a cooperative agreement with the National Science Foundation.

This publication makes use of data products from the Wide-field Infrared Survey Explorer, which is a joint project of the University of California, Los Angeles, and the Jet Propulsion Laboratory/Caltech, funded by the National Aeronautics and Space Administration.

Based on observation obtained as part of the VISTA Hemisphere Survey, ESO Program, 179.A-2010 (PI: McMahon). 

This publication has made use of data from the VIKING survey from VISTA at the ESO Paranal Observatory, program ID 179.A-2004. 

The VISTA Data Flow System pipeline processing and science archive are described in \cite{Irwin2004} and \cite{Hambly2008}.

The NVSS and VLASS data was taken by the NRAO Very Large Array.
The National Radio Astronomy Observatory is a facility of the National Science Foundation operated under cooperative agreement by Associated Universities, Inc.

This paper includes archived data obtained through the CSIRO ASKAP Science Data Archive, CASDA (http://data.csiro.au).
This scientific work uses data obtained from Inyarrimanha Ilgari Bundara / the Murchison Radio-astronomy Observatory. We acknowledge the Wajarri Yamaji People as the Traditional Owners and native title holders of the Observatory site. CSIRO’s ASKAP radio telescope is part of the Australia Telescope National Facility (https://ror.org/05qajvd42). Operation of ASKAP is funded by the Australian Government with support from the National Collaborative Research Infrastructure Strategy. ASKAP uses the resources of the Pawsey Supercomputing Research Centre. Establishment of ASKAP, Inyarrimanha Ilgari Bundara, the CSIRO Murchison Radio-astronomy Observatory and the Pawsey Supercomputing Research Centre are initiatives of the Australian Government, with support from the Government of Western Australia and the Science and Industry Endowment Fund. This paper includes archived data obtained through the CSIRO ASKAP Science Data Archive, CASDA (https://data.csiro.au).

LOFAR data products were provided by the LOFAR Surveys Key Science project (LSKSP; https://lofar-surveys.org/) and were derived from observations with the International LOFAR Telescope (ILT). LOFAR (van Haarlem et al. 2013) is the Low Frequency Array designed and constructed by ASTRON. It has observing, data processing, and data storage facilities in several countries, which are owned by various parties (each with their own funding sources), and which are collectively operated by the ILT foundation under a joint scientific policy. The efforts of the LSKSP have benefited from funding from the European Research Council, NOVA, NWO, CNRS-INSU, the SURF Co-operative, the UK Science and Technology Funding Council and the Jülich Supercomputing Centre.

This research made use of Astropy, a community-developed core Python package for Astronomy \citep{Astropy2018}.
\end{acknowledgements}

\end{appendix}
\end{document}